\documentclass[10pt,a4paper,twoside]{article}
\usepackage[utf8]{inputenc}
\usepackage[left=4.4cm,right=4.4cm]{geometry}
\usepackage{amsmath}
\usepackage{amsfonts}
\usepackage{amssymb}
\usepackage{bm}
\usepackage{graphicx}
\usepackage{subcaption}  
\usepackage{caption}
\usepackage{plain}  
\usepackage{nameref} 
\usepackage[round]{natbib}
\newcommand\ff{\frac{\Psi_{mk}}{2\pi G\sigma_0 a_R}}

\newcommand\frontmatter{
    \cleardoublepage
    \pagenumbering{roman}}
\newcommand\mainmatter{
    \cleardoublepage
    \pagenumbering{arabic}}
\newcommand{\drop}{1truecm}
\newcommand*{\titlenew}{\begingroup
\parindent=0pt
\vspace*{\drop}
\rule{12.2cm}{0.5mm}
\vskip 0.3truecm
{\Huge\bfseries Planetary Ring Dynamics}\\ [\baselineskip]
\vspace*{0.4truecm}
{\huge {The Streamline Formalism\\ 
{\LARGE {\itshape 1. From Boltzmann Equation to Celestial\\ \hspace*{0.6truecm} Mechanics}}
} 
}\par
\rule{12.2cm}{0.5mm}
\vfill
\begin{figure}[h]
    \centering
     \includegraphics[width=\textwidth]{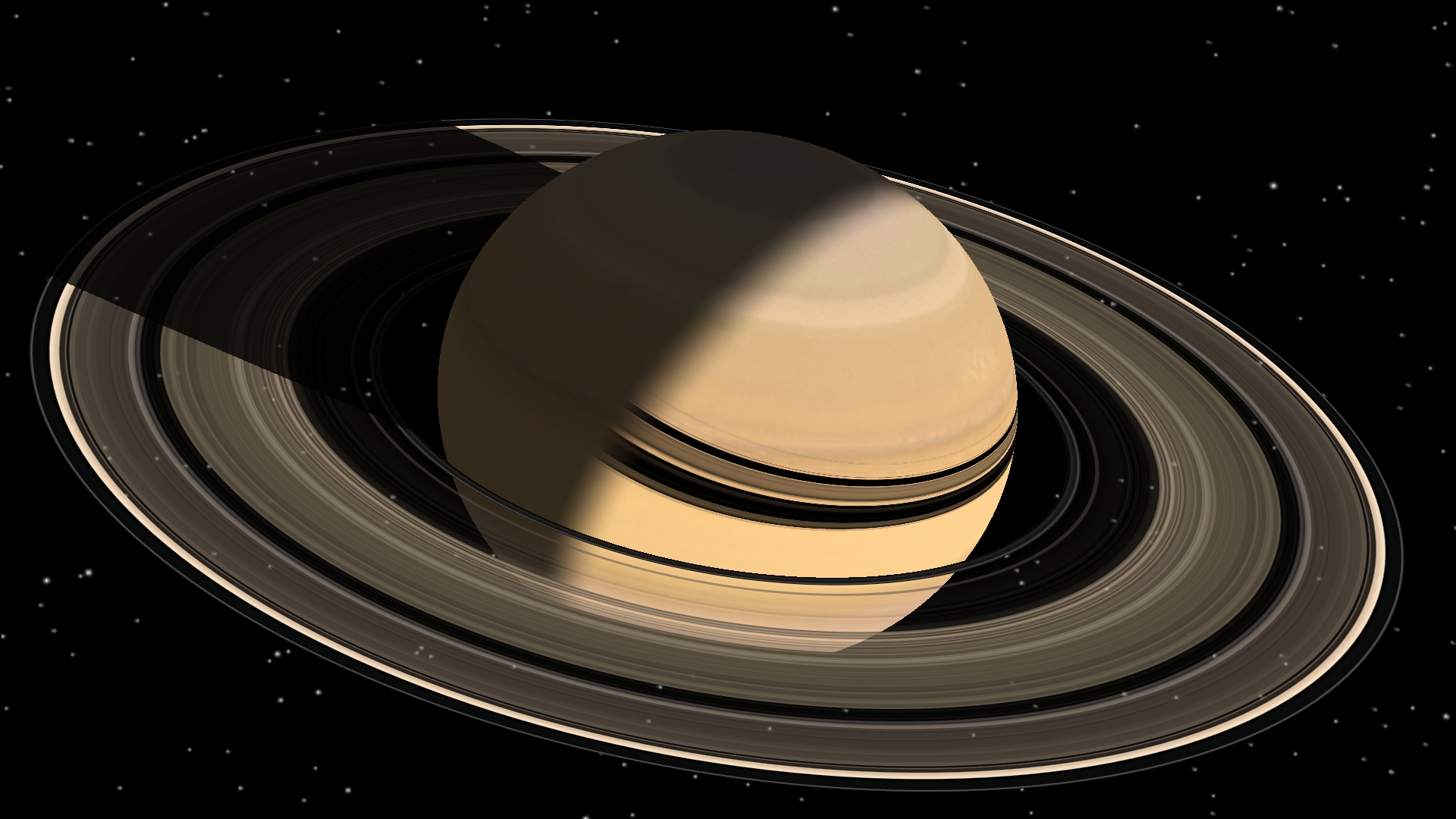}
\end{figure}
\vfill
{\Large{\itshape Pierre-Yves Longaretti}}\\ [\baselineskip] 
{\large IPAG, CNRS \& UGA\\}
\texttt{pierre-yves.longaretti@univ-grenoble-alpes.fr}\\
\rule{12.2cm}{0.3mm}
\begin{figure}[h]
    \centering
    \begin{subfigure}[b]{0.15\textwidth}
        \includegraphics[width=\textwidth]{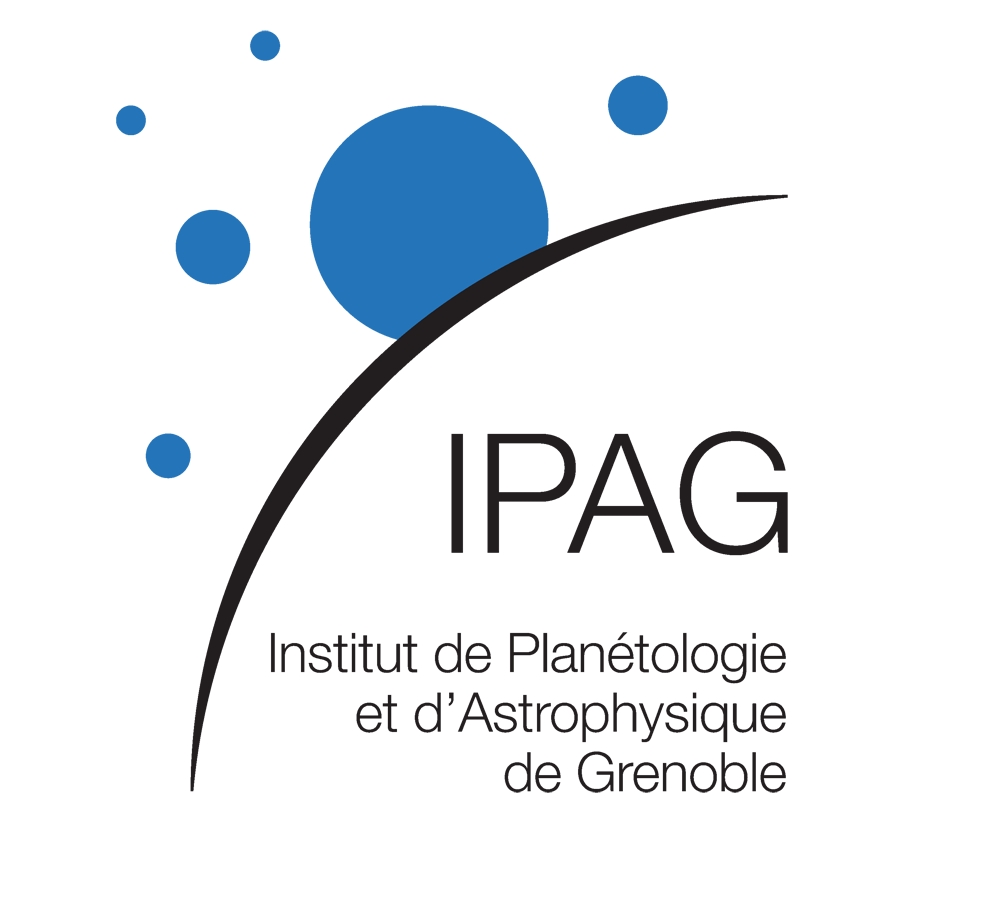}
    \end{subfigure}
    \hfill
    \begin{subfigure}[b]{0.13\textwidth}
        \includegraphics[width=\textwidth]{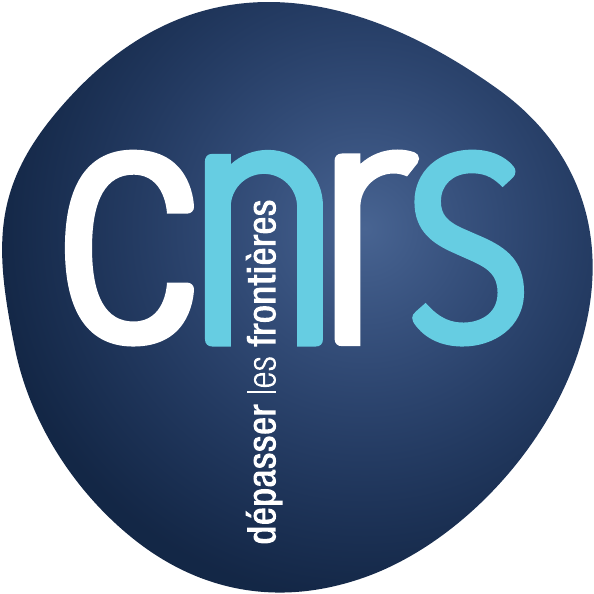}
    \end{subfigure}
    \hfill
    \begin{subfigure}[b]{0.2\textwidth}
        \includegraphics[width=\textwidth]{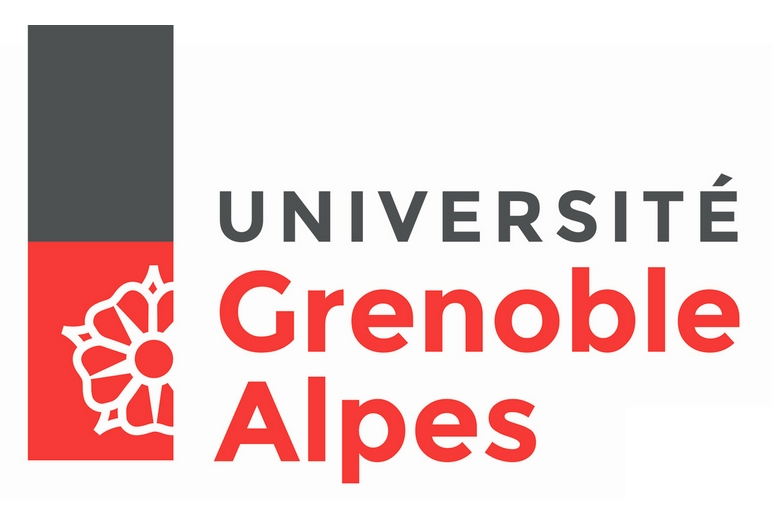}
    \end{subfigure}
\end{figure}
\endgroup}
\begin{document}
%
%
\titlenew
\thispagestyle{empty}
\clearpage
%
%
\newpage\null\thispagestyle{empty}\newpage
\frontmatter
\section*{}
\markboth{Contents}{}
\vspace*{-1cm}
\tableofcontents
\markboth{Contents}{}

\vfill
\noindent
Photo credits:\\
\noindent Cover: NASA\\
\noindent Backpage:
\begin{itemize}
\item Collection of early drawings of Saturn by various observers, from Huyghens' \textit{Systema Saturnium} (1659).
\item Guido Bonatti, from his \textit{De Astronomia Tractatus X Universum quod ad iudiciariam rationem nativitatum} (Basel, 1550)
\end{itemize}

\clearpage
\section*{Foreword}
\label{sec:foreword}
\addcontentsline{toc}{section}{\nameref{sec:foreword}}
\markboth{Foreword}{}

The present notes constitute a somewhat corrected and mildly extended --- through the addition of an appendix --- version of a set of lecture notes devoted to planetary ring dynamics and initially published in the \textit{Goutelas} series of volumes \citep{L92}. They were designed to form an overall but rather complete pedagogical introduction to their subject matter, i.e.\ to cover the material needed for a detailed understanding of the theoretical results published by Borderies, Goldreich and Tremaine --- but also Shu and coworkers to a smaller extent --- in a variety of research papers in the late 70s and in the 80s. However, some sections are more detailed than others.

These notes have never been widely distributed, and are almost impossible to find now. But although they are more than 20 years old, I believe or at least hope they might still be useful to a number of people in the field. The upcoming book (\textit{Planetary Ring Systems}), edited by Matt Tiscareno and Carl Murray and to be published later on this year at Cambridge University Press has been an incentive to post them on the Astrophysics Preprint Archive. In particular these notes are occasionally referenced in my own chapter of this book (\textit{Theory of Narrow Rings and Sharp Edges}). If you wish to refer to these notes, please quote the initial publication above along with their \textit{arXiv} number for accessibility.


Needless to say, these notes ignore recent advances in the understanding of ring macro- and microphysics, such as the existence of propellers, the more recent focus on viscous overstabilities, the existence and potential r\^ole of self-gravitational wakes, the now large albeit indirect body of evidence against the DEBs model of ring particles which was fashionable back in the 1980s, and of course the wealth of new observational constraints and theoretical challenges brought to light by the \textit{Cassini} mission. 

A number of these shortcomings are addressed in my chapter of the new ``ring book''. The interested reader will find in there both a complete although less detailed exposition than the present one of the streamline formalism and a thorough discussion of all the physics that has not found its way in the present notes, most notably the physics of narrow rings and sharp edges and associated modes, as indicated by the chapter's title. The literature review will be complete up to the time of publication. If permitted by the copyright agreement, this chapter will also be posted on \textit{arXiv} as part 2 of these notes. The two main topics covered in detail in the present notes and that have not been revisited in my ring book chapter are the ring microphysics theory and the theory of linear and nonlinear density waves.

\bigskip

Last point: please feel free to send me any feedback on these notes, if only to point out the unavoidable remaining mistakes.

\vskip 1truecm
\hfill \textit{Pierre-Yves Longaretti, June 1, 2016}

\clearpage
\newpage\null\thispagestyle{empty}\newpage
%
%
\mainmatter
\setcounter{figure}{0}    
\section{Introduction}
\label{sec:intro}

Ring systems are now known to exist around the four giant planets of the Solar System. However, they differ widely from one another. Jupiter's ring is very tenuous, and its constituting particles are
permanently destroyed and created. Saturn's rings constitute the
most extensive system: from the inside to the outside, one successively finds the D, C, B rings, the Cassini division, the A, F, G and E rings; rings D, F, G and E are rather tenuous, whereas the others are much more massive and more opaque. Uranus is surrounded by nine main rings of high optical depth, separated by dust bands. Finally, Neptune has a very peculiar system of very low optical depth rings, with some azimuthal structures known as ``ring arcs".

Rings are made up of particles of various sizes and compositions, and this property allows us to divide them roughly in two broad classes: the ``major" rings contain mostly ``big" particles (typically meter-sized); the ``etheral" rings are made of much smaller particles (typically microns). Saturn's main rings (A to C) and the nine main rings of Uranus ($\epsilon$, $\alpha$, $\beta$, $\gamma$, $\delta$, $\eta$, 4, 5, 6) belong to the first category. The other rings of Saturn and Uranus, as well as the rings of Jupiter and Neptune belong the second.

The two classes of rings thus defined exhibit very different dynamical behaviors, because light particles are submitted to important electromagnetic forces, which is not the case for meter-sized particles. In this lecture, we will restrict ourselves to the dynamics of the major ring systems. Of course, the distinction between the two types of rings is not absolute: ring particles have a broad size distribution, and micron-sized particles can also be found in the major rings; for example, they play an essential r\^ole in the well-known  ``spokes" phenomenon, mainly seen in Saturn's B ring. However, they dominate neither the mass nor the optical depth of these rings, and can therefore be ignored in discussions of their global dynamics.

The aim of these notes is to develop from first principles
a general approach to ring dynamical phenomena, usually referred to as  the ``streamline formalism". This formalism was introduced by \cite{BGT82} and used by these authors in their subsequent papers (e.g. \citealt{BGT83a,BGT83b,BGT85} and references therein). It is a fluid approach to ring dynamics, based on the first three moments of the Boltzmann equation, and making use of perturbation techniques developed in celestial mechanics; in this sense, this is a hybrid (but powerful) method. Quite a number of fundamental theoretical results were obtained in this framework, but its main assumptions, limitations, methods and domains of application are not easily extracted from the existing literature. These notes are designed to remedy these shortcomings, by providing the reader with all the building blocks of the formalism. Therefore, the writing is designed to be as self--contained  as possible, and the amount of material rather large. The reference list is purposely biased and limited, in the spirit of an introduction rather than a review. It is useful for  the readers to have some practical knowledge of celestial mechanics, and some notions of fluid dynamics and physical kinetics. Some background knowledge on rings is not strictly necessary, but certainly helpful.

In the next section, basic notions and orders of magnitudes are
presented. Section 3 introduces the Boltzmann equation and discusses the fundamental features of its application to ring systems. It is argued in these two sections that ring systems can be described with the Boltzmann moment equations, up to second order; that the velocity field and the corresponding variations of the ring density are mainly imposed  by the planet; and that the 
evolution of these two quantities takes place on a time-scale much
longer than the orbital time-scale, whereas the pressure tensor
(measuring the random motions) reaches steady-state on a time-scale
comparable to the orbital time-scale. Therefore, ring dynamical problems can be solved in several steps: first, the general form of the ring density and velocity field must be found. Then the  
ring pressure tensor corresponding to this form of the velocity field and ring density can be computed, using a (semi)-Lagrangian description. Finally, the long time-scale evolution of the velocity field and ring density can be found, by using standard
perturbation techniques inspired from Celestial Mechanics, as all the forces involved in the problem (the ring self-gravity, internal stress and satellite perturbations) are small compared to the gravitational attraction of the planet. This program is detailed in sections 4 through 7.  The basic concepts of ring kinematics and dynamics are elaborated in section 4. In particular, a simple and general parametrization of ring shapes, motions (streamlines) and densities is introduced, which constitutes the root of the streamline formalism. Solutions for the pressure tensor are given in Section 5, for both
dilute and dense systems, and some important general features of its behavior in rings are outlined. Section 6 presents the basic perturbation equations, borrowed from Celestial Mechanics, and provides a general discussion of ring mass, energy and angular 
momentum budgets. Some applications are described in Section 7, namely the now standard (but questioned) self-gravity model of narrow elliptic rings and the theory of density waves at Lindblad resonances. Section 8 presents a critical assessment of the scope and limitations of the formalism, as well as a quick overview of important issues in ring dynamics, and concludes these notes.

\section{Basic concepts and orders of magnitude}
\label{sec:basic}
  
The dynamics of the major rings, as we have defined them, is
dominated by the gravitational field of their parent planet. The motion of individual ring particles is  pertubed from an ellipse by interparticle collisions, the gravitational field of the other ring particles -- the disk self-gravity -- and the planet satellites.

Major rings all share a number of striking characteristics: they are flat and confined to the equatorial plane of their planet, they exhibit mostly axisymmetric features and the existing non-axisymmetric features show simple and regular patterns, they have complicated radial structures\dots In this section, we wish to develop some kind of heuristic understanding of the physical processes which generate these remarkable features.

The reason for the axisymmetry is easily understood. Any non-axisymmetric feature is quickly erased by the Keplerian shear supplied with a little diffusion. The existing non-axisymmetric features -- like density-waves, ring arcs, or the global eccentricity of some Uranian rings and some Saturn ringlets -- must have a dynamical origin (which is the reason for their regularity).

\subsection{Angular momentum axis}

Rings form thin circular disks because collisions dissipate energy 
while they conserve angular momentum. For an oblate planet, only the component of angular momentum parallel to the planet rotation axis is conserved, and the rings lie in the equatorial plane. A simple (although somewhat unrealistic) Celestial Mechanics argument can be made to illustrate this point.

Let us consider a situation in which the ring particles are 
initially confined to a plane different from the equatorial plane 
of the planet, on uninteracting, inclined circular orbits, as in Figure~1.

\begin{figure}[h]
\centering
\includegraphics[width=0.7\linewidth]{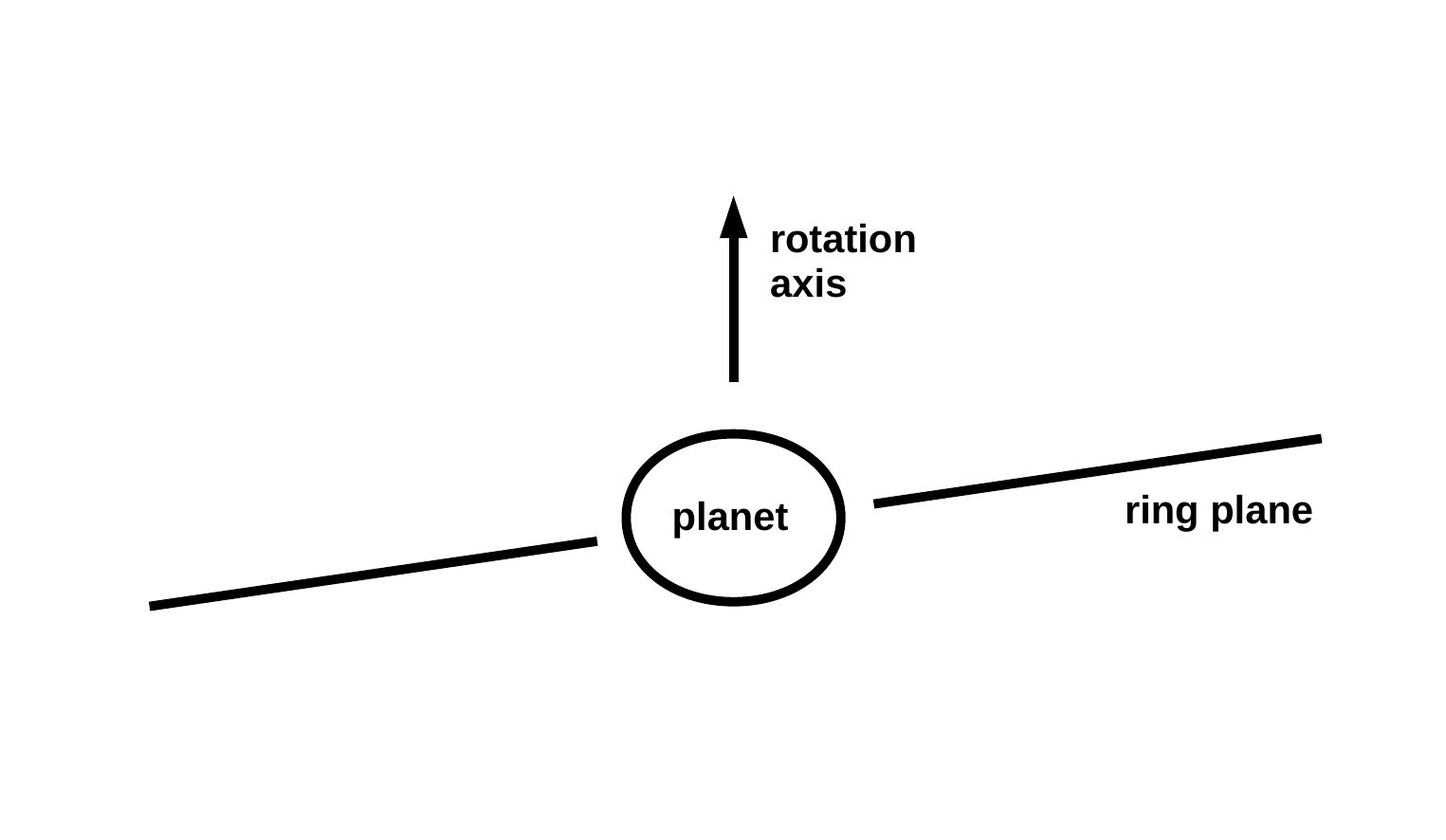}
\caption{\small{Sketch of an hypothetical ring initially inclined with respect to the equatorial plane of an oblate planet.}}
\label{fig:Fig1}
\end{figure}

In this situation the total angular momentum of the ring is not aligned with the axis of symmetry of the planet. Due to the planet-induced differential precession of the lines of
nodes, the system quickly falls into the stable situation represented on Figure~2. The total angular momentum of the disk has changed, and is now aligned with the symmetry axis of the planet. Because the system is isolated, some  angular momentum has been 
transferred to the planet, but the orientation of its spin axis has  basically not been affected, because it is much more massive than the rings. 

It is interesting to give an order of magnitude estimate of
the time-scale of this process, by considering the time required for two particle orbits of semimajor axes $a$, separated by 
the typical ring width $d$, to loose the correlation of their lines  of nodes\footnote{In narrow rings, this process can be prevented by the action of the ring self-gravity. However, the ring viscous stress associated with the Keplerian shear probably leads to the damping of the ring inclinations, as it does for the ring eccentricities, at least if no viscous overstability occurs.}. This differential precession time-scale $t_{pr}$ is then given by:

\begin{equation}
t_{pr}\sim\left[J_2 \Omega\left(\frac{d}{a}\right)\right]^{-1},\label{prec}
\end{equation}
where $\Omega$ is the angular frequency (mean motion) and
$J_2$ the oblateness of the planet.  It is interesting to have in mind typical orders of magnitude for the various quantities entering this formula: $J_2$ is of order $10^{-2}$ to $10^{-3}$, $d\sim a$ (in Saturn's rings), $a \sim$ $10^5$ km, and particle 
rotation periods around the planet are typically of the order of a few hours. For Saturn's rings, e.g., one obtains $t_{pr} \sim 10$ to $100$ days; this figure reaches $\sim 10^4$ to $10^5$ years for a 1~km wide ringlet.

\begin{figure}[h]
\centering
\includegraphics[width=0.7\linewidth]{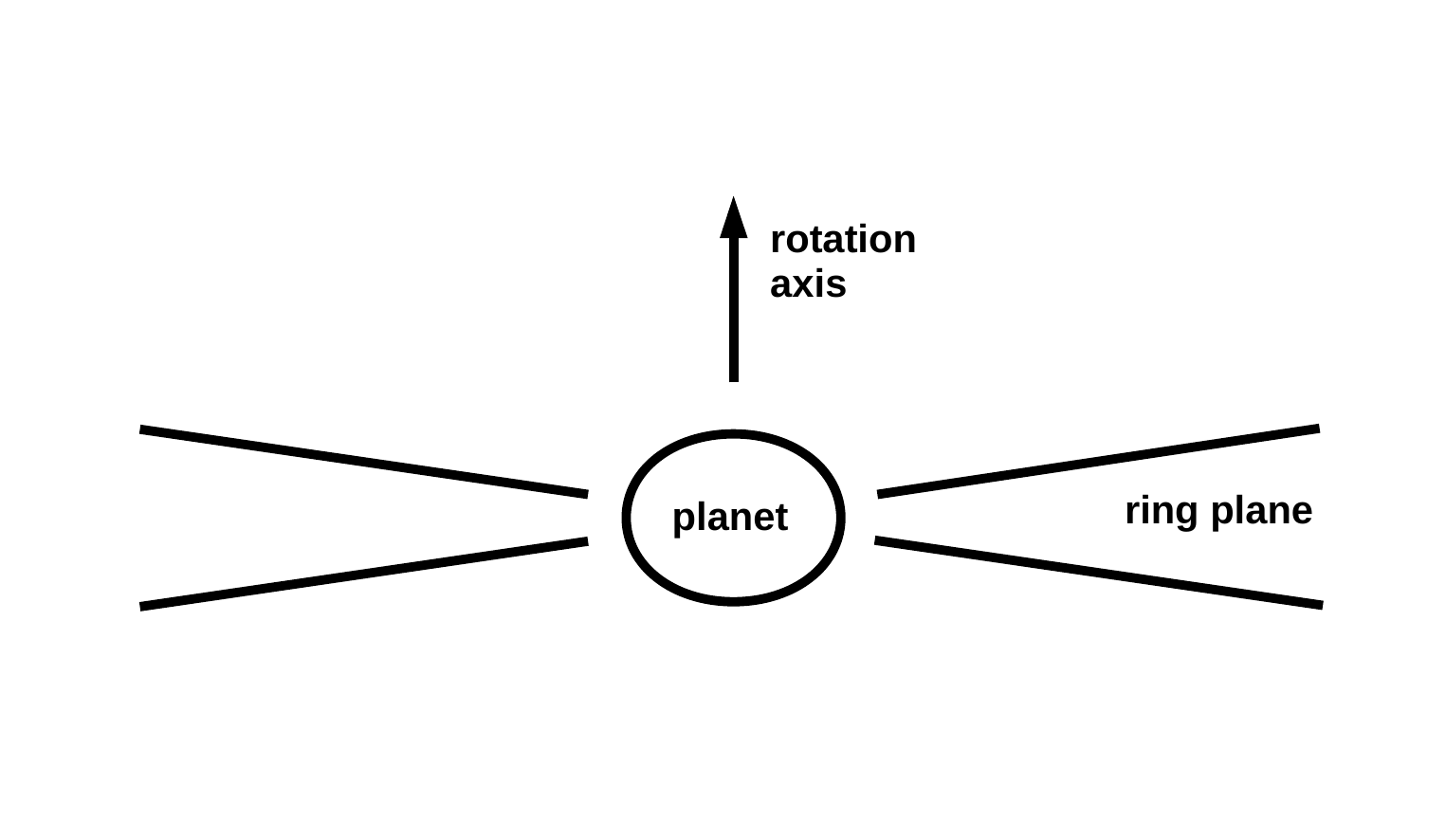}
\caption{Situation reached from the starting point of Figure~1 under the action of differential precession.}
\label{fig:Fig2}
\end{figure}

\subsection{Collisional quasi-equilibrium}
 
Assuming that an hypothetical ring has reached the state just described, there is no net vertical motion: the vertical velocity averaged over a large
number of ring particles is zero. The thickness of the ring measures the velocity dispersion of the ring particles\footnote{The velocity dispersion is in principle 
anisotropic. This anisotropy is ignored here.}. In reality, collisions affect this velocity dispersion and control the ring thickness, in various ways: 

\begin{enumerate}
\item Direct collisions are inelastic. Consider two particles on
a collision orbit with relative velocity $v_r$. During the collision, the relative velocity is reduced and one has 
\begin{equation}
v_r'\sim \epsilon v_r,\label{vbounce}
\end{equation}
where $v_r'$ is the post-collision relative velocity, and $\epsilon$ is the so-called ``{\it coefficient of restitution}\footnote{In fact one should define normal and tangential coefficients of restitution. The tangential coefficient is often ignored in the specialized literature, although it provides an important source of coupling between the translation and the spin degrees of freedom. However, such distinctions are not important for our order of magnitude estimates.}"; $\epsilon$ is a function of the relative velocity before encounter. Its
functional form depends strongly on the bulk and surface properties of the colliding materials, and is unfortunately not very well constrained for planetary ring particles; however it is usually a decreasing function of $v_r$, so that the energy lost is larger for larger relative impact velocities, as one would intuitively expect. The important point here is of course that the velocity dispersion of the particles (which is the source of their relative collision velocities) is damped due to the inelasticity of the collisions.
\item On the other hand collisions are as usual a source of random
motions: they scatter the particles. Furthermore in a sheared medium like planetary rings, collisions can transfer energy from the mean Keplerian motion to the random motions, thus increasing the velocity dispersion. This process appears macroscopically as the viscosity of the rings in the hydrodynamical approximation. 
\item As orbital energy is transferred into random motions some
of the ring material must fall on the planet. However as the total
angular momentum of the ring is conserved, some other fraction of the material must go away from the planet. In fact, collisions do actually transfer the orbital motion angular momentum from the inside to the outside, while most of the mass is transfered inwards, and the ring spreads.  Expressions for the rate of spreading will be given in Eq.~(2.11) and for the rate of angular
momentum transfer in sections 5 and 6.
\end{enumerate}

The net result of all these processes is that the velocity dispersion attains an equilibrium when the rate of transfer of energy from the orbital motion equals the rate of dissipation of energy during collisions. The equilibrium can be obtained in particular through the dependence of the coefficients of restitution on the relative collision velocities (but see also section 2.2.4). In the process, rings spread. Thus, very sharp or narrow features must be confined by some sort of dynamical agent. This is why the discovery of the complex radial structure of Saturn's and Uranus' rings was so surprising.

We are going to quantify somewhat these processes, but some important notions must be introduced first.

\subsubsection{Optical depth, collision frequency, and ring thickness}

Under the simplistic assumption that the ring particles all have the same radius $r$, the optical depth $\tau$ is related to the particle number density $n$, the ring thickness $H$ and the particle cross section $S\sim \pi r^2$ by:

\begin{equation}
\tau\sim n H S.\label{tau}
\end{equation}
Note that the particle number density $n$ is related to the ring surface density $\sigma$ by $nH\sim\sigma/m$ where $m$ is the mass of the particles.

One can also obtain an expression for the collision frequency as
follows. The number of collisions $N_c$ undergone by a given ring particle during a short time $\delta t$ is $N_c\sim nSc\delta t$ where $c$ is the velocity dispersion of the ring particles. Thus, the collision frequency $\omega_c$ (which is the inverse of the time $\delta t$ during which, on average, a particle experiences only one collision), is given by:

\begin{equation}
\omega_c\sim n S c.\label{collfreq}
\end{equation}

On the other hand, Figure~2 above shows that for small inclinations, the thickness $H$ of the ring is of order $a i$ where $i$ is the typical inclination of ring particles at distance $a$; furthermore, the typical vertical velocity of a ring particle on such an inclined orbit is typically of the order of the orbital velocity $\Omega a$ 
times the inclination $i$. Therefore, as the vertical velocity is of the order of the random velocity, one obtains:

\begin{equation}
c\sim \Omega a i\sim \Omega H.\label{veldisp}
\end{equation}
By combining Eqs.~\eqref{tau}, \eqref{vbounce} and \eqref{veldisp}, one obtains the following relation between the collision frequency and the optical depth\footnote{A more quantitative estimate is given in Section 5. It is also discussed there that the ring vertical self-gravity increases substantially the effective vertical oscillation frequency, leading to significantly higher collision frequencies in main ring systems}:

\begin{equation}
\omega_c\sim \Omega\tau.\label{collfreq2}
\end{equation}
It is again useful to have in mind some orders of magnitude for the
various quantities we have just introduced. The velocity dispersion is best estimated from the analysis of the damping of density and bending waves in Saturn's rings\footnote{In the hydrodynamic approximation, the damping is controlled by the ring 
viscosity, which can be related to the velocity dispersion; see below.}; this yields $c\sim$ a few mm/s. This corresponds to a ring thickness of a few tens of meters at most, which is $10^7$ times smaller than the typical radius of the rings; rings are {\it extremely} flat, and, correlatively, extremely cold\footnote{The ring Reynolds number is enormous, but they can't be turbulent, because their scale of granularity is comparable to the vertical scale; this is a major difference with other astrophysical disks.}. As the major rings have optical depths which are typically of order unity, the collision frequency is comparable with the orbital frequency (the orbital period is about a few hours).

\subsubsection{Ring viscosity}

The concept of viscosity is strictly speaking dependent on the
existence of a local stress-strain relation. It turns out that in rings, (and more generally for all fluids which cannot be described in the hydrodynamic limit) no such relation exists in general, but one can be found in the special (but important) case of axisymmetric flows. However, the concept of viscosity is convenient to discuss dissipation phenomena, and a heuristic derivation of the form of the viscous coefficient, due to \cite{GT78a}, is provided here. Microscopically, the kinematic viscosity coefficient can be expressed in terms of the collision frequency and the particles mean free path $l$ as [see, e.g., \citet{R65}]:

\begin{equation}
\nu \sim \omega_c l^2.\label{vis}
\end{equation}
If the optical depth $\tau$ is larger than unity, there are several
collisions per rotation period. The particles follow more or less
rectilinear trajectories between two collisions, and the mean free path is simply given by $l\sim c/\omega_c\sim c/\Omega\tau$. On the other hand, if the optical depth is smaller than unity, collisions occur only once in several rotation periods, and the particle paths are curved between two collisions. Note also that in this case, the mean free path and the transport coefficients
should become anisotropic and one should define viscosity coefficients for all directions. However, such complications are again ignored here. Then, according to what we just said, the mean free path is of the order of the radial excursion of a particle on 
its elliptic motion, i.e., the ring thickness: $l\sim H\sim c/\Omega$ [see Eq.~(2.5)]. Both regimes can be included in the same simple interpolation formula, which reads:

\begin{equation}
l\sim \frac{c}{\Omega(1+\tau^2)^{1/2}};\label{freepath}
\end{equation}
Replacing this expression in Eq.~\eqref{vis} and making use of Eq.~\eqref{collfreq2}
yields:

\begin{equation}
\nu\sim \frac{c^2\tau}{\Omega(1+\tau^2)};\label{vis2}
\end{equation}
Note that the same expression holds for the transverse viscosity 
of a plasma in a magnetic field, provided that the mean motion 
$\Omega$ is replaced by the plasma gyrofrequency.

\subsubsection{Velocity dispersion quasi-equilibrium}

We can now derive an approximate equation of evolution of the ring
velocity dispersion. The internal energy of the ring (kinetic energy of random motions) per unit mass is $c^2$. The rate of viscous transfer of energy per unit mass from the orbital motion to the random motions is $\sim\nu(ad\Omega/da)^2\sim\nu \Omega^2$ (for a simple justification, see \citealt{LL87}, p.~50, Eq.~(16.3), with the Keplerian circular velocity field, which has zero divergence). The rate of change of $c^2$ due to the energy loss during collisions is comparable to the change of the squares of the relative velocities of the colliding particles, i.e., to $v_r^2-v_r'^2\sim (1-\epsilon^2)v_r^2\sim (1-\epsilon^2)c^2$ [see Eq.~\eqref{vbounce}]. The typical collision time is of order
$1/\omega_c$, so that the rate of damping of $c^2$ is of order
$\omega_c(1-\epsilon^2)c^2$. Putting these two results together and
making use of Eq.~\eqref{collfreq2} leads to the desired equation of evolution of the velocity dispersion:

\begin{equation}
\frac{dc^2}{dt} \sim -\alpha \Omega\tau c^2(1-\epsilon^2)+\beta\frac{\Omega
c^2\tau}{1+\tau^2},\label{veldispeq}
\end{equation}
where coefficients of order unity, $\alpha$ and $\beta$, have
been introduced. This equation of evolution has two important
consequences:

\begin{enumerate}
\item There is only one time scale involved: the orbital time-scale.
Thus, the velocity dispersion reaches equilibrium on a time-scale
comparable to the particle orbital period, i.e. a few hours. This is 
exceedingly short. When the velocity dispersion equilibrium is reached,
the ring equilibrium thickness is also reached: thus the flattening
of the rings takes place in a few hours. This is much shorter than
the spreading time-scale $t_{sp}$, which is basically the time for 
a particle to random-walk across the ring:
\begin{equation}
t_{sp}\sim (R/l)^2\omega_c^{-1},\label{spread} 
\end{equation}
where $R$ is the ring radial extent. 

\item At equilibrium ($dc^2/dt=0$), the equation of evolution
yields a relation between the coefficient of restitution $\epsilon$ and
the optical depth $\tau$:
\begin{equation}
\epsilon^2\sim 1-\frac{\gamma}{1+\tau^2},\label{coefrest}
\end{equation}
where $\gamma$ is another coefficient of order unity ($\gamma\sim 
0.5$; see section 5). Note that $\epsilon$ is an increasing
function of $\tau$, the minimum of which is obtained for $\tau=0$; this
minimum is rather high (0.6 or 0.7) and is even higher when coupling
with the spin degrees of freedom is considered (citealt{S84,AT86}). This relation is known as the $\epsilon-\tau$ relation, and was derived by \cite{GT78a}, both in the heuristic manner presented here and in a more formal way by solving Boltzmann's equation. At first glance, this relation does not seem to involve the ring velocity dispersion. Remember however that the coefficient of restitution is a function of
the particles' velocity: $\epsilon=\epsilon(c)$. This determines in principle the magnitude of the velocity dispersion once the optical depth and the functional form of $\epsilon$ are known (which, let us recall, is not the case). Notice finally that the equilibrium is stable if $\epsilon$ is a decreasing function of the impact velocity; this is the case for all known materials. This result can be understood as follows. The rate of transfer of energy from the orbital energy into random motions is
proportional to $c^2$; the rate of dissipation of random energy is proportional to $(1-\epsilon^2)c^2$. If $c^2$ is, e.g., increased from, its equilibrium value, the collisions become more inelastic, $\epsilon$ is decreased, and the rate of loss of random motions wins over the injection from the shear: the disk is returned to equilibrium. A similar reasoning holds if $c^2$ is decreased rather than increased.
\end{enumerate}

\subsubsection{Limitations and extensions of the previous model}

The preceding analysis ignores a number of complications. We are not going to present them with as many details, but just outline the problems and their solutions when they are known, and refer the reader to the specialized literature.

First, we have ignored the possibility of having perturbed flows, i.e. flows for which the mean velocity is not circular. Such flows occur for example in the vicinity of resonances with the planet satellites. This type of flow will be fully discussed from section 4 onwards. 

We have implicitly assumed that the particles have the same size. It appears that the equilibrium velocity dispersion is roughly independent of the particle size, provided that the coefficient of restitution $\epsilon$ does not depend too strongly on $v$ (see e.g. \citealt{SL84}). Note also that the upper bounds to the ring thickness derived from the Voyager data argues also strongly in favor of an equilibrium velocity dispersion independent on the particle size (equipartition would lead to a ring thickness of the order several tens of kilometers whereas the Voyager data imply one or two hundred of meters at most). The single particle size model poses however one difficulty: the effective particle size, optical depth, collision frequency... are not unambiguously defined (this issue will be addressed in section 5). Note also that in most models, the particles are assumed spherical.

Gravitational binary encounters have been neglected. They also act as an effective particle scatterer. A heuristic discussion of the effect of these encounters can be found in \cite{CDBH79} (see also \citealt{SS85}). The main effect is on small particles, which are scattered out of the ring plane by large particles, with a velocity dispersion a few times larger than the velocity dispersion of the large particles. Note that because of the thin disk geometry, the main contribution is not due to the encounters with large impact parameters, in opposition to the
standard situation in galactic dynamics.

The ring particles are usually assumed indestructible. This is a crude assumption, especially that accretion and erosion processes are likely
to be very important in planetary ring dynamics. The problem
of the collisional evolution of the particle size distribution has been
addressed by a few authors (see \citealt{WCDG84,L89}), but no self-consistent model taking into account both the evolution of the size distribution and the velocity dispersion is
available.

  The coupling between the velocity dispersion and the spin degrees of
freedom has been neglected. A heuristic discussion of these effects can
be found in Shukhman 1984. Basically, a rough equipartition of energy is established between the velocity dispersion and the energy of rotation.

  Finally, one can wonder what happens if the ring material does not
allow the coefficient of restitution $\epsilon$ to be larger than the
minimum value required for the equilibrium to exist. This is quite
likely to be the case if the particles are regolith-covered, or are the
loose aggregates called {\it Dynamical Ephemeral Bodies}, as has been
suggested in the past few years (\citealt{WCDG84,L89})). The velocity
dispersion then decreases, and the ring becomes more compact until a
different regime is reached, in which the particles are no longer at
mutual distances much larger than their sizes. In this case, the collisional processes are dramatically altered. For example, the viscous transfer of momentum occurs not because the particles carry the momentum on the mean free path, but because it is carried across the particle itself; this gives rise to a minimum viscosity 

\begin{equation}
\nu \sim \Omega d^2,\label{numin} 
\end{equation}
where $d$ is the typical particle size \citep{B77,S84,AT86}. The properties of particle flows in such conditions have been investigated by \cite{BGT85}.

The reader will notice that none of these complications alters the basic physical phenomena described earlier: the equilibrium velocity dispersion is established on a time-scale comparable to the orbital time-scale; in the process, energy is permanently drawn from the orbital motion into random motions, and then dissipated as heat; as orbital energy is dissipated while angular momentum is conserved, rings spread. The only important issue is to know whether the ring particle properties are compatible with a ``thick" equilibrium in which the viscosity is given by Eq.~\eqref{vis2}, or lead to a ``thin" equilibrium for which it is reduced to its minimum value given by Eq.~\eqref{numin}; this last possibility seems most likely in high optical depth ring systems.

\subsection{Angular momentum transport and the origin of the rings radial structures}

Angular momentum transport is a key concept to the understanding of gross and fine features of planetary ring dynamics. There are two major causes of angular momentum transport in planetary rings: the ring viscous stress and the gravitational perturbations, either due to the ring itself or to some external agents, e.g. satellites.

To the lowest order of approximation, as we have just argued, the ring viscous stress transports the angular momentum from the inside to the outside\footnote{This is true only of unperturbed flows. In perturbed flows, the viscous flow of angular momentum can be reversed; see section 5.}. However, some interesting other phenomena can occur. The viscous luminosity of angular momentum (the rate of transfer of angular momentum through a given radius of the ring due to the viscous stress) is proportional to $\nu\sigma$ (cf section 5), so that the viscous torque on a ringlet is proportional to $d\nu\sigma/dr$. The reader can check that if $\partial \nu\sigma/\partial \sigma <0$, fluctuations in the ring density are amplified, because the viscous torques tend to push material from regions of low optical depths to regions of high optical depths. This is the case for example for viscosities of the form of Eq.~\eqref{vis2}, if one uses $\sigma\propto\tau$. Such a phenomenon is at the origin of the viscous and thermal instabilities which have been invoked to explain the complex radial structure of planetary rings (for a review, see \citealt{SLB84}). However, it appears that this such a possibility does not occur in dense rings (see e.g. \citealt{AT86,WT88})

On the other hand the gravitational interaction between rings and satellites often results in the creation of density waves, or at least in the creation of some kind of wake. This generates a net angular momentum exchange between the disk and the satellite, the angular momentum flowing again from the inside to the outside (\citealt{GT80}; see also sections 6 and 7): if the satellite lies inside the ring, it gives angular momentum to the ring; this is the case for example with the inner ``shepherd" satellite of the $\epsilon$ ring; the reverse is true in the opposite case. Thus a satellite and a ring repel one another. This phenomenon was invoked as a
possible confining process to explain the outer edges of Saturn's A and B ring, and also to explain the confinement of narrow rings \citep{GT79a,BGT82, BGT89}. One also sees that a small satellite situated in a ring ``repels" the ring material around it, by virtue of these angular momentum exchanges. The existence of a collection of kilometer-sized particles randomly distributed in the rings was therefore
envisioned as a possible cause to their radial structure.

Let us stress that these mechanisms are not the only possible source of structure in the rings\footnote{Density and bending waves are also well-identified sources of structure in Saturn's A and B rings.}. In the analysis of dense rings performed by \cite{AT86}, these authors argued that no viscous instability of the type described here actually occurred, but instead that some type of ``phase transition" between zero shear (solid) and high shear (fluid) regimes was possible, and suggested that this could also account for the ring complex radial structure. However, angular momentum exchanges (and energy exchanges)  are of fundamental importance, because they regulate the long term dynamics of the rings.

\section{The Boltzmann equation and its moments}

Rings are composed of countless particles. This suggests various types
of approach to study their dynamics. For example, one can treat them as a
collection of independent test particles. This yields some crude
estimates of some basic dynamical properties, but as a given particle
experiences many collisions during its life-time, it cannot give a
realistic description of ring evolution. The statistical character of
the effects of collisions on ring particles dynamics is in fact appropriately 
described by the Boltzmann equation, which applies to the evolution of
the particle distribution function $f({\bf r, v}, m, t)$. By
definition, $fd^3{\bf r}d^3{\bf v}dm$ is the number of particles of mass
$m$ at position ${\bf r}$, with velocity ${\bf v}$ in the elementary
seven-dimensional volume of phase space $d^3{\bf r}d^3{\bf v}dm$. One should
in principle consider that the distribution function depends on the
particle spin as well. This would however unnecessarily complicate the
analysis without affecting the basic principles that we wish to expose.
Therefore, the coupling with the spin degrees of freedom is ignored from
now on. For the same reason, the parameters describing the particle
shapes and surfaces (which are important for the collisional dynamics) are
also ignored.

  The Boltzmann equation reads:
  
\begin{plain}
$${\partial f\over\partial t}+{\bf v}.{\partial f\over\partial {\bf r}}
+{\bf\nabla}\phi.{\partial f\over\partial {\bf v}}=\left(\partial f\over
\partial t\right)_c,\eqno(3.1)$$
\end{plain}
where the right-hand side represents the effect of the
particle collisions on the evolution of the distribution function $f$.
The form of this collision term does not need to be further specified yet. 
The potential $\phi$
includes all the dynamical agents acting on the particles except the collisional
forces: the planet potential $\phi_p$, the potential arising from the
disk self-gravity $\phi_d$, and the satellite perturbations $\phi_s$. 

  The Boltzmann equation without collisions expresses the fact that the
six-dimensional flow of particles of a given mass in phase space is 
incompressible (the flow is confined to constant mass hypersurfaces). It
has the form of a continuity equation. The collision term acts as a
local source and sink term in phase space.

\subsection{The moments of the Boltzmann equation}
  The Boltzmann equation, involving six-dimensional derivatives, is in
general rather difficult to solve. This is why one usually prefers to
work with its moments. It is necessary to define first various local (in
physical space) average quantities: the mean density of particles $N$,
the mean velocity ${\bf u}$, the pressure tensor\footnote{The pressure tensor is the opposite of the stress tensor, and the two expressions may indifferently be used.} $p_{ij}$. these
quantities are defined as follows:

\begin{plain}
$$N=\int fd^3{\bf v},\eqno(3.2)$$
$${\bf u}={1\over N}\int {\bf v} fd^3{\bf v},\eqno(3.3)$$
$$p_{ij}=\int (v_i-u_i)(v_j-u_j)fd^3{\bf v},\eqno(3.4)$$
\end{plain}
\noindent where the subscripts $i,j$ refer to a cartesian
inertial reference frame. The integrals are performed over all velocity space.
One sees that these mean quantities are indeed the velocity moments of
the distribution function. Note that these quantities are defined per
unit particle mass, except for the mean velocity; for example, to retrieve 
the usual number density of particles, one needs to integrate $N$ over all 
masses. Also, $p_{ij}\sim Nc^2$ where $c$ is the mean square one
dimensional velocity dispersion,
defined by $3Nc^2\equiv p_{xx}+p_{yy}+p_{zz}$.

  By multiplying the Boltzmann equation by $1, v_i, v_iv_j$ and
integrating over the velocity space, one obtains respectively, after some
manipulations:

\begin{plain}
$${\partial N\over\partial t}+{\partial\over\partial x_i}(Nu_i)=
\left(\partial N\over\partial t\right)_c,\eqno(3.5)$$
$$N\left({\partial u_i\over\partial t}+ u_j{\partial u_i\over\partial x_j}
\right)
=-N{\partial \phi\over\partial x_i} -{\partial p_{ij}\over
\partial x_j} + \left(\partial N u_i\over\partial t\right)_c,\eqno(3.6)$$
$${\partial p_{ij}\over\partial t}+p_{ik}{\partial u_j\over\partial x_k}
+p_{jk}{\partial u_i\over\partial x_k} +{\partial\over\partial x_k}
(p_{ij}u_k)+{\partial q_{ijk}\over\partial x_k}=\left(\partial p_{ij}\over
\partial t\right)_c,\eqno(3.7)$$
\end{plain}
\noindent where $q_{ijk}$ is the tensor of the third order velocity
moments of the distribution function.

  Note that the equation of evolution of the moments of any order
depends on the moments to the next order. Thus, one needs some external
information in order to close the hierarchy of moment equations.
Fortunately, the special characteristics of ring systems provide us 
with quite a number of simplifying assumptions which make the problem
tractable.

\subsection{The moment equations in ring systems}

  Rings constitute cold media: the mean velocity ${\bf u}$, of the
order of the orbital velocity (a few km/s), is several orders of magnitude 
larger than the velocity dispersion $c$ (a few mm/s). 
Thus, the term involving the third order
moments $q_{ijk}\sim N c^3$ can be neglected in comparison with the
other terms of Eq.~(3.7), except maybe the $z$-derivative terms, because
of the small vertical scale height of the rings. From a hydrodynamical 
point of vue, this
approximation is equivalent to killing the heat conduction terms. It is
justified because the equilibrium of the internal heat of the ring
particle fluid (i.e., its velocity dispersion) is controlled by the
input from the shear, and the loss due to the inelasticity of the
collisions, as argued in the previous section, and not by heat
conduction phenomena, which occur on a much longer time-scale, except
possibly in the vertical direction where on the contrary they tend to
make the disk isothermal\footnote{Horizontal gradients can
also lead to a non negligible heat conduction term when the flow is
sufficiently perturbed from the axial symmetry, but this phenomenon is
neglected throughout these lecture.}.

  We have argued that the velocity dispersion is more or less size independent.
Furthermore, all the forces which can generate the mean velocity 
are gravitational, and therefore insensitive to the particle mass. Thus,
in first approximation, both {\bf u} and $p_{ij}$ do not depend on the
particle mass, and one can integrate Eqs.~(3.5) through (3.7) on the mass
parameter. Notice also that $\int m (\partial N/\partial t)_c dm = 0$,
because the collisions conserve the total local mass (the size
distribution evolves by creation, destruction accretion and erosion of
particles, but in all these processes, the total mass is locally
conserved). Therefore, multiplying Eq.~(3.5) by $m$ and integrating over mass yields:

\begin{plain}
$${\partial \rho\over\partial t}+{\partial\over\partial x_i}(\rho u_i)=0,
\eqno(3.8)$$
\end{plain}
\noindent where $\rho\equiv\int mNdm$ is the local mass density. This equation
has the standard form of a continuity equation. Let us also introduce
here, for future use, various mass averaged quantities. The mean mass of the 
distribution is defined by $M=\int mNdm/\int Ndm$; in
first approximation, $M$ is independent of the location in the rings. 
With this definition, $\rho=Mn$, where $n=\int Ndm$ is the local
particle density. One can also define a mass weighted
pressure tensor ${\mathrm p}_{ij}\equiv\int mp_{ij}dm=Mp_{ij}$ and a mass 
weighted third order moment ${\mathrm q}_{ijz}\equiv\int mq_{ijk}dm =Mq_{ijk}$.
  
  Collisions are locally momentum conserving if the particle size is
smaller than the mean free path (even if coupling with the spin degrees
of freedom is included). If the particle size is of the same
order as or larger than the mean free path, there is a non-local
collisional contribution to momentum transport, across the particle
size, because the particles behave as an almost incompressible medium in
comparison with the rings. However, one can show
that this contribution can be expressed as the divergence of a second
rank tensor (see Shukhman 1984). Therefore, it can be
simply taken into account by a suitably redefinition of $p_{ij}$, and
one can assume, without loss of generality, that $(\partial N
u_i/\partial t)_c=0$. Multiplying Eq. (3.6) by $m$ and integrating 
over mass yields, after division by $\rho$:

\begin{plain}
$${\partial u_i\over\partial t}+ u_j{\partial u_i\over\partial x_j}
=-{\partial \phi\over\partial x_i} -{1\over\rho}{\partial{\mathrm p}_{ij}\over
\partial x_j},\eqno(3.9)$$
\end{plain}
\noindent which has the standard form of an equation of fluid motion.
The physical meaning of the pressure tensor term in Eq. (3.9) is
well-known from Fluid Mechanics. Let us ignore the term due to $\phi$,
and compute the rate of change of momentum per unit volume,
$\partial\rho{\bf u}/\partial t$ from the momentum equation and the
continuity equation. Integrating over some arbitrary fixed volume in
space, one obtains $\partial/\partial t\int\!\!\!\int\!\!\!\int\rho u_i\ dV=
-\int\!\!\int({\mathrm p}_{ij}+\rho u_i u_j)dS_j$, where $d{\bf S}$ is an elementary
surface vector. The right-hand side of this expression is the flux of
momentum through the surface bounding the volume considered. The second
term is the momentum that the fluid crossing the surface during the
motion carries with it (the so-called advective term), the first term
characterizes the fact that the fluid outside the volume considered
exerts a force equal to $-({\bf n}.{\bf e}_i){\mathrm p}_{ij}{\bf e}_j$ on the
fluid in the volume considered per unit area of the bounding surface,
where ${\bf n}$ is the outside normal to the surface, and $\{ {\bf e}_i,
i=x,y,z\}$ are the unit vectors of a cartesian system of coordinates.
The existence of this term is natural, because ${\mathrm p}_{ij}$ is a measure of
the correlation of the random velocities in the ($i,j$) directions, and
is therefore similar to the $\rho u_i u_j$ term. Note
also that this contribution is a consequence of the existence of random
velocities, and not of collisions, although random velocities are often
the result of collisions.

Let us now complete our mass integrations by noting that Eq.~(3.7) is 
unchanged by the mass weighting process, except for the third order 
moment term, which reduces to $\partial {\mathrm q}_{ijz}/\partial
z$, and, possibly, for the collisional term.

   For completeness, let us mention the existence of a few processes
which do not conserve mass nor momentum locally. For example,
micron-sized particles are permanently accreted onto large particles and
formed from them. As these micron-sized particles are submitted to
electromagnetic forces as well as gravitational ones, they evolve
differently from the large ones. Such effects could introduce mass and
momentum source and sink terms, and require a multi-fluid description. 
However, we have already mentioned that
the ring fraction of mass included in these particles is very small, 
so that this contribution can be safely neglected.  
Also, ballistic transport processes can result in  mass and
momentum exchanges across large distances in the rings (for a
review, see \citealt{D84}). These processes are neglected in this lecture.

Finally, let us point out that the mass integration just performed is
not completely innocuous. Very big particles are quite underpopulated
with respect to smaller ones in rings, so that their mean separation can be very large. If one considers dynamical phenomena having characteristic scales smaller than the mean separation of the bigger
particles, the mass integration is not very meaningful. 

  This problem is specific to ring systems and cannot be found 
in ordinary fluids, because the individual particles, usually molecules, 
all have roughly the same size, much smaller than any scale of interest
for the macroscopic motions. In ring
systems, the size distribution spans several orders of magnitude...
There is also another major difference between ordinary fluids and
rings. In ordinary fluids -- i.e., in the so-called hydrodynamical
approximation -- the divergence of the pressure tensor in Eq.
(3.9) is reduced to the sum of a pressure term and of a viscous term, and
the pressure tensor evolution, Eq. (3.7), is reduced to the evolution of
its diagonal trace, which represents the internal energy of the fluid.
The third order moments of the form $q_{iij}$ are expressed as heat flux
terms. The system of equation is finally closed with the provision of an
equation of state (see also section 5). No such manipulation is possible 
in rings, because the pressure tensor is not symmetric enough. This is due 
to the fact that the particle paths are curved between collisions, whereas in
ordinary fluids, the collision time is always much shorter than the
dynamical time (or equivalently that the distance travelled between two
collisions is smaller than the system characteric scales).

  It is necessary for future use to recast the equations just obtained
in Lagrangian form. Eqs. (3.8), (3.9) and (3.7) constitute a Eulerian 
description
of ring systems: the various moments, $n$, ${\bf u}$, $p_{ij}$, are
functions of position and of time, and describe the state of the rings
at a given place and at a given moment. In the Lagrangian description,
the attention is focused on a {\it fluid} particle, i.e., a 
collection of ring particles. This collection must be large enough so
that its size is much larger than the typical particle size and particle
mean separation, but much smaller than the typical length-scale of the
phenomena under consideration. Also, the concept of fluid
particle can be meaningful only if the typical time for a ring particle
to random-walk out of the fluid particle is much larger than the
dynamical time-scale. One sees again that the existence of a broad
particle size distribution can in certain circumstances invalidate these
assumptions. In order to have a complete description of the fluid, one
wants to know at each time the position, density... of all its fluid
particles. Let us call ${\bf r}$ the position of a fluid particle, and
${\bf r}_0$ its position at some initial time $t_0$. The position of 
the fluid particle
depends necessarily on time. Also, for a given flow, the fluid particles
paths do not cross at any given time, so that the path is completely
specified if the initial position is known; thus the fluid particle
position depends on its initial position as well: 

\begin{plain}
$${\bf r}={\bf r}({\bf r}_0,t).\eqno(3.10)$$
\end{plain}

 With these definitions, the velocity of the flow at a
given time and location is equal to the velocity of the fluid 
particle located at the same place at the same time, and tangent 
to the fluid particle path:

\begin{plain}
$$\left(\partial{\bf r}\over\partial t\right)_{{\bf r}_0}={\bf u}({\bf
r},t)={\bf u}({\bf r}({\bf r}_0,t), t),\eqno(3.11)$$
\end{plain}

  By the same token, any fluid quantity $X$ (scalar, vector or tensor)
can be considered as a
function of either ${\bf r}$ and $t$ or of the fluid particle to which
it belongs, i.e. as a function of ${\bf r}_0$ and $t$ through Eq. (3.10). 
This enables us to introduce the concept of substantial derivative,
noted $D/Dt$:

\begin{plain}
$${DX\over Dt}\equiv \left(\partial X({\bf r}_0,t)
\over \partial t\right)_{{\bf r}_0}
={\partial X({\bf r},t)\over\partial t}+{\bf u}.{\partial X({\bf r},t)
\over\partial {\bf r}}.\eqno(3.12)$$
\end{plain}

  This substantial derivative, commonly used in Fluid Dynamics, 
expresses the change with time
of a given quantity $X$ along the flow, i.e. as it is carried by a fluid
particle along its path. Note that we have implicitly assumed that all 
quantities are defined in a ``smooth" way at the initial time $t_0$; for 
example, the initial velocity is assumed to vary
smoothly from one fluid particle to the next (it is a continuous
function of ${\bf r}_0$).

  We are now in position to recast Eqs.~(3.8), (3.9) and (3.7) in
Lagrangian form. They read:

\begin{plain}
$${D\rho\over D t}+\rho{\partial u_i\over\partial x_i}=0,\eqno(3.13)$$
$${D^2 r_i\over Dt^2}={D u_i\over Dt}=-{\partial \phi\over\partial
x_i}-{1\over\rho}{\partial {\mathrm p}_{ij}\over\partial x_j},\eqno(3.14)$$
$${D {\mathrm p}_{ij}\over Dt}+{\mathrm p}_{ik}{\partial u_j\over\partial x_k}
+{\mathrm p}_{jk}{\partial u_i\over\partial x_k} 
+{\mathrm p}_{ij}{\partial u_k\over\partial x_k}
+{\partial{\mathrm q}_{ijz}\over\partial z}
=\left(\partial {\mathrm p}_{ij}\over \partial t\right)_c.\eqno(3.15)$$
\end{plain}
\noindent where again all derivatives with respect to $x_i$ are evaluated
at ${\bf x}={\bf r}$ (this will be implicitly assumed in the remainder of
these notes).

  Further progress can be made by taking advantage of the extreme
thinness of ring systems, and of the dominance of the planet attraction
over the potential of the disk self-gravity, the perturbations of
the satellite\footnote{This is true even near a resonance with
a satellite.}, and the disk pressure tensor, except for the
determination of the ring vertical structure. Thus the
horizontal component of the velocity, which is mainly driven by the
planet, is nearly independent of the
vertical coordinate in the ring plane. A somewhat different result holds
for the vertical component, for the following reason. Let us consider for
example a ring in circular motion in the equatorial plane of 
its parent planet but with non
zero thickness. Typically, $H/r\sim 10^{-7}$. Thus, the variation of the 
vertical component of the planet force across the ring thickness is very
small, comparable to the vertical component of the pressure
tensor force, and even smaller than the ring self-gravity; the pressure 
tensor can then prevent the crossing of the fluid particle paths that 
the planet and the ring self-gravity would tend to impose in
the vertical direction. This physical constraint is often expressed
through a condition of vertical hydrostatic equilibrium both in ring
dynamics and in accretion disk theory. In this case, the vertical
velocity is nearly independent of the vertical coordinate; it 
essentially is equal to zero,
except for example when inclined satellites excite coherent vertical
motions of the ring plane (this is the case for bending waves).
Alternatively, when the ring particles
are in a ``thin" equilibrium (in the sense defined in section 2), i.e.
when the ring particles have mean separations comparable to their 
radii, the
hydrostatic condition has to be replaced with the constraint of
incompressibility of the three dimensional flow (see \citealt{BGT85} and section 5). 

  In any case, because of the remarkable thinness of the rings, 
it is useful to vertically integrate the preceding equations, and 
forget about the precise vertical structure. In this operation, all
fluid particles with different $z$ are replaced by an equivalent
fluid particle whose vertical coordinate is equal to the mean plane height.
Let us call ${\bf R}=$($R,\Theta, \mathcal{Z}$) the position of this equivalent fluid
particle; $\mathcal{Z}=\int z\rho dz/\sigma$, and the vertically integrated
equation of motion Eq. (3.14) reduces to the 
equation of motion of this equivalent particle. For
simplicity, we will assume here that the condition of vertical hydrostatic
equilibrium holds\footnote{One could think at first
glance that this assumption is not very good for perturbed flows,
because the horizontal and vertical orbit perturbations
seem to have the same period. However, this is not true: in the vertical
direction, the self-gravity of the ring is much larger than the
restoring force of the planet, which is not the case in the horizontal
direction. Therefore, the vertical effective epicyclic frequency is
about ten times larger than the horizontal one, and the vertical
structure adjusts on a time-scale much shorter than the horizontal
perturbation time-scale (see section 5).}; similar results are obtained
in the incompressible case (see 
section 5). The integration of the continuity 
equation and of the equations of fluid motion is greatly simplified
by the fact that the velocity is nearly independent on the vertical
coordinate in comparison with the density $\rho$. Therefore, in cylindrical 
coordinates ($r,\theta,z$), the vertically integrated continuity equation 
reads:

\begin{plain}
$${D\sigma\over Dt}+ \sigma{\partial u_r\over\partial r}+
{\sigma\over r}{\partial u_\theta\over\partial\theta}=0,\eqno(3.16)$$
\end{plain}
\noindent where $\sigma\equiv\int \rho dz$ is the surface density of the
ring.

Multiplying Eq. (3.14) by $\rho$, integrating over $z$ and dividing the
resulting equations by $\sigma$ yields:

\begin{plain}
$${D^2 R\over Dt^2}-R\left(D\Theta\over Dt\right)^2=-{\partial\phi_0\over
\partial r}-{1\over\sigma}\left[{1\over r}{\partial(rP_{rr})\over\partial
r}+{1\over r}{\partial P_{r\theta}\over\partial\theta}-
{P_{\theta\theta}\over r}\right],\eqno(3.17)$$
$${1\over R}{D\over Dt}\left(R^2{D\Theta\over Dt}\right)=
-{1\over r}{\partial\phi_0\over\partial\theta}-{1\over\sigma}
\left[{1\over r^2}{\partial(r^2 P_{r\theta})\over\partial r} +
{1\over r}{\partial
P_{\theta\theta}\over\partial\theta}\right],\eqno(3.18)$$
$${D^2 \mathcal{Z}\over Dt^2}=-\left(\partial\phi\over\partial z\right)_{z=\mathcal{Z}}
-{1\over\sigma}\left[{1\over r}{\partial (r P_{rz})\over\partial r}+
{1\over r}{\partial P_{\theta z}\over\partial\theta}\right],\eqno(3.19)$$
\end{plain}
\noindent where $P_{ij}\equiv \int {\mathrm p}_{ij}dz$ are the vertically
integrated components of the pressure tensor, and where
$\phi_0=\int\rho\phi dz/\sigma$ is the vertically averaged potential
(function of $\mathcal{Z}$). Due to the small thickness of the ring,
$\phi_0\simeq\phi$. In Eq. (3.19) we have used the constraint of vertical
hydrostatic equilibrium with respect to the mean plane, which reads
$\partial{\mathrm p}_{zz}/\partial z=-\rho[\partial\phi/\partial z - (\partial 
\phi/\partial z)_{z=\mathcal{Z}}]$, and $D\mathcal{Z}/Dt=\int u_z\rho dz/\sigma$, which
follows from Eqs. (3.13) and (3.16) (for details, see \citealt{SS85}).

  Finally, Eq. (3.15) yields:

\begin{plain}
$${D P_{ij}\over Dt}+P_{ik}{\partial u_j\over\partial x_k}
+P_{jk}{\partial u_i\over\partial x_k} 
+P_{ij}{\partial u_k\over\partial x_k}
=\left(\partial P_{ij}\over \partial t\right)_c.\eqno(3.20)$$
\end{plain}
\noindent where all terms involving $z$ derivatives have been removed by
the vertical integration. In Eqs. (3.17) through (3.20), all quantities
depending on the spatial coordinates are evaluated at ${\bf R}$.

Note that when there are no vertical motions ($D^2 \mathcal{Z}/Dt^2 =0$), the
rings are in principle symmetric with respect to the equatorial plane, so that $P_{rz}=0$ and $P_{\theta z}=0$, and Eq.~(3.19) is trivially
satisfied.

   It is useful to recast Eq.~(3.20) in cylindrical coordinates. There
are several ways to perform the change of variables. On can make it directly 
on Eq.~(3.20); or directly on the Boltzmann equation itself Eq.~(3.1) and
compute afterwards the moment equations; or use tensor calculus, by
replacing the partial derivatives in cartesian coordinates by covariant
derivatives in cylindrical coordinates and compute the Christoffel
symbols. The latter is probably the fastest. In any case, we give the
resulting equations for $P_{rr}, P_{r\theta}, P_{\theta\theta}$ and
$P_{zz}$ (we will see in section 5 that the remaining equations are not
needed for our purposes, but the interested reader can find them in the
appendix B of \cite{SS85}):

\begin{plain}
$${D P_{rr}\over Dt}+P_{rr}\left(3{\partial u_r\over\partial
r}+{u_r\over r}+{1\over r}{\partial
u_{\theta}\over\partial\theta}\right)+2P_{r\theta}\left({1\over
r}{\partial u_r\over\partial\theta}-{2\over r}u_\theta\right)$$
$$\hspace{5truecm}=
\left(\partial P_{rr}\over\partial t\right)_c,\eqno(3.21)$$

$${D P_{\theta\theta}\over Dt}+P_{\theta\theta}
\left({\partial u_r\over\partial r}+{3u_r\over r}+{3\over r}{\partial
u_{\theta}\over\partial\theta}\right) +{2\over r}P_{r\theta}
{\partial\over \partial r}(r u_\theta)$$
$$\hspace{5truecm}=
\left(\partial P_{\theta\theta}\over\partial t\right)_c,\eqno(3.22)$$

$${D P_{r\theta}\over Dt}+2P_{r\theta}\left({\partial u_r\over\partial
r}+{u_r\over r}+{1\over r}{\partial
u_{\theta}\over\partial\theta}\right)+
{1\over r}P_{rr}{\partial\over\partial r}(ru_\theta)+
P_{\theta\theta}\left({1\over
r}{\partial u_r\over\partial\theta}-{2\over r}u_\theta\right)$$
$$\hspace{5truecm} =
\left(\partial P_{r\theta}\over\partial t\right)_c,\eqno(3.23)$$

$${D P_{zz}\over Dt}+P_{zz}\left({\partial u_r\over\partial
r}+{u_r\over r}+{1\over r}{\partial
u_{\theta}\over\partial\theta}\right) +2P_{zr}{\partial\over\partial r}\left(D\mathcal{Z}\over Dt\right)+ 2P_{\theta z}{1\over
r}{\partial\over\partial\theta}\left(D\mathcal{Z}\over Dt\right)$$
$$\hspace{5truecm}=
\left(\partial P_{zz}\over\partial t\right)_c,\eqno(3.24)$$
\end{plain}

   Let us conclude this section with a final fundamental comment.
We have already pointed out that all forces are small compared to the
planet's, except in the vertical direction, but the question of the
vertical structure has just been evicted by the vertical integration. 
This means also that the dynamical time-scale imposed by the planet
is much shorter than the dynamical evolution time-scale due to the
disk self-gravity, the satellite perturbations, or the pressure tensor.
Furthermore, we have also argued in section 2 that the pressure tensor
reaches steady-state on a time scale comparable to the orbital time
scale. This means that {\it the pressure tensor components reach
steady-state for steady-state values of the mean velocity ${\bf u}$ 
and surface density $\sigma$ mainly imposed by the
planet}. Therefore, the theory of ring dynamics can be developed
according to the following scheme:

\begin{enumerate}
\item The motion is first solved when the planet force is the only
one acting on the rings.
\item The $P_{ij}$ are then found by solving Eq.~(3.21) with
steady-state ${\bf u}$ and $\sigma$ appropriately chosen. 
\item The others forces (self-gravity, satellite perturbations and
pressure tensor) are treated as perturbations. When they drives a slow
evolution of the surface density and velocity field of the rings, the
pressure tensor evolution is of course ``enslaved" to this evolution.
\end{enumerate}

  This program forms the basis of the streamline formalism and is
described in the next sections.

\section{Ring kinematics: streamlines}

Our primary objective is to find the general solution of Eqs.~(3.17) and (3.18) when the potential is reduced to the planetary potential, and when
the viscous stress terms are neglected. Thus, the problem at hand is
analogous to the motion of a test particle around a planet,
except that it applies to a fluid particle rather than to an individual
particle. We refer to this problem as to the ``fluid test particle
motion" in the remainder of these notes.  For convenience, we drop the
capital letter notation for the position of the equivalent fluid
particle.

  Furthermore, for simplicity, we
will restrict ourselves to motions confined to the equatorial plane of
the planet, i.e., $z=0$ and Eq.~(3.19) does not need to be considered.
This restriction eliminates the dynamics of bending waves and inclined
rings from the analysis. However, bending waves do not differ very much
from density waves, and inclined rings are similar in their dynamical
properties to eccentric rings, so that this restriction is not
essential while simplifying the exposition of the method.

  Note that the total derivative ($d/dt$) and the substantial derivative
($D/Dt$) have very similar meanings: the total time derivative of Classical 
Mechanics (resp. the substantial derivative) refers to the motion of a given 
point (resp. fluid) particle under given initial conditions. It is
therefore customary to write Eqs.~(3.17) through (3.19) with usual total
derivatives instead of substantial ones, and reserve the substantial
derivative symbol to express the time variation along the fluid particle
paths of quantities expressed in usual Eulerian coordinates. We will 
follow this custom in the remainder
of these notes, but the reader should not forget that we are solving a
fluid problem and not a point particle one.

\subsection{Fluid test particle motion}

  In the conditions just outlined, the equation of motion of the fluid
test particle reads:

\begin{plain}
$${d^2{\bf r}\over dt^2}=-\nabla\phi_p.\eqno(4.1)$$
\end{plain}
\noindent where ${\bf r}$ is restricted to the equatorial plane of the
planet. This equation is formally identical to the equation
of motion of a point particle in the planet potential; the only
difference is that ${\bf r}$ is a function of the initial position as
well as of time. This equation can therefore be solved with standard
techniques. It is usual in Celestial Mechanics to use the elliptic
solution of the two-body problem and treat the deviations of the planet
from spherical symmetry as perturbations. This procedure raises however
quite a number of subtle technical issues which will be briefly
described in section 4.2. We will therefore depart from 
this well-established custom; however, the solution used here is
closely related to the elliptic solution, and everyone already used to 
Celestial Mechanics techniques will feel at ease with it. Note that the
equation of motion
Eq.~(4.1) does not depend on the surface density, so that it can be solved
independently of the continuity equation.

  The solution we will use derives from the epicyclic theory, which was
initially developed for galactic dynamics. A fundamental feature of ring
problems is that although the density contrast can be strongly nonlinear,
the deviations from circular trajectories are always very small (see in
particular section 4.4). The analysis therefore relies on the
existence of purely circular solutions (the planet is axisymmetric), and
looks for general solutions in the form of small deviations around one
of these circular solutions, in successive approximations. 
Up to second order in deviation from
circularity, the radius $r$ and true longitude $\theta$ of a fluid test
particle on an equatorial orbit read \citep{LB91}:

\begin{plain}
$$r=r_0\left[1+{3\eta^2\over 2\kappa^2}\epsilon^2 -
\epsilon\cos\xi-{\eta^2\over 2\kappa^2}\epsilon^2\cos 
2\xi\right],\eqno(4.2)$$
$$\theta=\gamma+{2\Omega\over\kappa}\epsilon\sin\xi+
{\Omega\over 2\kappa}
\left( {3\over 2}+{\eta^2\over\kappa^2}\right)\epsilon^2
\sin 2\xi,\eqno(4.3)$$
\end{plain}

\noindent with 

\begin{plain}
$$\xi=\int_0^t\kappa dt+\delta=\kappa t +\delta,\eqno(4.4)$$
$$\gamma=\theta_0+\int_0^t\Omega\left[1+\left({3\over 2}-{3\eta^2\over\kappa^2}
\right)\epsilon^2\right]dt   
=\theta_0$$
$$\hspace{2truecm}+\Omega\left[1+\left({3\over 2}-{3\eta^2\over\kappa^2}
\right)\epsilon^2\right]t,\eqno(4.5)$$   
$$\Omega^2\equiv {1\over r_0}{d\phi_p\over dr}(r_0),\eqno(4.6)$$
$$\kappa^2\equiv \left[{3\over r}{d\phi_p\over dr}+
                       {d^2\phi_p\over dr^2}\right]_{r=r_0},\eqno(4.7)$$
$$\eta^2\equiv \left[{2\over r}{d\phi_p\over dr}-
                      {r\over 6}{d^3\phi_p\over 
                      dr^3}\right]_{r=r_0},\eqno(4.8)$$
\end{plain}

\noindent where $r_0, \epsilon, \theta_0$, and $\delta$ are 
the constants of integration of the problem, and are functions of the
initial position of the fluid particle ${\bf r}_0$ (do not get confused
between $r_0$ and ${\bf r}_0$ !!): $r_0$ is the radius of the
circular motion around which the solution is expanded (analogous to an
average radius); $\epsilon$ is the relative departure from circularity
(analogous to an eccentricity); in ring problems, $\epsilon$ is always a small quantity; $\theta_0$ and $\delta$ are initial
phases; $\Omega$ and $\kappa$ are the usual rotation and 
epicyclic frequencies, i.e. the frequencies of the motion around the
planet and of the radial oscillations, respectively; 
$\eta$ is an auxiliary quantity homogeneous to a
frequency; $\phi_p$ is the potential of the planet, 
including the $J_k$ terms (the coefficients of the expansion
of the planetary potential in spherical harmonics\footnote{In some
cases, it is useful to include the axisymmetric contributions of the
satellites and of the rings in the definition of $\phi_p$.}). In the limit of a 
spherical planet, $\Omega^2=\kappa^2=\eta^2=GM_p/r_0^3$, where $M_p$ 
is the mass of the planet, so that in general, the three 
frequencies differ only by terms of order $J_2$ and smaller. General expressions for these frequencies in terms of the $J_k$s can be found in \cite{BL87}. Expressions in terms of $J_2$ are given below.

These expressions constitute the epicyclic solution as it is generally
derived; however, the close analogy with the elliptic solution is much
more apparent if one makes use of a somewhat different set of epicyclic
elements\footnote{The usefulness of the change of variable
from $r_0$ to $a_e$ was pointed out by Phil Nicholson 1990 (private
communication).}: {$a_e$, $\epsilon$, $\varpi_e$, $ M_e\equiv\xi$ }, where 
$a_e$ and $\varpi_e$ are defined by:

\begin{plain}
$$a_e\equiv {r_0\over{1-\epsilon^2}},\eqno(4.9)$$
$$\varpi_e\equiv\gamma-\xi=\gamma- M_e.\eqno(4.10)$$
\end{plain}

  The change of notation from $\xi$ to $ M_e$ is adopted for mnemonic
reasons which will soon be made obvious. Keeping terms up to second 
order in $\epsilon$, the preceding formul\ae\ become:

\begin{plain}
$$r=a_e\left[1+\left({3\eta_a^2\over 2\kappa_a^2}-1\right)\epsilon^2 -
\epsilon\cos  M_e -{\eta_a^2\over 2\kappa_a^2}\epsilon^2\cos 
2 M_e\right],\eqno(4.11)$$
$$\theta=\varpi_e+ M_e+{2\Omega_a\over\kappa_a}\epsilon\sin  M_e+
{\Omega_a\over 2\kappa_a}
\left( {3\over 2}+{\eta_a^2\over\kappa_a^2}\right)\epsilon^2
\sin 2 M_e,\eqno(4.12)$$
\end{plain}

\noindent where now the frequencies are evaluated at $a_e$, so 
that\footnote{Note that the terms of order
$\epsilon^2\kappa_a t$ are not explicitly given in Eq.~(4.13). Such terms
cannot be self-consistently computed from a second order epicyclic
theory, because a third order expansion produces a nonlinear frequency 
correction of the same magnitude to $ M_e$. This correction
actually kills the contribution of order $\epsilon^2$ to $\kappa_a$, so
that the remaining contribution is of order $J_2\epsilon^2$. Note also that
in this case, $\Omega_a$ and $\kappa_a$ differ by a term of order 
$J_2\epsilon^2$. The knowledge of this contribution is important for
some ring problems, but is not needed in these notes.}:

\begin{plain}
$$ M_e=\int_0^t\kappa_a dt+\delta+O(J_2\epsilon^2),\eqno(4.13)$$
$$\gamma=\theta_0+\int_0^t \Omega_a\left[1+\left({7\over 2}-
{3\eta_a^2\over\kappa_a^2}-{\kappa_a^2\over\Omega_a^2}
\right)\epsilon^2\right]dt,\eqno(4.14)$$   
$$\Omega_a^2\equiv {1\over a_e}{d\phi_p\over dr}(a_e),\eqno(4.15)$$
$$\kappa_a^2\equiv \left[{3\over r}{d\phi_p\over dr}+
                       {d^2\phi_p\over dr^2}\right]_{r=a_e},\eqno(4.16)$$
$$\eta_a^2\equiv \left[{2\over r}{d\phi_p\over dr}-
                      {r\over 6}{d^3\phi_p\over 
                      dr^3}\right]_{r=a_e}.\eqno(4.17)$$
\end{plain}

For definiteness, let us give expressions for $\Phi_p$, $\Omega_a$ and $\kappa_a$ in terms of $J_2$ ($\eta_a$ is not needed):

\begin{eqnarray}
\Phi_p(a_e) & = &\frac{G M_p}{a_e} \times  \left[-1+\frac{1}{2}\left(\frac{R_p}{a_e}\right)^2 J_2 \right],\nonumber\\
\Omega_a(a_e) & = & \quad n_a\ \times  \left[\ \  1+\frac{3}{4}\left(\frac{R_p}{a_e}\right)^2 J_2\right],\nonumber\\
\kappa_a(a_e) & = & \quad n_a\ \times  \left[\ \ 1-\frac{3}{4}\left(\frac{R_p}{a_e}\right)^2 J_2\right],\nonumber
\end{eqnarray}

\noindent where $M_P$ and $R_p$ are the planet mass and radius. For future use, we have also introduced an effective elliptic mean motion defined by $n_a=(GM_p/a_e^3)^{1/2}$.
 
The analogy is apparent if one expands the elliptic solution to second
order in eccentricity:

\begin{plain}
$$r=a\left[1+{1\over 2}e^2 -e\cos M-
                  {1\over 2}e^2\cos 2M\right],\eqno(4.18)$$
$$\theta=\varpi+M+2e\sin M + {5\over 4}e^2\sin 2M,\eqno(4.19)$$
\end{plain}

\noindent with standard notations for the elliptic elements. If one
makes the following formal identifications:

\begin{plain}
$$a\rightarrow a_e,$$
$$e\rightarrow \epsilon,$$
$$M\rightarrow  M_e,$$
$$\varpi\rightarrow\varpi_e,$$
\end{plain}

\noindent one sees that the two types of expansions are formally nearly
identical; the only difference comes from the ratios of frequencies 
in the epicyclic solution, which differ from unity by terms of order
$J_2$. In the epicyclic solution, the precession due 
the deviations of the planet from sphericity is of course
already included: $\dot\varpi_e = \Omega_a -\kappa_a \sim J_2 \Omega_a$. 
For convenience, we will refer to the elements ($a_e, \epsilon,
 M_e, \varpi_e$) as to the epicyclic semi-major axis, eccentricity, mean
motion and periapse angle respectively. These elements are also, of
course, function of the initial position of the fluid particle. They
represent the average of the epicyclic elements of the individual ring
particles composing the fluid particle under consideration. The
individual particle epicyclic eccentricities are made up of two
contributions: this mean part,
and a random contribution which is connected to the velocity second
moments (the pressure tensor) of the fluid particle.

  To conclude this subsection, let us write down the perturbation
equations of the epicyclic elements. Such equations are necessary as we
have decided to treat all forces but the planet's as perturbations.
Due to the formal analogy between the point particle equations of motion and the Lagrangian equations we have derived, the perturbation equations can be obtained with standard variation of the constants techniques. They read \citep{LB91}:

\begin{plain}
$${da_e\over dt}={2\over\kappa_a}\left[R\epsilon\sin M_e+{\Omega_a\over
\kappa_a}S\left(1+\epsilon\cos M_e\right)\right]+O(\epsilon^2),\eqno(4.20)$$
$${d\epsilon\over dt}={1\over \kappa_a a_e}\left[R\sin  M_e +
                      2{\Omega_a\over\kappa_a}S\cos 
                       M_e\right]+O(\epsilon),\eqno(4.21)$$
$${d\varpi_e\over dt}=\Omega_a - \kappa_a
+{1\over\kappa_a a_e\epsilon}\left[-R\cos M_e
+2{\Omega_a\over\kappa_a}S\sin M_e\right]+O(\epsilon^0),\eqno(4.22)$$
$$\eqalignno{{d M_e\over dt}=\kappa_a+&{1\over \kappa_a a_e\epsilon}
\biggl[R\biggl.\left(\cos M_e -3{\eta_a^2\over\kappa_a^2}\epsilon+{\eta_a^2\over\kappa_a^2}
\epsilon\cos 2 M_e\right)-\cr 
&{\Omega_a\over\kappa_a}S\left(2\sin M_e+
\left({1\over 2}+{\eta_a^2\over\kappa_a^2}\right)\epsilon\sin
2 M_e\right)\biggl.\biggr]+O(\epsilon),&(4.23)\cr}$$
\end{plain}

 The $\Omega_a-\kappa_a$ term in Eq.~(4.22) represents 
the effect of the planet oblateness on the precession of the apses. For comparison, let us write down the equations of perturbation of the
elliptic motion (see, e.g., \citealt{MD99}), to first order in eccentricity:

\begin{plain}
$${da\over dt}={2\over n}\left[Re\sin M + S\left(1+e\cos
M\right)\right]+O(e^2),\eqno(4.24)$$
$${de\over dt}={1\over na}\left[R\sin M + 2S\cos 
M\right]+O(e),\eqno(4.25)$$
$${d\varpi\over dt}={1\over nae}\left[-R\cos M + 2S\sin 
M\right]+O(e^0),\eqno(4.26)$$
$$\eqalignno{{dM\over dt}=n+{1\over nae}\biggl[ \biggr.
&R\left(\cos M -3e +e\cos 2M\right)\cr 
&-S\left(2\sin M +{3\over 2}e\sin 2M\right)\biggl. \biggr]+O(e),&(4.27)\cr}$$
\end{plain}

\noindent with standard notations. In both sets of equations,
$R$ and $S$ are the radial and tangential components of the perturbing
acceleration, as usual. Notice the similar roles played by
 $n$ and $\kappa_a$. One sees here again that the perturbation equations for
elliptic and epicyclic variables are formally nearly identical, except
for the various ratios of frequencies in the epicyclic equations, which
differ from unity by terms of order $J_2\ (\ll 1)$.

In what follows these relations are used to first order in eccentricity at most. As a consequence, $d\varphi/dt=d(\varpi_e+M_e)/dt$ is negligible compared to $d\varpi_e/dt$, and only the first free equations are needed in practice. Because $R,S$ are small (compared to the planet's acceleration) and $\epsilon\ll 1$, one can replace all frequency ratios with $1$ in these relations; we also replace everywhere $\kappa_a$ and $\Omega_a$ by the effective mean motion $n_a$ except in $\Omega_a - \kappa_a$, to the same level of precision. In this limit, these relations become formally identical to their elliptic counterpart, except for the notable fact that $a_e$ and $\epsilon$ differ from the osculating elliptic ones by terms of order $J_2$. This simplification is usually made from now on.

The comparison of the expressions of the specific
energy and angular momentum in terms of the elliptic and epicyclic
elements is also useful.
In elliptic variables, the specific energy and angular 
momentum are related to the semimajor axis and the 
eccentricity by the well-known formul\ae:

\begin{plain}
$$E=-{G M_p\over 2 a},\eqno(4.28)$$
$$H=\sqrt{G M_p a(1-e^2)}= na^2\left(1-{e^2\over 2}\right)+O(e^4)
,\eqno(4.29)$$
\end{plain}

\noindent whereas in epicyclic variables, we have:

\begin{plain}
$$E=\phi_p(a_e)+{\Omega_a^2a_e^2\over 2}+O(\epsilon^4),\eqno(4.30)$$
$$H=\Omega_a a_e^2\left[1-{1\over 2}\left(\kappa_a\over\Omega_a\right)^2
\epsilon^2\right]+O(\epsilon^4).\eqno(4.31)$$
\end{plain}

It is easily seen that Eqs.~(4.28) and (4.29) are recovered in the case of
a spherical planet. Note that $r_0$ (resp. $a_e$) is the radius of the
circular orbit having the same angular momentum (resp. the same energy
to order $\epsilon^4$) as the non-circular epicyclic orbit of
Eqs.~(4.2) and (4.3) [resp.\~(4.11) and (4.12)].

\subsection{Epicyclic versus elliptic elements}

Although elliptic elements are much more widely known 
and almost exclusively used in Celestial Mechanics in general 
and in the ``ring community" in particular, epicyclic elements
are much more adapted to the observational and theoretical 
descriptions of planetary rings for the three following 
reasons:

\begin{enumerate}
\item First, the fluid particle trajectories  in a circular ring 
are described by the simple equation $r={\mathrm constant}=a_e$ in the 
streamline formalism (see section 4.3).
However, it is well-known that although the ring fluid particles
follow a circular trajectory, their elliptic osculating eccentricity 
$e_0$ is non zero; $e_0=3/2 J_2 (R/r)^2$ to leading order in 
the gravitational coefficients of the planet potential. 
Furthermore, the eccentricities 
which are typically considered in some ring problems, e.g. {\it the 
mean eccentricities} which are involved in the description of 
density waves in Saturn's rings, {\it can be orders of magnitudes smaller
than the osculating eccentricities}, which are then of order $e_0$
\citep{LB86}. Similarly,
the osculating semi-major axis $a_0\simeq r[1+3/2J_2(R/r)^2]$ is substantially
different in absolute value from its ``mean" value $r$, although not
in relative value. Second, for non-circular motions, the osculating elements 
exhibit short-period variations due to the harmonic 
coefficients of the planet, whereas the elements 
used in the streamline formalism, as well as in data fits, are supposed to be 
time-independent, or, to the very least, to vary only on
much longer time scales than the orbital period. This time variation 
cannot be ignored: it is at the origin of this sometimes very large
discrepency between the osculating and observed elements. 

\item Epicyclic elements are ``more constant" 
than elliptic elements: the non--sphericity of the
planet is readily taken into account in the epicyclic 
formul\ae\ whereas, as already pointed out, it leads to short-period 
variations of the elliptic elements.  Furthermore, the
non-sphericity of the planet is the most important 
source of short-period variations. The main effects of the 
shepherd satellites, and of the ring self-gravity and pressure 
tensor (which perturb both elliptic and epicyclic elements)
occur on much longer time-scales (their short-period contributions are
negligible). The argument 
developed in this paragraph can be rephrased and summarized 
in a different way: {\it the mean (short-period averaged) and 
the osculating elliptic elements of a ring particle can be substantially
different, whereas its mean (short-period averaged) and osculating epicyclic 
elements are always identical or nearly identical.}

\item No approximation with respect to the harmonic 
coefficients of the planet is involved, which is not the 
case for elliptic elements. On the other hand, elliptic 
formul\ae\ are valid to all orders in eccentricity, but in 
practice, expansions in eccentricity (and inclination) are
always required, and the exactitude of the elliptic formul\ae\
turns out to be no great 
advantage, because epicyclic formul\ae\ can easily be obtained 
to the order needed for a good description of the data.
\end{enumerate}

  Up to now the equations which have been used both in data fits and in
dynamical analyses are the equations of the elliptic motion, although
they were applied to elements which were assumed to be
constant at least on the short time-scale, and known to be quite
different from the elliptic osculating elements in some circumstances.
It is therefore legitimate to
wonder why this procedure was valid, especially at the light of the
comments above. However, we have shown [Eqs.~(4.11) to (4.27)] that the
elliptic elements are formally nearly identical to a suitably defined
set of epicyclic element. This argument combined with the three points
exposed above shows that (i) the elements
obtained from the observations are indeed the epicyclic elements $a_e,
\epsilon,  M_e, \varpi_e$ and (ii) the application of elliptic 
formul\ae\ and equations to these epicyclic elements generally
constitutes an acceptable approximation.

\subsection{Ring streamlines and kinematics}

 In fluid dynamics, the streamlines are the
lines of the velocity field of the fluid. In the streamline formalism,
the word is sometimes used in a different way. Most of the times, it 
designates the actual streamlines of the flow, at least in some suitably 
defined rotating frame. For example, this is the case for density waves 
and eccentric rings, but the $m=0$ mode of the $\gamma$ ring of Uranus 
is a notable exception. In all cases, the streamlines provide a
description of ring shapes: more precisely, they designate the curve that 
an infinitely thin ring would define in space. Therefore, to conform with 
past usage, we will use the word ``streamline" to designate both ring
velocity field lines and ring shapes, keeping in mind that when the two
concepts do not overlap, the word refers to the latter, in opposition to
the more common usage.

It is customary when treating a fluid dynamics problem to look for solutions with a specific space and time dependence. For example, in the analysis of fluid stability in the linear approximation, one often looks for oscillating solutions with phases of the form $kx -\omega t$ for a one-dimensional problem, $s$ being the space variable. The same approach is assumed in ring problems, with two notable subtleties attached:
\begin{enumerate}
\item The celestial mechanics perturbation technique adopted in the streamline formalism implies to define not only the streamline shapes, but also the defining quantities of each individual particle in a given streamline. As the unperturbed background is not at rest but is constituted by circular motions around the planet, this makes the \textit{a priori} specification of the shape of the motion somewhat more involved than usual.
\item One looks for \textit{nonlinear} solutions in general, but with a specific form of nonlinearity that makes them analytically tractable to a larger extent. In disk systems in general, one can identify two types of nonlinearity: large radial extension, and large density variations. The two are not necessarily coupled, and it turns out that, in rings, deviations from circularity are always quite small, but the associated density contrast can be quite large. This will be discussed in section \ref{sec:surf}
\end{enumerate}

  Let us start with the trivial case: a circular ring, in which the fluid
particles are in circular motion\footnote{We
have argued in section 2 that, because interparticle collisions are
dissipative, orbital energy is permanently lost, so that no fluid
particle can be exactly in circular motion, but let us ignore this
complication for the time being, as we have not yet included the effect
of the viscous stress in our analysis. In any case, this is a small
effect, occurring on the longest of all the time-scales of interest in
ring dynamics.}. The fluid particles' positions
reduce to:

\begin{plain}
$$ r = a_e,\eqno(4.32)$$
$$ \theta = \varphi \equiv \varpi_e +  M_e.\eqno(4.33)$$
\end{plain}

\noindent where $\varphi$ is the epicyclic mean longitude. One sees also that at any given time, to any given fluid particle with
initial position ${\bf r}_0$ corresponds a unique set ($a_e,
\varphi$). This suggests that these quantities
  can be chosen as (semi-)Lagrangian 
labels instead of the initial position ${\bf r}_0$ if needed.
This is also obviously true for the more general eccentric solution, 
and in the rest of these notes, it is considered when needed that
the epicyclic elements $\epsilon$, $\varpi_e$ and $ M_e$ are functions 
of $a_e$ and $\varphi$ considered as Lagrangian 
labels (naturally, $\varpi_e$ and $ M_e$ are functions of time as well
in the fluid test particle solution). Of course, $a_e$ and $\varphi$ 
are independent Lagrangian variables: the change of variable from 
${\bf r}_0$ to ($a_e, \varphi$) is nowhere singular at any given
time. Note that in non steady-state flows, the change of variable from ${\bf
r}_0$ to ($a_e, \varphi$) may be time dependent.

\subsubsection{Eccentric rings}

  Let us now consider some more general cases. Usually, in Eulerian fluid 
dynamics,
 one is interested in special solutions of the Navier-Stokes equations.
The situation is similar here, because it is observationally found that
rings can be well described by some special form of the general
solution we have just written down. For example, the shape of the
elliptic Uranian rings is known from the analysis of stellar
occultation data (for a review, see \citealt{EN84}). 
Their mean eccentricity is typically of the order of
$10^{-3}$ to $10^{-4}$; they also present a difference of eccentricity 
between the inner and outer edges, of the order of $10^{-4}$ to 
$10^{-5}$ \citep{FEL86,Fetal88}. The 
analysis of the data strongly suggests that the ring fluid
particles having the same (epicyclic) semimajor-axis also have the same
(epicyclic) eccentricity and the same (epicyclic) periapse angle. This
means that the ring shape, which in this case
coincide with the ring fluid particle streamlines -- is
parametrized as follows, to first order in eccentricity [combine Eqs.
(4.11) and (4.12)], as in Figure~3 below:

\begin{plain}
$$r(a_e,\phi)=a_e\left[1-\epsilon(a_e)\cos\left(\theta-\varpi_e(a_e)\right)\right].
\eqno(4.34)$$
\end{plain}

\noindent In this type of solution, the epicyclic elements $\epsilon$ and $\varpi_e$ 
do not depend on the Lagrangian coordinate $\varphi$. Note that if such
a dependence existed, it would generate an increased shear, and
therefore, in usual situations, be quickly erased by viscous
forces, unless it were maintained by some dynamical agent.
This is also true for eccentric rings, which tend to generally tend to
become circular under the action of the ring viscous stress.
Thus, the eccentricities of elliptic rings need generally
speaking to be maintained by some external agents (for example the
shepherd satellites), unless the viscous stress has such an unusual form
that viscous instabilities can take place and generate them (see section 7.1).

   Note also that the dependence of the precession rate 
on the fluid particle semimajor axis implies that the inner 
edge streamline of elliptic rings tends to precess faster than the outer one
under the action of the planet. In the absence of other
forces, the alignment  of the inner and outer apses would be very quickly
destroyed, streamlines would cross and the ring would become circular. 
Therefore, this differential precession must be balanced by some
dynamical agent, e.g. the ring self-gravity (\citealt{GT79a,GT79b}; see section 7.1).

\subsubsection{Density waves}

  It is also instructive to consider the case of density waves. Such
waves are excited by the satellites near resonance locations, and
propagate away from the resonance. They are sustained by the
self-gravity of the disk, arise from the coherent radial excursions
of the ring fluid particles, and look stationary in a particular
rotating frame. The eccentricities involved are typically
of order $10^{-3}$ to $10^{-5}$. The form of the fluid particle paths
has been known for a long time in galactic dynamics from the work of
Lindblad on spiral galaxies. The usual parametrization is most conveniently 
understood from the following elementary analysis, reproduced from 
\cite{GT82}.

  Let us consider the forced linear response of a test particle in
circular orbit around an oblate planet, perturbed by a satellite. Expressing
the radius and longitude of the test particle as $r=r_0+r_1$ with
$r_1\ll r_0$ and
$\theta=\theta_0+\Omega t +\theta_1$, where $r_0$ is the radius of the
circular orbit and $\Omega$ is the orbital frequency, given by Eq.~(4.6),
the linearized equations of motion for $r_1$ and $\theta_1$ read:

\begin{plain}
$${d^2 r_1\over dt^2}+r_0\left(d\Omega^2\over dr\right)_{r_0}r_1-
2r_0\Omega_0{d\theta_1\over dt}=-
\left(\partial\phi_s\over\partial r\right)_{{\bf r}_0},\eqno(4.35)$$
$$r_0^2{d^2\theta_1\over dt^2}+2r_0\Omega_0{dr_1\over dt}=-
\left(\partial\phi_s\over\partial\theta\right)_{{\bf r}_0},\eqno(4.36)$$
\end{plain}

\noindent where $\phi_s$ is the satellite potential. The
satellite is supposed to orbit in the equatorial plane of the planet.
Let us call $a_s$ its (epicyclic) semimajor axis, $e_s$ its eccentricity, 
and ${\varpi_s}$ its periapse angle; $\kappa_s$ is the epicyclic 
frequency evaluated at $a_s$. At any given time, the satellite
potential is of course periodic in azimuth. Furthermore, in a frame
rotating at $\dot\varpi_s$, the satellite orbit is closed of period
$2\pi/\kappa_s$. Thus, the satellite potential can be expanded in a
double Fourier series, one in time, and one in azimuth; this yields\footnote{The phase of the satellite has been taken equal to 0 at
$t=0$ (the origin of time $t=0$ is chosen when the satellite is at periapse); also, corotation resonances with $a=a_s$ are excluded from the
expansion of Eq.~(4.37): $r<a_s$ is assumed.}:

\begin{plain}
$$\phi_s(r,\theta,t)=\sum_{m=0}^\infty\sum_{k=-\infty}^\infty
\Phi_{mk}\left({r/a_s}\right)\cos\left[m(\theta-\dot\varpi_s t)-
(m+k)\kappa_s t\right].\eqno(4.37)$$
\end{plain}

The Fourier coefficients $\Phi_{mk}$ are expressed in terms of Laplace
coefficients (\citealt{MD99}; a
particularly synthetic and convenient derivation can be found in the
appendix A of \citealt{Sh84}):

\begin{plain}
$$b_{1/2}^m (\alpha)={2\over\pi}\int_0^{\pi}{\cos mu\ du\over(1-2\alpha
\cos u +\alpha^2)^{1/2}}.\eqno(4.38)$$
\end{plain}

The coefficients $\Phi_{mk}$ are of order 
$e_s^{|k|}$, so that only
the Fourier coefficients with small $k$ are important in practice,
because usually $e_s\ll 1$. Let us define $\alpha=r/a_s$, and call $M_s$
the mass of the satellite. The terms of order $|k|\le 1$ read:

\begin{plain}
$$\Phi_{m0}=-{GM_s\over a_s}{b_{1/2}^m(\alpha)-\delta_{m1}\alpha\over
1+\delta_{m0}},\eqno(4.39)$$
$$\Phi_{m,\pm1}=-{GM_s e_s\over a_s}{\left[{1\over 2}\left(1\pm 2m
+\alpha{d\over d\alpha}\right)b_{1/2}^m(\alpha)-\alpha\delta_{m1}(1\pm
1)\right]\over 1+\delta_{m0}},\eqno(4.40)$$
\end{plain}

\noindent where $\delta_{ij}$ is the Kronecker delta symbol (the
contribution of the indirect term is included).

  For a single Fourier component, the solution of the linearized
equations of motion reads:

\begin{plain}
$$r_1=\left\{ {\cos[m(\Omega-\Omega_p)t+m\theta_0]\over{m^2(\Omega
-\Omega_p)^2-
\kappa^2}}\left({d\Phi_{mk}\over dr}+{2m\Omega\over m(\Omega-\Omega_p)r}
\Phi_{mk}\right)\right\} _{r_0},\eqno(4.41)$$
\end{plain}

\begin{plain}
$$\eqalignno{\theta_1=&-\left\{ {\sin [m(\Omega-\Omega_p)t+m\theta_0]\over
m^2(\Omega-\Omega_p)^2-\kappa^2}\left({2\Omega\over m(\Omega-\Omega_p)r}
{d\Phi_{mk}\over dr}\right.\right.+\cr 
& \left.\left.\left[{4\Omega^2-\kappa^2\over
m^2(\Omega-\Omega_p)^2}+1\right]{m\Phi_{mk}\over
r^2}\right)\right\}_{r_0},&(4.42)\cr}$$
\end{plain}

\noindent where $\Omega_p = \Omega_s + k/m\ \kappa_s$ is the so-called
pattern speed: it is the angular speed of the frame in which the $(m,k)$
potential component is stationary. The linear response is singular
either when $\Omega_p=\Omega_0$ (corotation resonance) or when $\kappa_0
=\pm m(\Omega_0-\Omega_p)$ (Lindblad resonance). The corotation
resonances that lie within a ring arise because the satellite orbit is
eccentric, and have $|k|\ge 1$; the $k=0$ corotation resonance (the
strongest) occurs at the satellite radius. The inner Lindblad resonance 
occurs inside the corotation resonance, and corresponds to the positive
sign; the other one is the outer Lindblad resonance, and lies outside.
If the multipole moments of the planet potential are neglected, the
Lindblad (resp. corotation) resonance condition reduces to 
$\Omega_0/\Omega_s=(m+k)/ (m\mp 1)$ (resp. $\Omega_0/\Omega_s=m+k/m$).
This ratio is often used to label a resonance. For example,
the $k=0$, $m=2$ inner Lindblad 
resonance is called a $2:1$ resonance (the outer edge
of the Saturn's B ring corresponds to such a resonance with the
satellite Mimas). 

  The resonance condition implicitly defines the resonance radius $r_R$,
which is the only radius for which the condition is satisfied. Near a
resonance, the radial perturbation $r_1$ can be expressed as a function
of $\theta$ and of the distance to the resonance $\Delta r=r_0-r_R$. For
a corotation resonance, we have:

\begin{plain}
$$r_1\simeq {A^c_{mk}\over \Delta r}\cos m(\theta-\Omega_p t),\eqno(4.43)$$
\end{plain}

\noindent with

\begin{plain}
$$A^c_{mk}=\left({4\Phi_{mk}\over 3\Omega^2}\right)_{r_R},\eqno(4.44)$$
\end{plain}

\noindent whereas for a Lindblad resonance, one obtains:

\begin{plain}
$$r_1\simeq {A^L_{mk}\over \Delta r}\cos m(\theta-\Omega_p t),\eqno(4.45)$$
\end{plain}

\begin{plain}
$$A^L_{mk}= \left[{1\over 3\Omega^2(1\mp m)}\left(r{d\Phi_{mk}\over dr}
\pm 2m\Phi_{mk}\right)\right]_{r_R},\eqno(4.46)$$
\end{plain}

  Note that this solution describes the circulating orbits at corotation
resonances, and the librating orbits at Lindblad resonances.
Density waves at corotation resonances will not be discussed
in this lecture, and from now on, only Lindblad resonances are
considered. Notice also that the amplitudes $A_{mk}^L/\Delta r\sim 
r^2 M_s/M_p(r-r_R)$, and that in a region of width $\Delta r\sim
r(M_s/M_p)^{1/2}$, these test particle orbits intersect. Therefore,
collective effects are expected to be important in this region. Note finally that the relative sign of $r_1$ and $\theta_1$ is preserved at OLR with respect to ILR upon the substitution of $m(\Omega - \Omega_p)=\pm \kappa$ (preserving the usual direction of epicyclic motions at both resonances).

  Although this test particle solution is somewhat 
unrealistic\footnote{In any case, 
this solution breaks down too close to the
resonance where the condition $r_1\ll r_0$ is no longer satisfied.}, it
indicates that ring streamlines can be chosen, to first order in eccentricity, 
as sinusoidal functions of the basic angle $m(\theta-\Omega_p t)$ when
one considers a density wave driven by the $(m,k)$ component of the
satellite potential. This choice reflects the fact that the ring fluid
particles behave like a forced oscillator, responding with the time
and angular dependence of the forcing. Notice also that all particles 
with the same
semimajor axis have the same eccentricity and periapse angle in this
simple test particle solution. Similarly, density waves are special
solutions for which the eccentricities of the fluid
particles are function of the semimajor axis only. For density waves at
Lindblad resonances, we are therefore motivated to assume that 
the streamlines are parametrized by (see Figure~3):

\begin{plain}
$$r(a_e,\phi)=a_e\{1-\epsilon\cos [m(\theta -\Omega_p t) +m\Delta]\},\eqno(4.47)$$
\end{plain}

\noindent where $\epsilon\ll 1$ everywhere in the wave 
region\footnote{Collective effects prevent the divergence seen in the test
particle solution at the resonance (see section 7). Notice also
that, although $\epsilon\ll 1$, the density contrast can be highly
nonlinear (see section 4.4) so that Eq.~(4.47) describes both linear 
and nonlinear
density waves.}. Finding solutions of this type will demonstrate {\it a
posteriori} that our assumption is correct\footnote{A
formal justification of Eq.~(4.47) is also provided from first principles
by \cite{Sh84} in an {\textit ab initio} analysis of nonlinear
density waves.}.

Combining Eqs.~(4.11) and (4.12) necessarily yields 
$r=a_e[1-\epsilon\cos (\varphi -\varpi_e)]$ to lowest order in
eccentricity. This is compatible with Eq.~(4.47) only if
$m(\varphi-\Omega_p t)+m\Delta = \varphi -\varpi_e$, i.e., if

\begin{plain}
$$m(\Omega_a-\Omega_p)=\Omega_a-\dot\varpi_e,\eqno(4.48)$$
$$\varpi_0=\varphi_0(1-m)-m\Delta,\eqno(4.49)$$
\end{plain}

\noindent where $\varpi_0$ and $\varphi_0$ are the periapse angle and
mean longitude of the fluid particle at $t=0$. Note that in Eq.~(4.48)
$\dot\varpi_e$ includes the precession rates due to the perturbations;
the contribution of the perturbations to $d\varphi/dt$ 
is negligible in front their contribution to
$\dot\varpi_e$, to leading order in $\epsilon$ [see Eqs. (4.12), (4.22) and
(4.23)]. The second relation expresses
the condition that the initial periapse angles and initial azimuths 
of particles having the same semi-major axis must satisfy in order for
these particles to belong both to the streamline Eq. (4.47) and to their
epicyclic orbit around the planet (it generalizes the equivalent condition 
for elliptic rings,
which is that all fluid particles with the same semimajor axis have the
same periapse angle). Eq.~(4.48) expresses the Lindblad resonance condition,
and is required if the ring fluid particles are to belong to the
streamline and to their natural epicyclic orbits at all times, and not
only the initial one. These
requirements follow from the fact that all forces are small in
comparison with the planet attraction, so that the streamlines of the
the flow cannot differ very much from the natural fluid particle
orbits. 

  The phase angle $\Delta$ has been added for the following reason. 
Note that the trajectories of the fluid particles having
epicyclic semimajor axes equal to the resonance radius and eccentricity
$\epsilon$, which are elliptic in an inertial
reference frame, appear as the $m$-lobe shape of Eq.~(4.47)
when viewed in a frame rotating at $\Omega_p$. This is a purely
geometric and kinematic effect; Eq. (4.49) then states that the
trajectories of all such particles will be identical in the rotating
frame.  However, for fluid test particles, this is true only at the 
resonance, and density waves cannot exist, as the precession rate 
$\dot\varpi_e$ is imposed by the planet only: for example, just outside 
the resonance [i.e., for fluid test particles of semi-major axis
$a>a_R$, where $a_R$ is the semimajor axis at resonance, defined by
$\kappa_{a_R}=m(\Omega_{a_R}-\Omega_p)$], $m(\Omega_a-\Omega_p)
\not=\kappa_a$, and the streamlines appear to have an 
angular velocity $\kappa_a-m(\Omega_a-\Omega_p)$ with respect to the 
streamline at the resonance, quickly leading to streamline crossing and
to the destruction of any density wave pattern, as the one shown on Figure~4. Thus, fluid test particles cannot 
support free density wave.
But the ring self-gravity can produce a contribution to the precession rate
$\dot\varpi_{sg}$ which actually cancels this secular drift, so that the
resonance condition is satisfied throughout the wave region. As
the ring gravity is nevertheless a very small force, it can only reach
the right magnitude when the phase shift between adjacent streamlines is
large enough, i.e., when the WKBJ, or tight-winding condition

\begin{plain}
$$ma_e\left|{d\Delta\over da_e}\right|\gg 1,\eqno(4.50)$$ 
\end{plain}

\noindent is satisfied. This is why density waves are so
tightly-wound in rings\footnote{Waves in rings are forced
density waves. However, the coupling of the wave with the forcing
potential occurs at the resonance on a small fraction of the wave 
zone, and the wave
propagates essentially as a free wave on most of its radial extent.}.
In spiral galaxies, as the self-gravity of the disk
dominates the gravity from the central bulge, the spiral arms 
appear much more open. We will return to the discussion of
density waves and justify these assertions 
later on in section 7, after having introduced the
dynamical tools of the streamline formalism.

\subsubsection{Eccentric modes}

  As a last example, let us consider the case of the $\gamma$ and
$\delta$ rings of Uranus, whose streamlines are not described by the
simple elliptic shape Eq.~(4.34). 

  Actually, the $\delta$ ring is
well-fitted by Eq.~(4.47), with $m=2$, and with $\Delta$ almost constant
across the ring. This last characteristic allows us to introduce the notion of
{\it mode}: a mode is a global sinusoidal oscillation of a ring, with
streamlines given by Eq.~(4.47) where $\epsilon$ and $\Delta$ depend on $a_e$, 
and where $\Delta$, as we just said, is more or less
constant across the ring; a mode is characterized by its number of lobes
$m$ and its pattern speed $\Omega_p$. Elliptic rings enter 
this definition, with $m=1$ and $m\Delta-m\Omega_pt=-\varpi_e$; 
note that in this case, Eq.~(4.49) implies as required that the periapse 
angle depends only on semimajor axis, and not on $\theta_0$. Note also 
that spiral density waves are excluded from this definition (but not
standing waves). In these notes only 
single mode motions are considered. The generalisation to multimode motions 
is somewhat discussed in \cite{L89b}.

\begin{figure}
    \centering
    \begin{subfigure}[b]{0.3\textwidth}
        \includegraphics[width=\textwidth]{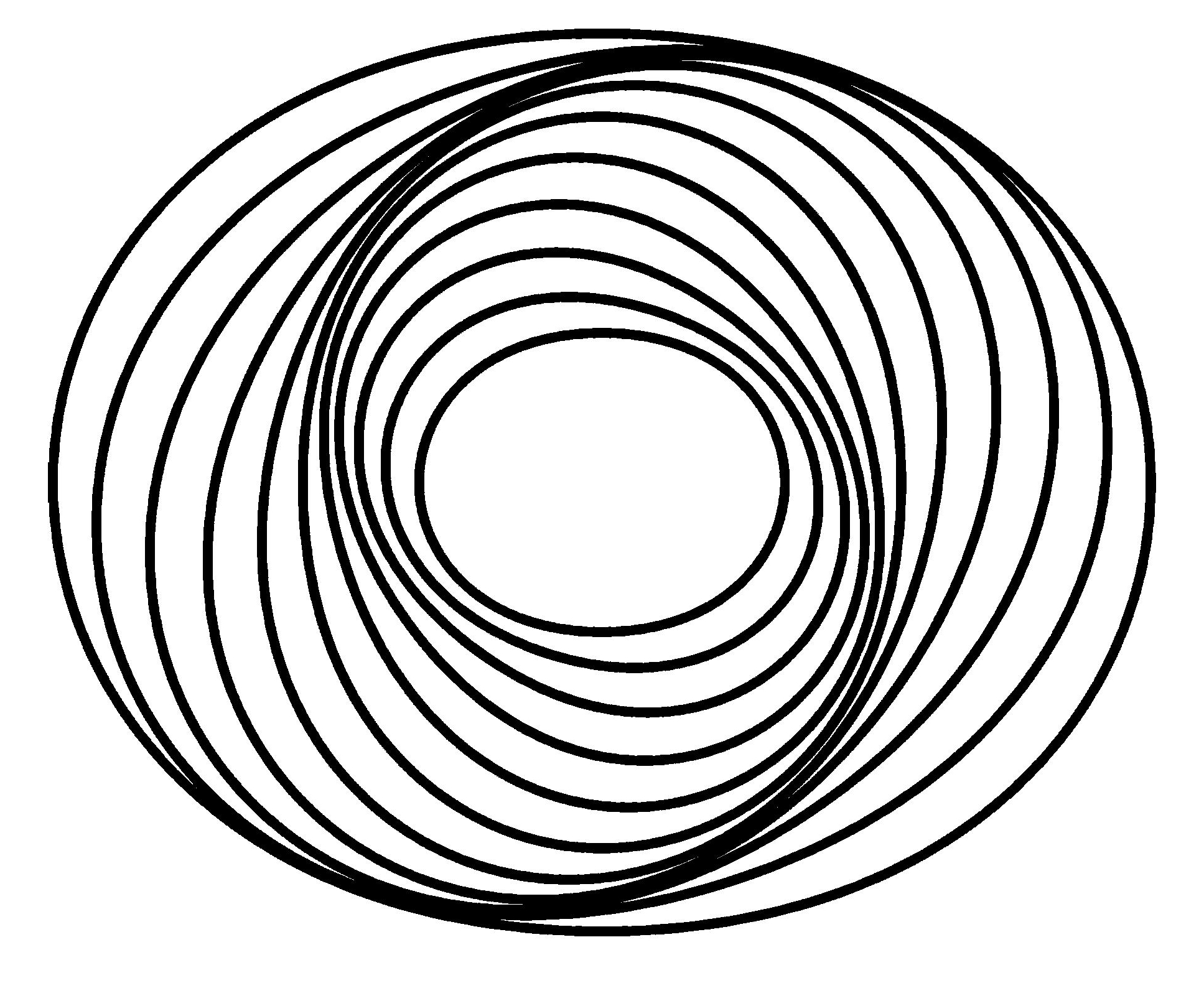}
        \caption{$m$=2 density wave}
        \label{densitywave}
    \end{subfigure}
    \hfill
    \begin{subfigure}[b]{0.21\textwidth}
        \includegraphics[width=\textwidth]{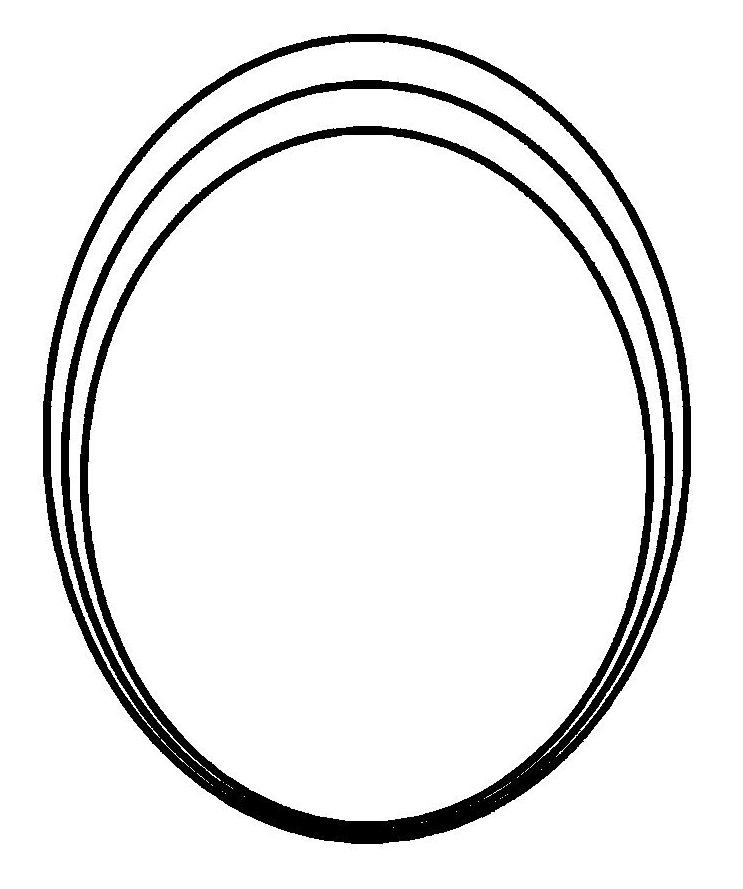}
        \caption{$m$=1 ring}
        \label{m1ring}
    \end{subfigure}
    \hfill
    \begin{subfigure}[b]{0.25\textwidth}
        \includegraphics[width=\textwidth]{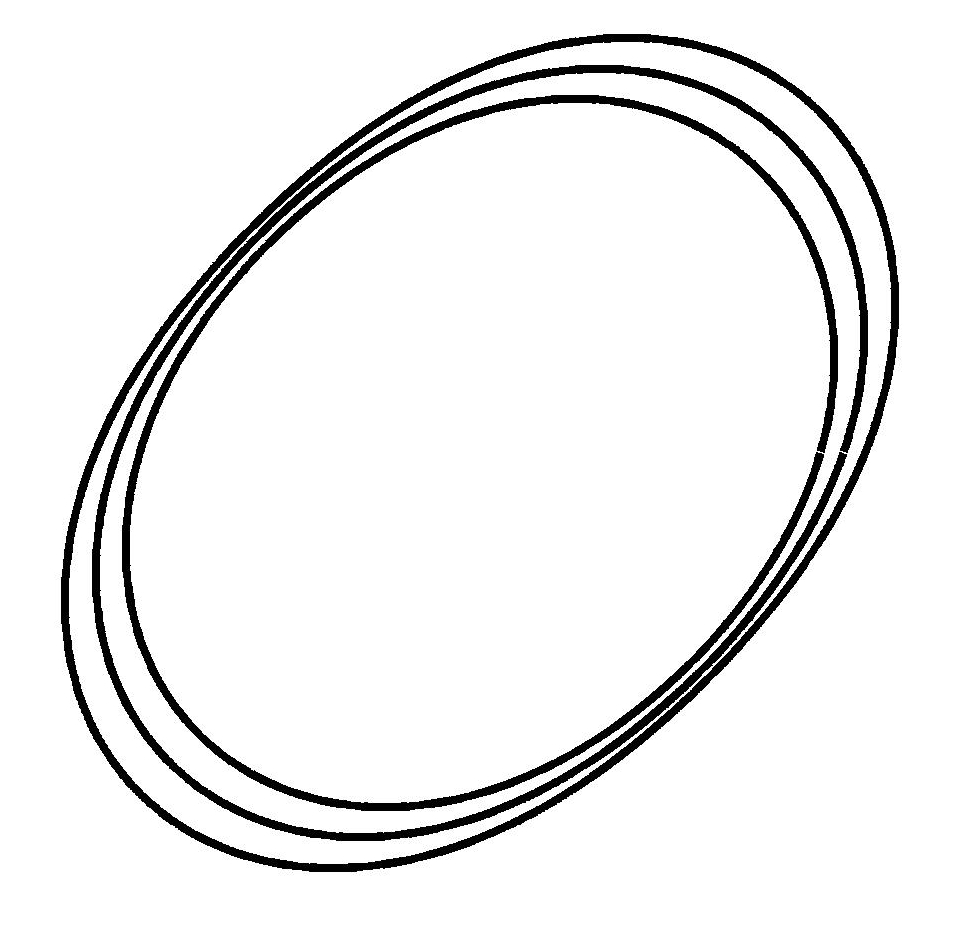}
        \caption{$m$=2 ring}
        \label{m2ring}
    \end{subfigure}
    \caption{\small{Examples of eccentric fluid motion in rings.}}\label{fig:ecc}
\end{figure}

  Because all forces are much smaller than the planet's, and because
the radial extent of the rings in which these modes are found is very
small, the contributions of the perturbations to the frequencies and
precession rates are much smaller than the planet's, and the pattern speed 
of the modes obeys the condition:

\begin{plain}
$$m(\Omega_a-\Omega_p)-(\Omega_a-\dot\varpi_{plan})\ll
\Omega_a-\dot\varpi_{plan}.\eqno(4.51)$$
\end{plain}

  Note that for density waves, the precession rate is not limited in
theory, because the winding can, in principle, be as high as needed.
However, the winding is nevertheless meaningful only as long as the wavelength
is larger than the typical particle size, and in any case, the wave is
damped before this limit is reached, so that in practice 
Eq.~(4.51) applies to density
waves as well. The streamlines of an $m=2$ density wave, and of $m=1,2$ 
ringlets are displayed on Figure~3 for comparison.

  Let us now consider the case of the $m=0$ mode of the $\gamma$ ring.
To understand the basic kinematic properties of this mode, let us
consider an infinitely thin ringlet whose fluid particles orbit on
eccentric trajectories, with the same semimajor axis $a_e$ and the same
eccentricity $\epsilon$. The periapse angles are evenly distributed, and
the phases of the ring particles on their orbits are initially 
all identical, so that at any
given time, all the fluid particles are at the same radial distance from
the planet. This situation is schematically depicted on Figure~4 where the
positions of the fluid particles are represented by dots; the particles
belong both to their epicyclic orbit and to the circular ring; particles never collide.
   Note that this situation is substantially different from the
case of the other modes. In Figure~3 for example, ring fluid particles are
present all along any given eccentric orbit (the fluid particle orbits and streamlines are identical in the rotating frame). Here, the orbits are essentially empty and the fluid particles are confined to a special
point along the orbit. At any given time, the ring appears circular. Its
radius oscillates sinusoidally at the orbital frequency: $r=a_e(1-
\epsilon\cos\zeta)$ where $d\zeta/dt=\kappa_a$. As the number
of lobes $m$ of the other modes refer to their azimuthal structure (it
is the azimuthal wavenumber), this type of motions actually corresponds
to the case $m=0$ in Eulerian analyses, i.e., a purely radial motion. 

\begin{figure}
\centering
\includegraphics[width=0.4\linewidth]{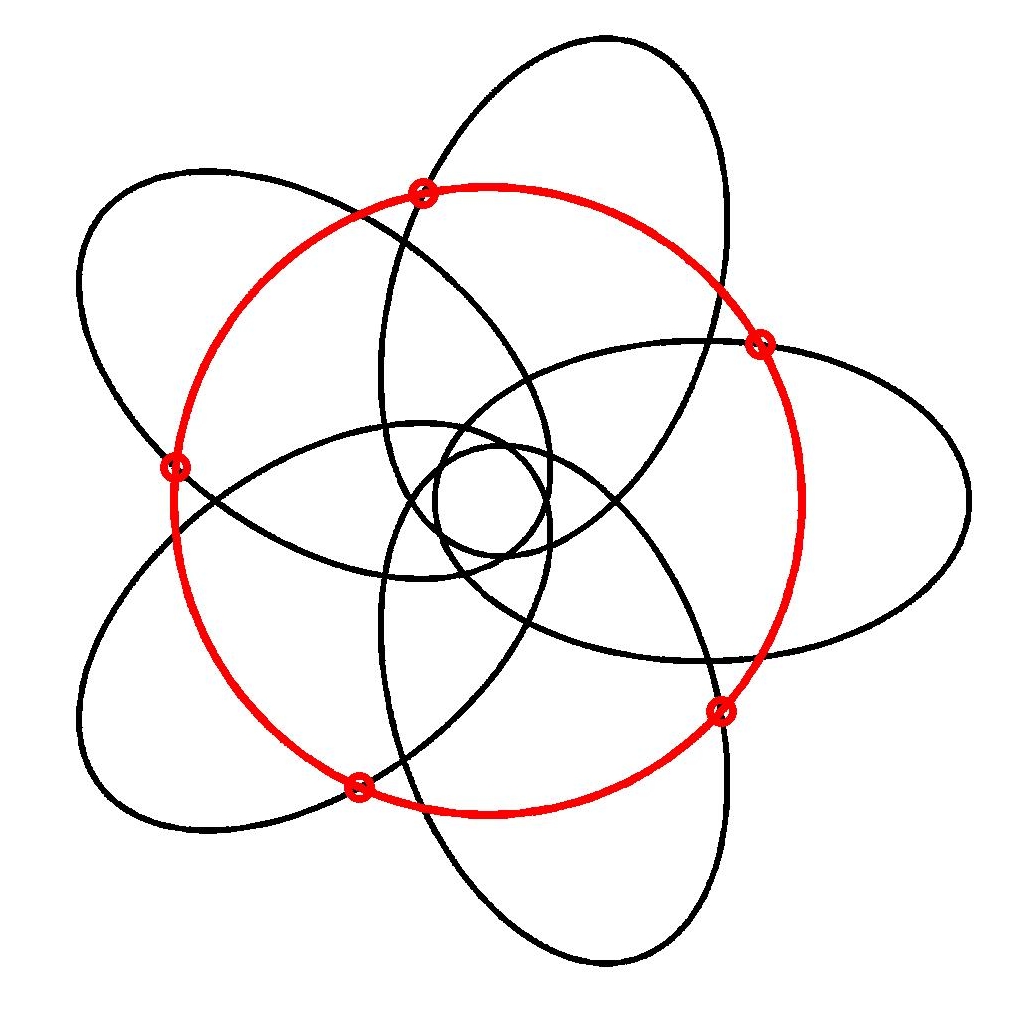}
\caption{\small{Sketch of an infinitely narrow $m=0$ mode. All fluid particles belong to the same circle oscillating radially, while occupying a given but identical azimuthal location along their individual orbits. This mode is kinematically different from $m\neq 0$ modes, for which orbits and streamlines are identical in the rotating frame.}}
\label{fig:M0}
\end{figure}

   Turning back to the case of a real ring (not infinitely thin),
one sees that the ring streamlines can be chosen as:

\begin{plain}
$$r=a_e\left[1-\epsilon(a_e)\cos(\Omega_p t +
\Delta(a_e))\right].\eqno(4.52)$$
\end{plain}

  Here again because all perturbing forces are such weak forces,
Eq.~(4.52) is compatible with $r=a_e[1-\cos(\theta-\varpi_e)]$
only if $\Omega_p - (\Omega_a-\dot\varpi_{plan}) \ll \Omega_a - 
\dot\varpi_{plan}$. However, we shall
see that the perturbing forces, although very weak, are nevertheless
essential, because they can counteract
the action of the Keplerian and precession shear, and 
therefore allow the mode to exist.

\subsubsection{Summary}

  Let us summarize and complete the results obtained so far on ring
kinematics. The results are discussed assuming $m\neq 0$, but can be
transposed to the $m=0$ case by replacing $-m\Omega_p$ by $\Omega_p$ and
$m\Delta$ by $\Delta$ in the following equations [compare Eqs.~(4.47)
and (4.52)]. Unperturbed ring fluid particles travel on 
epicyclic orbits, which read\footnote{From now on, 
we keep terms up to first
order in eccentricity only, as second order terms are generally not
accessible from the data.}:

\begin{plain}
$$r=a_e[1-\epsilon\cos M_e],\eqno(4.53)$$
$$\theta=\varpi_e+ M_e+2{\Omega_a\over\kappa_a}\epsilon\sin M_e,\eqno(4.54)$$
\end{plain}

\noindent where ($a_e, \epsilon,\varpi_e, M_e$) are the ``constants" of
the motion, and are functions of the fluid particle initial position
($\varpi_e$ and $ M_e$ depend on time as well). They can equivalently be
considered as functions of $a_e$
and $\theta_0$ (the fluid particle initial phase), or of $a_e$ and
$\varphi\equiv M_e +\varpi_e$; $ M_e$ is given by
Eq.~(4.13), $\varpi_e=\varpi_0+\int_0^t(\Omega_a-\kappa_a)dt$,
and $\Omega_a$ and $\kappa_a$ are defined in Eqs.~(4.15) and
(4.16). This fluid test particle solution is perturbed by the ring
self-gravity, the ring viscous stress, and the satellites. These
perturbations induce a time dependence of the ``constants" of the motion
which is expressed by Eqs.~(4.20) through (4.23). Differentiating 
Eqs.~(4.53) and (4.54)  with respect to time yields the 
velocity field of the unperturbed motion:

\begin{plain}
$$u_r={dr\over dt}=a_e\epsilon\kappa_a\sin M_e,\eqno(4.55)$$
$$u_\theta=r{d\theta\over dt}=a_e\Omega_a[1+\epsilon\cos M_e].\eqno(4.56)$$
\end{plain}

  As usual Eqs.~(4.51) through (4.56) apply to the perturbed fluid particle
motion as well: they give its osculating position and velocity.

On the other hand, the collective motion of ring fluid particles is \textit{assumed} to constitute an $m$-lobe mode in a frame rotating with an angular velocity noted $\Omega_p$:

\begin{plain}
$$r(a_e,\varphi,t)=a_e\{1-\epsilon(a_e,t)\cos(m(\varphi-\Omega_p t)+ m\Delta(a_e,t))\}.\eqno(4.57)$$
\end{plain}

\noindent In these relations, ($a_e,\varphi$) represent the circular motion the fluid particle would have in the absence of perturbation, and are used as (semi-)Lagrangian labels. This choice is clearly motivated by observations, and the dynamical equations self-consistently specify the conditions of existence of such motions. Quite often, the $m$-lobe shape is stationary and $\epsilon$ and $\Delta$ are independent of $t$; a time-dependence occurs, e.g., due to viscous overstabilities or in the relaxation phase to stationary in which case the time dependence is transient.

Eqs.~(4.57) and (4.53) can be satisfied simultaneously only if 
   
\begin{plain}
$$M_e=m(\varphi-\Omega_p t)+m\Delta,\eqno(4.58)$$
\end{plain}

\noindent i.e., if the following relations are satisfied

\begin{plain}
$${dm\Delta \over dt}=-m(\Omega_a-\Omega_p)+\Omega_a-\dot\varpi_e,\eqno(4.59)$$
$$\varpi_0=\varphi_0(1-m)-m\Delta_0,\eqno(4.60)$$
\end{plain}

\noindent  where $\varpi_0$, $\theta_0$ are the periapse angle and azimuth of the fluid particle at $t=0$ and $\Delta_0$ the phase at the same time [remember that the contribution
of the perturbations to $d\varphi/dt$ is negligible; see the 
discussion after Eq.~(4.49)]. Note that for stationary patterns, equation (4.59) implies that 

\begin{plain}
$$\eqalign{\dot\varpi_{pert.} & = -m(\Omega_a-\Omega_p)+(\Omega_a-\dot\varpi_{plan.})\cr
& \simeq\left[\frac{3}{2}(m-1)+{21\over 4}\left(1+{m-1\over 2}\right)\left({R_p\over a_R}\right)^2 J_2\right] \times \left({G M_P\over a_R^3}\right)\left({a-a_R\over a_R}\right)}\eqno(4.61)$$
\end{plain}

\noindent where $a_R$ is the resonance radius, and where 
$\dot\varpi_{pert.}$ and $\dot\varpi_{plan.}$ are the
contributions of the perturbing forces and of the planet to the
precession rate, respectively. 

We have up to now encountered two basic time-scales:

\begin{enumerate}
\item The short or orbital time-scale.
\item The intermediate or ``synodic" time-scale arising 
from the secular drift of
test particle streamlines with respect to one another in the vicinity of
the ``resonance" radius $a_R$ implicitly defined by the relation
$m(\Omega-\Omega_p)=\kappa$. This time scale
is of order $[\Omega_a\delta a/a]^{-1}$ (or $[J_2\Omega_a\delta
a/a]^{-1}$ if $m=1$) where $\delta a$ is the width of
the perturbed region; also, as argued in the paragraph around Eq.~(4.51), 
$\delta a/a \ll 1$.
\end{enumerate}

One sees that the perturbing accelerations 
must produce a secular variation of
the line of the apses on the intermediate time-scale. 
Because the perturbing forces are weak forces,
$\dot\varpi_{pert}\ll\Omega_a-\dot\varpi_{plan.}$ (or equivalently
$a_e-a_R\ll a_R$), and the motion is mainly imposed by the planet. 
This precession
rate is in general provided by the ring self-gravity. Note also that for
an elliptic ring ($m=1$), the required contribution of the perturbations
to the precession rate is down by a factor $J_2$.
Therefore the $m=1$ mode is easier to maintain than other modes, and
elliptic rings more common; this is a
natural result, because ellipses are the natural form of oscillations of
the fluid particles around the planet. In spiral galaxies, as the
central bulge does not dominate the gravity, the most common mode of
oscillation corresponds to $m=2$, as can be expected for objects with a
flat rotation curve. Thus, two-arms spiral galaxies tend to be more common.

  Let us conclude this section with a final comment. Generally,
unperturbed rings appear circular, and are described by $r=a_e$; the
motion of ring fluid particles reduces to $r=a_e, \theta=\varphi$. When
the ring is perturbed, it is useful to express the perturbed position
($r,\theta$) of a fluid particle in terms of the unperturbed position
($a_e,\varphi$) it would have in the absence of perturbation. This is
easily performed from Eqs.~(4.53), (4.54) and (4.58), and yields:

\begin{plain}
$$r=a_e[1-\epsilon\cos(m(\varphi-\Omega_p t)+ m\Delta)],\eqno(4.62)$$
$$\theta =\varphi+2\left(\Omega_a\over\kappa_a\right)
\epsilon\sin[m(\varphi-\Omega_p t) 
+m\Delta].\eqno(4.63)$$
\end{plain}

  Note that these equations define at any given time a Eulerian change 
of variables from ($r,\theta$) to ($a_e,
\varphi$). This property will be sometimes used in the remainder of
these notes. Note also that differentiation of Eqs.~(4.62) and (4.63)
yields the following expressions for the velocity field:

\begin{plain}
$$u_r= ma_e\epsilon(\Omega_a-\Omega_p)\sin[m(\varphi-\Omega_p
t)+m\Delta],\eqno(4.64)$$
$$u_\theta=a_e\Omega_a\left[1+\epsilon\left(2{m(\Omega_a-\Omega_p)\over
\kappa_a}-1\right)
\cos(m(\varphi-\Omega_p t)+m\Delta)\right].\eqno(4.65)$$
\end{plain}

  The ``kinematic" set, Eqs. (4.62), (4.63), (4.64) and (4.65) differs
from the ``osculating" set, Eqs. (4.53), (4.54), (4.55) and (4.56) by
terms of order $\epsilon \delta a_e/a_e\ (\ll\epsilon)$, where 
$\delta a_e$ is the distance to the resonance radius, implying that the
osculating elements differ from the (correct) kinematic ones by terms of
the same order\footnote{Note that the epicyclic
frequency appearing in these equations is defined by Eq.~(4.7) and does
not include the effect of the perturbations on the periapse angle.}. This difference is accounted for by the short period terms of the osculating elements. However, such terms are usually negligible, and the difference between the two types of elements does not have to be specified.

  In the derivation of these equations, as well as in the remainder of
these notes, we have assumed that $a_e$, $\epsilon$, and $\Delta$ are
time independent. However, most results, if not all, are still valid if
these quantities are time-dependent, provided that they vary on
time-scales much longer than the orbital period, as is the case in
section 7.1.

\subsection{Ring surface density}\label{sec:surf}

   Having solved the equation of motion, we wish now to consider the
equation of mass conservation [Eq.~(3.16)]. In fact, we are not going to
solve directly this differential equation, but instead present a general and well-known Lagrangian solution
directly from the constraint of mass conservation between the perturbed
(elliptic) and unperturbed (circular) states of a ring. The following
argument should be more properly developed in integral form, but we will
just present it in differential form, as this does not affect the
result. 

Let us consider an elementary mass element $\delta M$, defined as the mass 
between the two streamlines of semimajor axis 
$a_e$ and $a_e+da_e$, and the two
unperturbed azimuths $\varphi$ and $\varphi+d\varphi$:

\begin{plain}
$$\delta M = \sigma_0(a_e) a_e da_e d\varphi,\eqno(4.66)$$
\end{plain}

\noindent where, by definition, $\sigma_0$ is the surface density in the
unperturbed state. It is a function of $a_e$ only, as any nonaxisymmetric
feature should be quickly erased by the Keplerian shear and by diffusion
\footnote{Ring arcs are not discussed in these notes.}. 
  Let us now perturb this flow, which becomes elliptic. We assume, as
argued in the previous sections, that the fluid particle positions are
given most generally by Eqs.~(4.62) and (4.63) so that the ring streamlines
form $m$-lobe shapes in a frame rotating at $\Omega_p$. The mass element is now given by:

\begin{plain}
$$\delta M = \sigma(r,\theta) r dr d\theta,\eqno(4.67)$$
\end{plain}

\noindent where $\sigma$ is the surface density of the fluid particle in
($r, \theta$). Let us introduce the Jacobian $J$ of the change of
variable from the unperturbed flow to the perturbed flow:

\begin{plain}
$$J=\left|{r\over a} {\partial(r,\theta)\over\partial(a_e,\varphi)}\right|,
\eqno(4.68)$$
\end{plain}

\noindent so that $rdrd\theta=J adad\varphi$. We can express the perturbed 
surface density in terms of the
unperturbed one and of this Jacobian, using the fact that the fluid
element of mass is the same in the two states:

\begin{plain}
$$\sigma={\sigma_0\over J}.\eqno(4.69)$$
\end{plain}

  In order to evaluate the Jacobian, we need to evaluate partial
derivatives of the (actual) perturbed position with respect to the
(fictitious) unperturbed position. Let us start with $\partial
r/\partial a_e$. As $\epsilon$ and $m\Delta$ are function of $a_e$ only,
one obtains:

\begin{plain}
$${\partial r\over\partial a_e}=1-q\cos[m(\varphi-\Omega_p t)+m\Delta +
\gamma],\eqno(4.70)$$
\end{plain}

\noindent where $q$ and $\gamma$ are defined by:

\begin{plain}
$$q\cos\gamma = {d a_e\epsilon\over d a_e},\eqno(4.71)$$
$$q\sin\gamma = ma_e\epsilon{d\Delta\over da_e}.\eqno(4.72)$$
\end{plain}

   We have just introduced another fundamental parameter: $q$. In
general, although $\epsilon \ll 1$, $q$ can be of order unity.
Streamlines cross if $|q|>1$. This follows from
the fact that the distance between two streamlines, to lowest order in
eccentricity, is given by $\Delta r = {\partial r/\partial a_e}\Delta
a_e$ where $\Delta r$ and $\Delta a_e$ are the differences in radii and
semimajor axes of the two streamlines; this shows also that the distance
between two adjacent streamlines varies with azimuth, and so does the
width of an elliptic ring, as observed for all the elliptic Uranian rings. 
Usually, the pressure tensor (if nothing else) will diverge as 
$q\rightarrow 1$ or before, preventing streamline crossing.

\begin{figure}[h]
\centering
\includegraphics[width=0.5\linewidth]{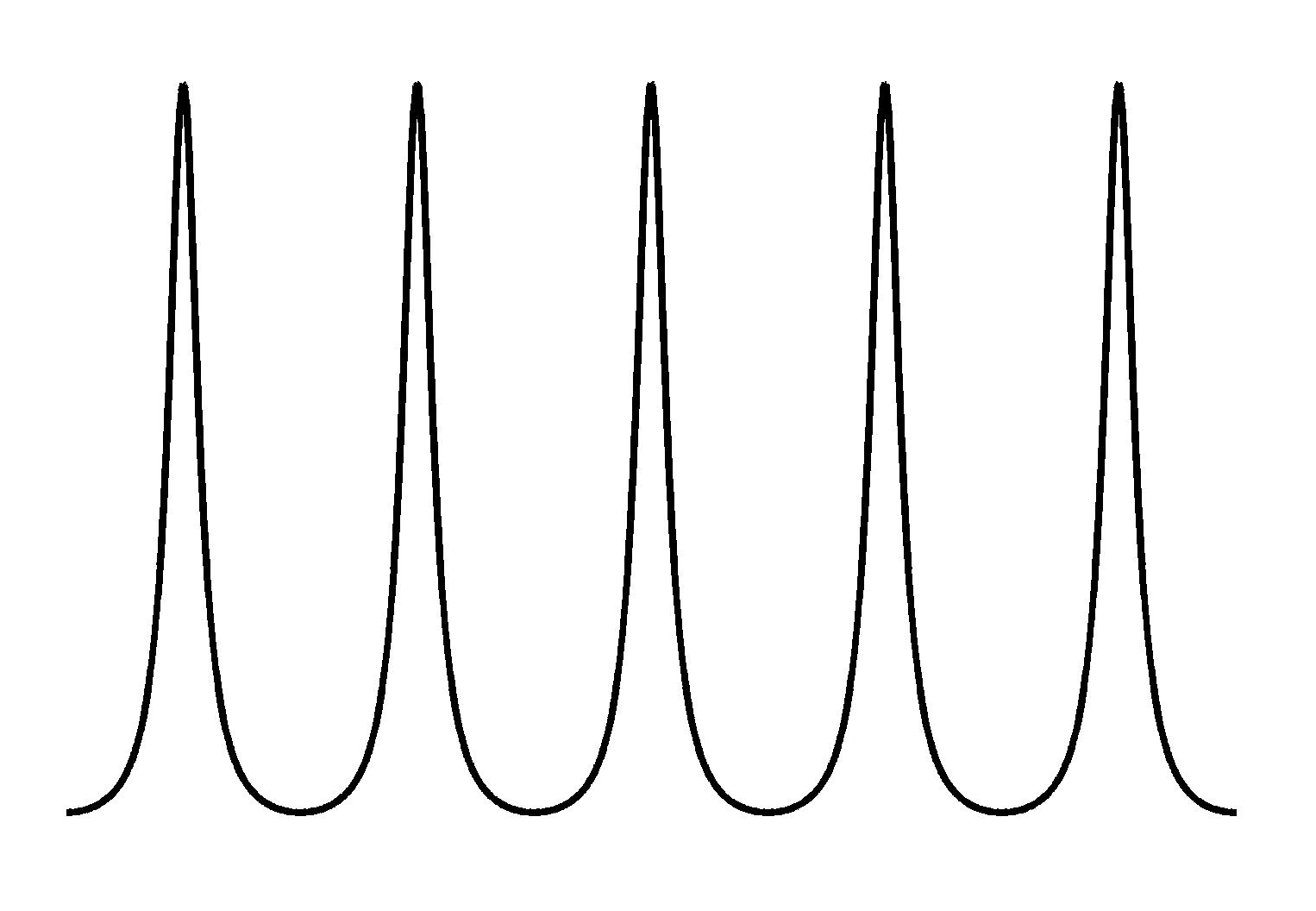}
\caption{\small{Schematic representation of the perturbed ring surface density as a function of azimuth for $q = 0.9$.}}
\label{fig:sig}
\end{figure}

   It is easy to check that $\partial\theta/\partial a_e \sim O(q)$,
$\partial r/\partial\varphi\sim O(\epsilon)$, and
$\partial\theta/\partial\varphi\sim 1+ O(\epsilon)$, so that to lowest
order in $\epsilon$, the Jacobian reads:

\begin{plain}
$$J={\partial r\over\partial a_e}=1-q\cos[m(\varphi-\Omega_p t) + m\Delta
+\gamma],\eqno(4.73)$$
\end{plain}

\noindent The
corresponding variation of the surface density with azimuth is
schematically represented on Figure~\ref{fig:sig}.

  One sees that the surface density azimuthal variations are
characterized by narrow peaks and broad troughs. An elliptic ring is
narrower and denser in the vicinity of its periapse than at its apoapse.
This trough-peak behavior is also seen
in the optical depths profiles of density waves in Saturn's rings. It is
an unavoidable consequence of the basic kinematics discussed in the
previous subsections. 

  A major kinematic difference between elliptic rings and density waves
is that in elliptic rings, the values of $q\gg\epsilon$ arise because of the
gradient of eccentricity across the ring, while the phase $\Delta$
remains constant or nearly constant. In density waves, the situation is
exactly reversed: the eccentricity is nearly constant in the wave
region, while the phase $\Delta$ varies very quickly with increasing
semimajor axis [see the tight-winding condition Eq.~(4.50)]. For elliptic
rings, $\gamma\simeq 0$, whereas for density waves,
$\gamma\simeq\pi/2$.

  The surface density can reach very high contrasts, although the
eccentricity remains very small: this type of nonlinearity is not
connected to large deviations from the circular motion, but rather from
the differences of deviations between neighboring streamlines. Thus we
have the possibility of developing nonlinear theories while still
linearizing the motion with respect to $\epsilon$ (but of course not
with respect to $q$), which is always a small parameter in ring
problems.

  We have solved the continuity equation, in the sense that we have found
the surface density of any ring fluid particle, in terms of
three unknown functions: $\sigma_0, q, \gamma$. The reader can check
that the {\it nonlinear} Eq. (3.16) is satisfied for any choice of these 
three functions, to
lowest order in eccentricity. It is impossible to go further on purely
kinematical grounds: the magnitude and form of these functions (except
maybe $\sigma_0$) is determined by the perturbations.

\section {Ring pressure tensor}
   The number of papers on the solution to the Boltzmann second-order
moment equations in
unperturbed flows is rather numerous. Some of them have already been
quoted in section 2. But there are only three studies to date
giving solutions for the second-order moments in perturbed flows. \cite{BGT83b} solve the Boltzmann second-order moment equations
with a collision term of the Boltzmann form, modified to take the
inelasticity of the collisions into account, assuming identical
indestructible spherical particles characterized by a normal coefficient
of restitution (there is no coupling with the spin degrees of freedom).
This work is an extension of the paper by \cite{GT78a}. \cite{SS85} and \cite{SDLYC85} use instead a Krook collision
term, also modified to account for the inelasticity of the collisions.
These first two analyses are very similar in
spirit and scope, but are however mostly
restricted to dilute systems (i.e. systems in which the particle size is
much smaller than the particle mean separations). On the other hand, 
\cite{BGT85} present a heuristic analysis 
of the pressure tensor behavior in dense systems (where the
particle size is much bigger than the interparticle distances), which is
not based on the Boltzmann second-order moment equations, but on a
hydrodynamical approach. This analysis is expected to apply to high
optical depth rings. Taken together, these papers give an interesting
insight into the pressure tensor behavior under two opposite sets 
of physical conditions, which more or less span the conditions relevant
to ring systems. Both dilute and dense systems will be discussed here. 

\subsection{Dilute systems}
The approach based on the Krook model has a number of advantages: first,  as the Krook model is simpler than the Boltzmann collision term, the 
analysis is somewhat simplified. Also, various generalizations like the
inclusion of gravitational encounters are more easily performed. Furthermore, 
the Krook model can be looked at as a model equation for more complicated 
collisional operators when its parameters are appropriately chosen. 
Therefore, only the Krook model will be described here. The material of
this section is taken from \cite{SS85}, and from \cite{SDLYC85}, unless otherwise specified.

  In this model the collision term of the Boltzmann equation is
expressed as

\begin{plain}
$$\left(\partial f\over\partial t\right)_c=\nu_c(f_I-f),\eqno(5.1)$$
\end{plain}

\noindent where $\nu_c$ is the mean collision frequency of the system under
consideration, and $f_I$ is a Maxwellian distribution with the same local number
density $\rho$ and mean velocity ${\bf u}$ as $f$:

\begin{plain}
$$f_I={\rho/M\over(2\pi c_I^2)^{3/2}}\exp \left(-{({\bf v}-{\bf u})^2\over
2c_I^2}\right),\eqno(5.2)$$
\end{plain}

\noindent where $M$ is the mean mass of the distribution [see the
discussion after Eq.~(3.12)]. To allow for the effects of inelastic
collisions, we let $c_I$ differ from $c$. Assuming that a
collision reduces the magnitude of the component of the relative velocity
along the lines of the centers of the two colliding particles
by a factor\footnote{An index {\textit
r} has been added to the coefficient of restitution to prevent confusion
with the epicyclic eccentricity.} $\epsilon_r < 1$, while it preserves 
the two components of the tangential velocity, allows us to require

\begin{plain}
$$3c_I^2=(2+\epsilon_r^2)c^2,\eqno(5.3)$$
\end{plain}

\noindent where the coefficient of restitution $\epsilon_r$ has been
averaged over all encounters, and can be regarded as a function of $c$.

  The Krook collision term has the following physical meaning. As we
follow the (individual) particles in orbit around the planet, per unit
volume of phase space, inelastic collisions remove particles at
a rate $\nu_c f$ and restore them with a Maxwellian distribution at a
rate $\nu_c f_I$, i.e. isotropically around the mean velocity ${\bf u}$.
This model does not examine in any detail the microphysics of the
collisions. It just expresses the fact that, independently of the
mechanics of individual collisions, collisional processes always tend to
make the distribution isotropic, and more specifically Maxwellian, in a
time-scale comparable to the collisional time-scale.

\subsubsection{Collision frequency, effective particle size, and effective optical depth}
  The collision frequency can be estimated with the following arguments,
mostly reproduced from the appendix A of Shu and Stewart 1985. Let us first
compute the collision frequency for a collection of particles of
identical masses $m$ and sizes $R$, assuming that the distribution
function is a maxwellian of velocity dispersion $c$:

\begin{plain}
$$f={\rho/m\over(2\pi c^2)^{3/2}}\exp\left(-{{\bf w}^2\over
2c^2}\right),\eqno(5.4)$$
\end{plain}

\noindent where ${\bf w}={\bf v}-{\bf u}$ is the velocity with respect to the
local mean velocity ${\bf u}$, and the vertical distribution of the 
mass density $\rho$ is assumed isothermal:

\begin{plain}
$$\rho={\sigma\mu_0\over(2\pi c^2)^{1/2}}\exp\left(-{\mu_0^2 z^2\over
2c^2}\right),\eqno(5.5)$$
\end{plain}

\noindent with $\sigma$ equal to the surface mass density. This
distribution of density expresses the vertical hydrostatic equilibrium
of the ring, i.e. the equilibrium between the planet and disk
self-gravity forces on one side, and the pressure on the other.
Therefore, the vertical epicyclic frequency $\mu_0\equiv (\partial^2
\phi/\partial z^2)_{z=0}$ entering Eq.~(5.5)
must include the contribution of the disk self-gravity, which can be
much larger than the restoring force of the planet in the vertical
direction, as will be seen in section 5.2.

  Computing the rate at which a particle with random velocity ${\bf
w}_1$ is hit by particles of all velocities ${\bf w}_2$, and
averaging over the distribution of ${\bf w}_1$ yields the local averaged
collision frequency as:

\begin{plain}
$$\langle \nu_c\rangle=4\pi R^2{\rho\over m}
\int\int{d^3w_1d^3w_2\over (2\pi c^2)^3}
|{\bf w}_1-{\bf w}_2|\exp\left(-{{\bf w}_1^2+{\bf
w}_2^2\over2c^2}\right).\eqno(5.6)$$
\end{plain}

  The integrals are most easily computed by performing the change of
variable from $({\bf w}_1,{\bf w}_2)$ to $\{{\bf W}={\bf w}_1-{\bf w}_2,
{\bf U}=({\bf w}_1+{\bf w}_2)/2\}$, which yields

\begin{plain}
$$\langle\nu_c\rangle={16\rho c R^2\over\pi^{1/2}m}.\eqno(5.7)$$
\end{plain}

  We can now compute the vertically averaged collision frequency:

\begin{plain}
$$\langle\langle\nu_c\rangle\rangle\equiv{1\over\sigma}\int\langle\nu_c
\rangle\rho dz = {8\over\pi}\mu_0\tau,\eqno(5.8)$$
\end{plain}

\noindent where we have introduced the normal optical depth $\tau=\sigma
\pi R^2/m$.

  We can now generalize this expression to a general distribution of
particle sizes. Let $N(R)dR$ be the number of particles per unit disk
area having radii between $R$ and $R+dR$. The surface density is related
to the particle size distribution by

\begin{plain}
$$\sigma=\int_{R_1}^{R_2}mN(R) dR,\eqno(5.9)$$
\end{plain}

\noindent with $m=4\pi \rho_p R^3$, for particles of bulk density $\rho_p$.
The two quantities $R_1$ and $R_2$ are the two cutoff radii of the
distribution. Typically in Saturn's rings, the distribution $N(R)\propto
R^{-3}$, and $R_1\sim 1$ cm and $R_2\sim 5$ m. 

  Following \cite{SS85}, we wish to define an effective binary
collision frequency which possesses the following properties: (1) is
symmetric with respect to the interchange of the members of the
colliding pair, (2) is proportional to the geometric cross section
$\pi(R+R')^2$, (3) is weighted by the reduced mass $mm'/(m+m')$ of the
colliding pair, and (4) is equivalent to the preceding expression for a
$\delta$-function distribution of particle sizes.
All these requirements seem natural, but might nevertheless need 
some comments. We have pointed out that the mean velocity and
the pressure tensor are independent of the particle size which allows us
to write the mass-dependent distribution function $f(m,{\bf r}, {\bf v})$
as $n(m)f_0({\bf r}, {\bf v})$. Also, Eq.~(5.1) should more properly be
written $(\partial f(m)/\partial t)_c=\int dm' \alpha(m,m') \nu_c(m,m')
n(m) ({f_0}_I-f_0)$, where $\nu_c(m,m')$ is the frequency of collision
of a particle of mass $m$ with particles of mass $m'$\footnote
{Note that at this point, $\nu_c$ is {\textit not} symmetric with 
respect to the interchange of $m$ and $m'$: if there are many more particles
of mass $m'$ than of mass $m$, any particle of mass $m$ will collide
much more often with a particle of mass $m'$ than the reverse.}. This would 
indicate first that the
collisional change for particles of one mass depends on the collisions
with particles of all other masses, and second that not all pair of
masses have the same efficiency in relaxing the distribution of a given
mass to a Maxwellian, as indicated by the factor $\alpha$. Indeed, in a
collision between two particles of mass $m$ and $m'$, the conservation
of momentum requires that the change of
momentum of each particle is $\mu O(c)$ where $\mu$ is the reduced mass
of the colliding pair. Therefore, the change of velocity of the particle
of mass $m$ is $\mu/m O(c)$, and one can take $\alpha\propto\mu/m$. Now,
remember that we have multiplied the Boltzmann moment equations by the
mass and integrated over mass, which shows that the integrated collision
term is proportional to $\int dm\ m \int dm' n(m)\mu(m,m')\nu_c(m,m')
\alpha(m,m')$, and that
the collision frequency must indeed be weighted by the reduced mass of
the colliding pairs. Furthermore, Eq. (5.6) can be generalized to give 
the collision frequency of a particle of mass $m$ with
particles of mass $m'$, resulting in the change of the factor $4\pi R^2
\rho/m$ into $\pi(R+R')^2 n(m')$. This series of argument would give the
effective collision frequency exactly, if it weren't for the 
efficiency factor $\alpha$ which is
only determined to within a multiplicative constant. We can
therefore only conclude that the effective collision frequency 
$\nu_c\propto\int\int dmdm'n(m)n(m')\pi(R+R')^2mm'/(m+m')$. The
coefficient of proportionallity is constrained by the last requirement.

  Finally, after getting rid of the vertical dependence of the
$n(m)n(m')$ factor by a vertical integration, we can write down 
the desired collision frequency by inspection as

\begin{plain}
$$\nu_c={4\mu_0\over\pi\sigma}\int\int \pi(R+R')^2\left(mm'\over m+m'\right)
N(R)N(R')dRdR',\eqno(5.10)$$
\end{plain}

\noindent where we have dropped the double-average notation. For
definiteness, let us take $N(R)=C R^{-3}$. If we assume that $R_2\gg
R_1$, then $C=3\sigma/4\pi\rho_pR_2$ from Eq. (5.9),
and the double integral can be computed:

\begin{plain}
$$\nu_c={6\mu_0\sigma\over{\sqrt 3}\rho_p R_2}.\eqno(5.11)$$
\end{plain}

  We can put this result under a form similar to (5.8) by defining an
effective optical depth and an effective particle size. We obtain

\begin{plain}
$$\nu_c={8\over\pi}\mu_0\tau_e,\eqno(5.12)$$
\end{plain}

\noindent with 

\begin{plain}
$$\tau_e={\sigma\pi R_e^2\over m_e}={3\sigma\over 4\rho_p R_e},\eqno(5.13)$$
$$R_e={{\sqrt 3}R_2\over \pi}.\eqno(5.14)$$
\end{plain}

  Note that this effective optical depth differs from the actual optical
depth obtained for $N(R)\propto R^{-3}$. In fact,

\begin{plain}
$$\tau=\int\pi R^2 N(R)dR={3\sigma\over 4\rho_p
R_2}\ln(R_2/R_1),\eqno(5.15)$$
\end{plain}

\noindent which is larger than $\tau_e$ by a factor ${\sqrt
3}/\pi\ln(R_2/R_1)$; this factor amounts to 3 or 4 if $R_1\sim 1$ cm and
$R_2\sim 500$ cm; $\tau_e$ is the quantity which should
enter Eqs. (2.9), (2.10) and (2.12), and particular attention should be
paid to this point when comparing theoretical results with observations.

  Let us conclude this section with a few general comments. First, note
that with the provision of Eqs.~(5.3) and (5.12), the Krook model is completely specified: we have
expressed its two parameters $c_I$ and $\nu_c$ in terms of the other
variables of the problem. The reader should however notice that instead
of looking at the Krook model as to a physical model in its own right,
one could as well consider Eq.~(5.1) as a model equation for more
complicated collision terms by keeping some freedom in the choice of
$c_I$ and $\nu_c$. For example, \cite{SS85} show that the
Boltzmann collision term for smooth identical spheres can be reproduced
by adopting $c_I^2=[(1+\epsilon_r)/(3-\epsilon_r)]c^2$ and
$\nu_c=(4\mu_0/3\pi)(3-\epsilon_r)(1+\epsilon_r)\tau_e$ instead of Eqs.~(5.3) and (5.12). Finally, let us note that corrective factors can be
added to Eq.~(5.12) to account for the effects of the anisotropy of the
distribution and of the possible close-packing of the ring 
particles\footnote{If the particles are close-packed, 
i.e. if their radii are not small in comparison with their mean distances, 
the particles have to travel much less than their relative distance
before a collision takes place, resulting in an increase of the
collision frequency. This effect will be further discussed in section
5.2.}. However, \cite{SDLYC85} show that the anisotropy correction does
not produce important modifications. The effect of the close-packing
correction has not been analyzed in the literature. We will therefore
not include it here. We merely note that Eq.~(5.12) applies only to
dilute systems, i.e. systems in which the filling factor is much smaller
than unity.

\subsubsection{Quasi-equilibrium of perturbed and unperturbed ring systems}
  We can pursue the program outlined at the end of section 3: 
we can now look for the steady-state solution of the pressure tensor 
equations with the velocity field of Eqs.~(4.64) and (4.65), and the ring surface density of Eqs.~(4.69) and (4.73).

  Let us first recast Eqs.~(3.21) through (3.24) in an appropriate form.
First, notice that to lowest order in $\epsilon$, $\partial/\partial r=
(1/J)\partial/\partial a$, and $d/dt\equiv \partial/\partial t+
\Omega\partial/\partial\theta +u_r\partial/\partial r= \Omega\partial/
\partial M'$ in steady-state where $M'= M+\gamma=m(\varphi-\Omega_pt)+m\Delta
+\gamma$ [Eq. (4.58) has been used]. Notice that the steady-state
condition $d/dt=\Omega\partial/\partial M'$ can be obtained either in
an inertial frame or in a frame rotating with angular speed $\Omega_p$.
In such a rotating frame, the pressure tensor is time-independent
because ${\bf u}$ and $\sigma$ are time-independent. The steady-state
condition expresses the fact that the pressure tensor depends on time
only through $M'$. Any other time dependence disappears in a
time-scale $\sim \Omega^{-1}$, in accordance with the heuristic argument
developed in section 2. Then, keeping only the leading
terms in $\epsilon$, one obtains:

\begin{plain}
$${dP_{rr}\over dM'}+{3q\over J}\sin M' P_{rr}-4P_{r\theta}=
{1\over\Omega}\left(\partial P_{rr}\over\partial t\right)_c,\eqno(5.16)$$
$${dP_{r\theta}\over dM'}+{P_{rr}\over 2J}+{2q\over J}\sin M' P_{r\theta}-
2P_{\theta\theta}={1\over\Omega}\left(\partial P_{r\theta}\over\partial t
\right)_c,\eqno(5.17)$$
$${dP_{\theta\theta}\over dM'}+{P_{r\theta}\over J}+{q\over J}\sin M'
P_{\theta\theta}={1\over\Omega}\left(\partial
P_{\theta\theta}\over\partial t\right)_c,\eqno(5.18)$$
$${dP_{zz}\over dM'}+{q\over J}\sin M'P_{zz}={1\over\Omega}\left(\partial
P_{zz}\over\partial t\right)_c.\eqno(5.19)$$
\end{plain}

  These equations show that the pressure tensor components depend on
azimuth only through the angle $M'$. From Eq. (5.1), one obtains the
right-hand sides of the preceding equations as

\begin{plain}
$${1\over\Omega}\left(\partial P_{ij}\over\partial t\right)_c=
{\nu_c\over\Omega}\left(\sigma c_I^2\delta_{ij}-P_{ij}\right),\eqno(5.20)$$
\end{plain}

\noindent where $\nu_c$ is defined by Eqs.~(5.12) through (5.14), and
$c_I$ by Eq.~(5.3). In performing the vertical integration, we have used
the fact that $c$ is independent of $z$ (see the discussion at the
beginning of section 3.2). Note also that by definition of
the velocity dispersion

\begin{plain}
$$3\sigma c^2=P_{rr}+P_{\theta\theta}+P_{zz}.\eqno(5.21)$$
\end{plain}

  If the relation $\epsilon_r(c)$ were known, Eqs (5.16) through (5.21)
supplemented by Eqs. (5.3) and (5.12) through (5.14) would yield a closed
set of equations for $P_{rr}, P_{r\theta},P_{\theta\theta}$, and $P_{zz}$
(notice that these quantities uncouple from the other pressure tensor
components). However, as this relation is not very well
constrained, one usually prefers to keep the absolute scale of the
pressure tensor as a free
parameter, and solve rather for the relative magnitude of the pressure
tensor components and for the relation $\epsilon_r(\tau_e)$ that the
equilibrium requires. In doing so, the dependence of $\nu_c$ on azimuth
is usually taken into account, but $\epsilon_r$ is usually taken to be
constant with azimuth, as well as $c^2$ and $c_I^2$ ($\nu_c$ depends on 
azimuth because it is proportional to $\sigma$).

  In order to illustrate these points, let us first look at the solution
for unperturbed flows. In this case, the velocity field and the surface
density are purely axisymmetric, so that $q=0$, and the pressure tensor
is independent of azimuth, as well as the collision frequency. Therefore, 
Eqs.~(5.16) through (5.19) reduce to

\begin{plain}
$$P_{rr}=\sigma_0 c_I^2\left[1+{6\Omega^2\over\nu_c^2+4\Omega^2}\right],
\eqno(5.22)$$
$$P_{r\theta}=\sigma_0 c_I^2{\nu_c\over 2\Omega}
{3\Omega^2\over \nu_c^2+4\Omega^2},\eqno(5.23)$$
$$P_{\theta\theta}=\sigma_0 c_I^2\left[1-{3\Omega^2\over 2(\nu_c^2+
4\Omega^2)}\right],\eqno(5.24)$$
$$P_{zz}=\sigma_0 c_I^2.\eqno(5.25)$$
\end{plain}

  In these equations, the pressure tensor is scaled to $\sigma_0 c_I^2$,
which we take as a free parameter. Combining Eqs.~(5.3) and (5.21) with 
the ``solution" we have just written down yields after some algebraic 
manipulations the required relation between the
coefficient of restitution and the effective optical depth

\begin{plain}
$$\epsilon_r^2=1-{9/11\over 1+(128/11\pi^2)\tau_e^2}.\eqno(5.26)$$
\end{plain}

  The reader will notice that this relation is in good agreement with
Eq. (2.12), which we had derived with heuristic arguments. Again,
equating this $\epsilon_r(\tau_e)$ relation to some $\epsilon_r(c)$
relation would give $c(\tau_e)$, and therefore would determine
the magnitude of the
pressure tensor in Eqs.~(5.22) through (5.25). Furthermore, summing Eqs.~(5.16), (5.18) and (5.19) yields the equation of equilibrium of the
velocity dispersion, which reads

\begin{plain}
$$4\Omega {P_{r\theta}\over\sigma_0}=\nu_c(1-\epsilon_r^2)c^2.\eqno(5.27)$$
\end{plain}

\noindent In the hydrodynamical approximation (see section 5.2, or \cite{LL87}, chapter 2), $P_{r\theta}/\sigma_0\sim \nu \Omega$ for
the Keplerian velocity field; also, Eq.~(5.12) shows that $\nu_c\sim
\Omega\tau_e$, so that Eq.~(5.27) is in agreement with Eq.~(2.10). We
have already pointed out in section 2 that this equation expresses the
equilibrium between the excitation due to the input from the shear of
the mean motion, and the damping due to the inelasticity of the
collisions.

  This completes our discussion of unperturbed flows. In what concerns
perturbed flows, no analytic solution of Eqs.~(5.16) through (5.19) is
available\footnote{Analytic expressions do exist in
the limit of small {\textit q}, but won't be given here. The interested
reader is referred to the paper by \cite{SDLYC85} for their
derivation.}. The set has been solved numerically, either by integrating
the differential equations from $M'=0$ to $2\pi$, subject to the
constraint that the functions are periodic in $M'$ \citep{BGT83b}, or the pressure tensor components are
expanded in Fourier series, and the various coefficients computed \citep{SDLYC85}. The pressure tensor components are computed as functions of
$q$, $\tau_{e0}\equiv J\tau_e=3\sigma_0/4\rho_pR_e$ and $M'$ (the weak
dependence of $\Omega$ and $\mu_0$ on $a_e$ is ignored).

\begin{figure}[h]
\centering
\includegraphics[width=0.7\linewidth]{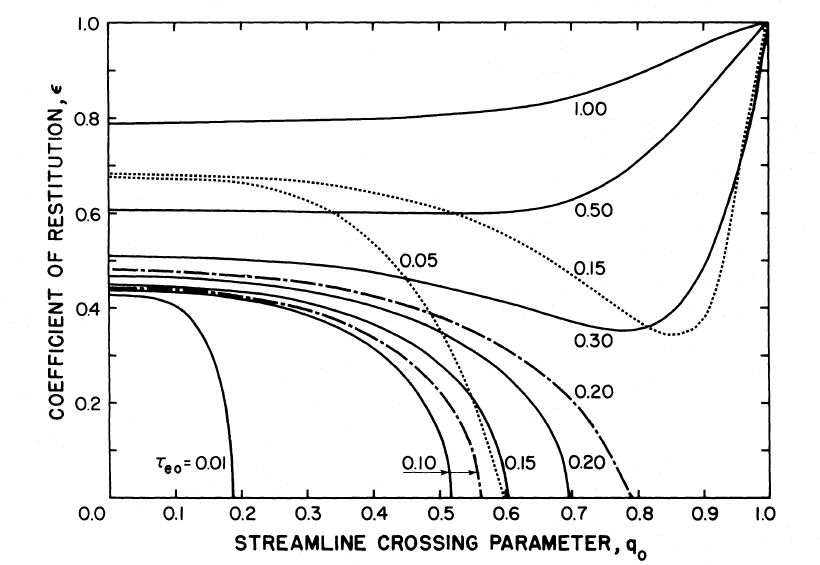}
\caption{\small{The relation between the particle coefficient of restitution and the rate of perturbation of the ring for several values of the effective optical depth; the three different types of lines (solid, dashed, dotted) correspond to three different forms of the Krook model (from \cite{SDLYC85}.}}
\label{fig:Shu}
\end{figure}

  The preceding $\epsilon_r(\tau_e)$ of unperturbed flows generalizes to
an $\epsilon_r(\tau_{e0},q)$ relation, as is depicted on Figure~\ref{fig:Shu} (adapted
from \citealt{SDLYC85}): the
curves show the variation of $\epsilon_r$ with $q$, parametrized by the
values of $\tau_{e0}$. These curves exhibit a very important feature: if
the unperturbed optical depth $\tau_{e0}$ is smaller than a critical
value (here of order 0.2 or 0.3, but its exact value is sensitive to the
details of the collisional model), the coefficient of restitution {\it
must} become zero for a finite value of $q=q_m$, here of order 0.7. On
the other hand, if $\tau_{e0}$ is larger than this critical value, $q$
is not limited\footnote{Except by $q<1$. This
constraints arises because as $q\rightarrow 1$, the streamlines tend to
cross, and therefore the pressure tensor, if nothing else, tends to
diverge to prevent this crossing.}, and $\tau_{e0}\rightarrow 1$ as 
$q\rightarrow 1$. This result has the following physical interpretation. 
Let us consider a ring perturbed
by some outside agent (a satellite near a resonance, for example), and
let us increase slowly the strength of this perturbation so that the
rate of perturbation of the flow, measured by $q$, tends to increase. To
dissipate the energy input of the perturbation, $\epsilon_r$ must adjust
according to the requirements of Figure~\ref{fig:Shu}. For small mean effective
optical depths, the collision frequency is also small, and $\epsilon_r$
goes to zero to make maximum use of the rare collisional events. As a
consequence the velocity dispersion increases without bound (because the
coefficient of restitution is a decreasing function of the impact
velocity), and the viscous damping increases [Eq.~(2.9)], preventing
further growth of the rate of perturbation of the flow $q$. On the other
hand, in regions of high $\tau_{e0}$, the collision frequency is always
high enough to keep the dispersion velocity low. Therefore, in high
optical depth regions, for a given strength of the outside perturbation,
the viscous damping is much smaller than it would be in small optical
depth regions. \cite{SDLYC85} mainly attribute to this effect the
difference in damping length-scales of density waves in Saturn's A (low
optical depth) and B (high optical depth) rings.

  Let us now consider the behavior of the pressure tensor components 
themselves. Actually, we will see in section 6 that the pressure
tensor does not affect the mean flow directly, but only through three
quantities, $t_1$, $t_2$ and $a_{r\theta}$, defined by\footnote{The present definitions differs by a factor $\sigma_0$ from the
equivalent definitions of \cite{BGT86}.}:

\begin{plain}
$$t_1=s_{rr}+2c_{r\theta},\eqno(5.28)$$
$$t_2=2s_{r\theta}-c_{rr},\eqno(5.29)$$
$$c_{ij}+is_{ij}=\langle\exp(iM')P_{ij}(M')\rangle,
\eqno(5.30)$$
$$a_{ij}=\langle P_{ij}(M')\rangle,\eqno(5.31)$$
\end{plain}

\noindent where the bracket notation stands for the azimuthal average, 
so that, for any quantity $X$,

\begin{plain}
$$\langle X\rangle\equiv {1\over 2\pi}\int_0^{2\pi}XdM'.\eqno(5.32)$$
\end{plain}

  Note that $\langle\sigma\rangle=\sigma_0/(1-q^2)^{1/2}$.  The quantities just
defined depend only on $q$ and $\tau_{e0}$, or equivalently
on $q$ and $\langle \tau_e\rangle=\tau_{e0}/(1-q^2)^{1/2}$.
In dilute systems, the pressure tensor is of the order of $\sigma c^2$, 
so that it is useful to define

\begin{plain}
$$v^2={\langle\sigma c^2\rangle\over\langle\sigma\rangle},\eqno(5.33)$$
\end{plain}

\noindent and, following \cite{BGT83a},
to introduce dimensionless quantities  of order unity $Q_{ij}$ such that

\begin{plain}
$$P_{ij}=\langle\sigma\rangle v^2 Q_{ij}(q,\langle\tau_e\rangle,M').
\eqno(5.34)$$
\end{plain}

  Due to the definitions of $v^2$ and $c^2$, the variables $Q_{ij}$ must 
satisfy the normalisation constraint $\langle Q_{rr}+Q_{\theta\theta}+Q_{zz}
\rangle=3$. The value of $v^2$ is essentially a free parameter in the
absence of constraint on the form of the $\epsilon_r(c)$ relation,
although one knows from various observations that $v^2\sim c^2\sim$ a
few mm/s (see section 2). Similarly, one can define ${\cal C}_{ij}$, 
${\cal S}_{ij}$, ${\cal T}_i$, and ${\cal A}_{ij}$ such that

\begin{plain}
$$\langle\sigma\rangle v^2{\cal C}_{ij}=c_{ij},\eqno(5.35)$$
$$\langle\sigma\rangle v^2{\cal S}_{ij}=s_{ij},\eqno(5.36)$$
$$\langle\sigma\rangle v^2{\cal A}_{ij}=a_{ij},\eqno(5.37)$$
$$\langle\sigma\rangle v^2{\cal T}_i=t_i.\eqno(5.38)$$
\end{plain}

\begin{figure}[th]
\centering
\includegraphics[width=0.7\linewidth]{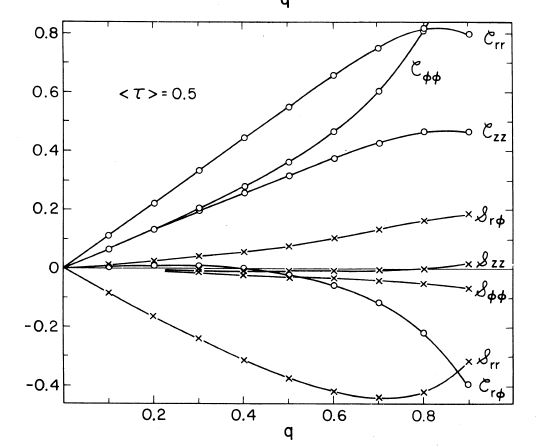}
\caption{\small{The behavior of the viscous coefficients defined in the text as a function of the perturbation rate of the ring q, for an effective optical depth of 0.5 (from \cite{BGT83a}.}}
\label{fig:BGT}
\end{figure}

  The behavior of ${\cal C}_{ij}$ and ${\cal S}_{ij}$ as functions of 
$q$ and for
$\langle \tau_e\rangle=0.5$ is represented on Figure~\ref{fig:BGT}, taken from\footnote{These curves were computed with a collision term of
the Boltzmann form instead of the Krook form, but this does not affect
much the results, at least for the purpose of these notes.} \cite{BGT83a}.

  The behavior for other values of $\langle \tau_e\rangle$ is quite
similar. It is only important to remark that for dilute systems, both
$t_1$ and $t_2$ are negative. The discussion of the effect of these
coefficients on the mean motion is differed to section 6 and 7. The
discussion of the behavior of $a_{r\theta}$ is differed to section 5.3.

\subsection{Dense systems}
  Another important consequence of the relation displayed on Figure~\ref{fig:Shu} is
that for moderately opaque rings (let's say, $\tau \gtrsim 1$),
the equilibrium can be obtained only for rather elastic materials 
($\epsilon_r$ is always close to unity). Reversely, if the ring particle
material is not elastic enough (which is quite likely, because
in any case ring particles are most probably covered with some more or
less fluffy layer of regolith), the equilibrium described in Figure~\ref{fig:Shu}
cannot be sustained, implying that the filling factor cannot be small
and that the ring must collapse to a close-packed configuration (see the
heuristic discussion of section 2.2.4). This argument ignores various
complicating physical effects: first, gravitational encounters tend to
increase the effective value of $\epsilon_r$, because they are
completely elastic; second, the coupling with spin degrees of freedom
tends to decrease this effective value, since energy is lost by
tangential friction; finally, irregular particle surfaces also tends to
reduce $\epsilon_r$. However, these effects could most certainly not
suppress the possibility of reaching a close-packed configuration, which
is likely to be relevant for Saturn's B ring as well as for the major
Uranian rings, which have mean optical depths as high as 3 or 4.

  In this section, we will therefore model the rings 
as a collection of particles with typical size $\sim d$. The typical 
separation distance of ring particle surfaces will be $\sim s$, with 
$s\ll d$. Considering that the particles have random velocity $\sim c$,
the collision frequency is $\omega_c\sim c/s$, i.e. larger than the
collision frequency that the dilute approximation would yield by a
factor $d/s$. In such
physical conditions, two important conditions can be drawn. First, the
collision frequency being greatly enhanced, the system can most
certainly be described in the framework of the hydrodynamical
approximation\footnote{The hydrodymical approximation
can be derived from Boltzmann equation, see, e.g., \citealt{SS85}.
However, it can also be derived directly from first physical principles,
with no implicit or explicit
reference to any specific form of the collision term, and therefore its
range of application does not exactly overlap that of the Boltzmann
moment equations.}. But as ring particles do not resemble the molecules of a
fluid in many respects, some attention must be paid to the computation
of the pressure and the viscosity of the system. Second, as $d\gg s$,
the medium is
nearly incompressible, and one can assume that the divergence of the
three-dimensional flow is zero. The computation of the pressure tensor
in these physical conditions is the subject of this subsection. 
Unless otherwise specified, the material is extracted from
the paper by \cite{BGT85}.

\subsubsection{Macroscopic equations}
  The macroscopic equations we will use are the continuity and the
Navier-Stokes equation, i.e. Eqs.~(3.8) and (3.9), with the pressure
tensor reducing to
 
\begin{plain}
$$p_{ij}=p\delta_{ij}-2\eta u_{ij},\eqno(5.39)$$
\end{plain}

\noindent where $p$ is the pressure, $\eta$ the coefficient of dynamic
viscosity and $u_{ij}$ is the strain tensor:

\begin{plain}
$$2 u_{ij}={\partial  u_i\over\partial x_j}+{\partial u_j\over\partial
x_i}.\eqno(5.40)$$
\end{plain}

The tensor $\sigma_{ij}=2\eta u_{ij}$ is known as the (viscous-)stress 
tensor and $p_{ij}$ as the internal stress tensor in
hydrodynamics. The validity of the hydrodynamical approximation relies
on the existence of an isotropic pressure force, and on a
linear stress-strain relation. These conditions are usually satisfied
when the collision time is much smaller than the dynamical times
involved in the problem.

  As the medium under consideration is incompressible, $\rho$ is
constant (in space as well as in time), and the continuity equation
reduces to 

\begin{plain}
$$\nabla . {\bf u}=0.\eqno(5.41)$$
\end{plain}

  From our discussion of section 3, we know that the pressure forces
play essentially no r\^ole in determining the horizontal motion, so that
the incompressibility condition Eq.~(5.41) can be combined with Eqs.~(4.64) and (4.65) to yield the vertical velocity:

\begin{plain}
$$u_z=-z{\Omega q\over J}\sin M'.\eqno(5.42)$$
\end{plain}

  In this equation, we have used $m(\Omega-\Omega_p)\simeq\kappa\simeq\Omega$. 
The surface and volumetric ring density are related by

\begin{plain}
$$\sigma=2\rho h,\eqno(5.43)$$
\end{plain}

\noindent where $h$ is the ring thickness. From $\sigma=\sigma_0/J$, one
has $h=h_0/J$, where $h_0$ is the unperturbed ring thickness.

  One can plug the velocity field in the vertical component of the
momentum equation, with $\phi=\phi_{plan.}+\phi_{sg}$, which then reads:

\begin{plain}
$$z{\Omega^2 q\over J^2}\left(q\sin^2 M' +q - \cos M'\right)=
-F_2\Omega^2 z-{1\over\rho}{\partial p\over\partial z}-2{\partial
\eta\over\partial z}{\Omega q\over J}\sin M',\eqno(5.44)$$
\end{plain}

\noindent where one has approximated the planet by a central mass
(neglected the gravitational harmonic coefficients), kept the first
order term in the the $z$ expansion of the resulting force, and used
Gauss theorem to compute the contribution of the self-gravity force (the
horizontal variation of thickness occurs on a scale much larger than the
ring thickness). Thus, $F_2$ is defined by

\begin{plain}
$$F_2=1+{4\pi G\rho\over\Omega^2}.\eqno(5.45)$$
\end{plain}

\noindent Note that $F_2$ can be of order ten or larger in planetary
rings (for ice particles, and a ring filling factor $\sim 0.5$), so 
that the vertical component of the ring self-gravity is much
larger than the planet force in this direction, and the effect of the
self-gravity on the vertical structure cannot be ignored. The vertical
component of the momentum equation shows that $p$ and $\eta$ have to be
quadratic in $z$. Furthermore, they must vanish at the top and bottom of
the rings, so that we can write:

\begin{plain}
$$p=p_0\left(1-{z^2\over h^2}\right),\eqno(5.46)$$
$$\eta=\eta_0\left(1-{z^2\over h^2}\right),\eqno(5.47)$$
\end{plain}

\noindent and recast Eq. (5.44) as

\begin{plain}
$${\Omega^2 q\over J^2}\left(q \sin^2 M' +q -\cos M'\right)+F_2\Omega^2=
{2p_o\over\rho h^2}+{4\eta_0\over\rho h^2}{\Omega q\over J}\sin
M'.\eqno(5.48)$$
\end{plain}

  For certain combinations of parameters, this equation might require
$p_0$ to be negative, at least in some locations in the ring. In these
conditions, the ring material ``splashes" in the vertical direction, so
that our assumption of incompressibility fails. We will assume here that
such conditions do not occur (formal requirements can be derived; see
\citealt{BGT85}.

  Finally, the rate of transfer of macroscopic energy into random
motions per unit volume is (see section 5.3)

\begin{plain}
$$\left(\partial{\cal E}\over\partial t\right)_{trans}=2\eta W^2,\eqno(5.49)$$
\end{plain}

\noindent where $W$ is the shear, i.e.,

\begin{plain}
$$\eqalignno{W^2&=u_{ij}u_{ij}\cr
&={\Omega^2\over 8J^2}\left(16q^2-15+24J\right).&(5.50)\cr}$$
\end{plain}

  This completes the required set of macroscopic equation.

\subsubsection{Microscopic equations}
   To proceed further, we need some microscopic equations for the
pressure, the dynamic viscosity and the rate of collisional dissipation
of energy. These quantities will be computed  keeping in mind the simple 
physical model described at the introduction of section 5.2, following
the derivation by \cite{H83}.

   To compute the pressure, any ring particle is imagined to vibrate
with average velocity $c$ in a cell of typical size $d$, and to exert some
pressure $p$ on the surrounding particles. In a collision, the typical
momentum transfer is of the order of $mc$. The pressure being the
momentum transfer per unit time and unit surface, one obtains therefore

\begin{plain}
$$p\sim\omega_c {mc\over d^2}=g_1{\rho d c^2\over s},\eqno(5.51)$$
\end{plain}

\noindent where $g_1$ is a dimensionless constant of order unity, and
$\rho\sim m/d^3$.

  The coefficient of dynamic viscosity is computed in the following way.
Assume that the ring particle fluid is submitted to some shear flow, in
which for definiteness the velocity $u$ is assumed to lie in the $x$
direction and the gradient of $u$ in the $y$ direction. Two adjacent
``layers" of ring particles will on the average move with velocities
differing by an amount $\Delta u\sim (du/dy)\ d$. When collisions occur
between particles belonging to the two layers, the average transfer of
$x$-momentum in the $y$ direction in $m\Delta u$. Therefore, the shear
stress, being the rate of transfer of momentum per unit time and across
a unit surface [see the discussion after Eq. (3.9)] is

\begin{plain}
$$\sigma\sim \omega_c{m\Delta u\over d^2}.\eqno(5.52)$$
\end{plain}

  As by definition $\sigma=\eta du/dy$, one finally obtains

\begin{plain}
$$\eta=g_2{\rho d^2 c\over s},\eqno(5.53)$$
\end{plain}

\noindent where $g_2$ is another factor of order unity.

  Finally, the rate of dissipation of kinetic energy of random motions
per unit volume is (see the discussion at the beginning of section
2.2.3)

\begin{plain}
$$-\left(\partial{\cal U}\over\partial t\right)_c\sim 
(1-\epsilon_r^2)\omega_c\rho c^2
=g_3{\rho c^3\over s},\eqno(5.54)$$
\end{plain}

\noindent where $g_3$ is the last dimensionless constant of order unity
of the problem. 

  We have now gathered all the pieces of the puzzle, and we can turn our
attention to the computation of the pressure tensor components.

\subsubsection{The pressure tensor components}
  Our first task is to find two relations between $p_0$ and $\eta_0$, in
order to obtain expressions for these two quantities. One such relation
is obviously Eq.~(5.48). The other one is obtained in the following way.
First, equating the rate of transfer of energy from the orbital motion
to random motions, Eq.~(5.49) with the rate of dissipation of energy in
collisions, Eq.~(5.54), yields the following constraint (which will be
further discussed in section 5.3):

\begin{plain}
$$2\eta W^2=g_3{\rho c^2\over s}.\eqno(5.55)$$
\end{plain}

\noindent Combining this relation with the expression of $p^2/\eta$, computed
from Eqs.~(5.51) and (5.53) gives us the required relation between $p_0$
and $\eta_0$, which reads

\begin{plain}
$$\eta_0= F_1{p_0\over W},\eqno(5.56)$$
\end{plain}

\noindent where $F_1=(g_2 g_3/2g_1^2)^{1/2}$ is a dimensionless factor
of order unity.

We can finally compute $p_0$ and $\eta_0$ from this relation and Eq.~(5.48). This yields
  
\begin{plain}
$$p_0=\rho h_0^2\Omega^2{A_1 A_2\over 2 J^2 A_3},\eqno(5.57)$$
$$\eta_0=\rho h_0^2\Omega\ 2^{1/2}F_1{A_1\over J A_3},\eqno(5.58)$$
\end{plain}

where we have defined

\begin{plain}
$$A_1=F_2+{q\over J^2}\left(q\sin^2 M' +q -\cos M'\right),\eqno(5.59)$$
$$A_2=(16 q^2 -15 + 24J)^{1/2},\eqno(5.60)$$
$$A_3=A_2+2^{5/2}F_1 q\sin M'.\eqno(5.61)$$
\end{plain}

  Note that in order of magnitude $W^2\sim \Omega^2$ so that Eqs. (5.55)
and (5.53) imply $c\sim \Omega d$. Furthermore, Eqs. (5.58) and (5.53)
imply $d^3\sim h_0^2 s$, so that the condition $s\ll d$ requires $h_0^2\gg
d^2$, i.e., the rings must be many particle thick.

  We are now in position to compute the vertically integrated pressure
tensor components $P_{ij}$. First, the vertical integration of Eq.
(5.39) yields

\begin{plain}
$$P_{ij}={4\over 3}h(p_0\delta_{ij}-2\eta_0 u_{ij}),\eqno(5.62)$$
\end{plain}

\noindent except for $P_{rz}$ and $P_{\theta z}$ which vanish. 
  Introducing 

\begin{plain}
$$A_4=A_2-2^{5/2}F_1q\sin M',\eqno(5.63)$$
$$A_5={2 A_1\over 3J^3},\eqno(5.64)$$
\end{plain}

\noindent one finally obtains the nonvanishing pressure tensor component
to lowest order in $h/a$ as

\begin{plain}
$$P_{rr}=\rho h_0^3\Omega^2{A_4 A_5\over A_3},\eqno(5.65)$$
$$P_{r\theta}=-\rho h_0^3\Omega^2{2^{1/2}F_1 A_5(1-4J)\over
A_3},\eqno(5.66)$$
$$P_{\theta\theta}={4\over 3}hp_0,\eqno(5.67)$$
$$P_{zz}=\rho h_0^3\Omega^2 A_5,\eqno(5.68)$$
\end{plain}

\begin{figure}[th]
\centering
\includegraphics[width=0.7\linewidth]{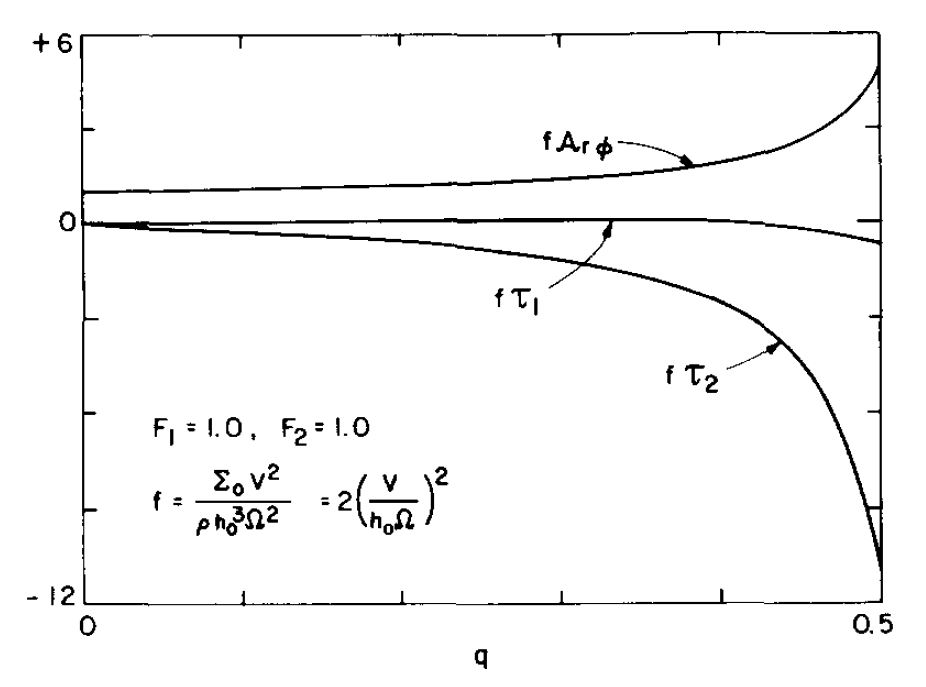}
\caption{\small{The behavior of the azimuthally averaged viscous coefficients as a function of the rate of perturbation of the ring streamlines q (from \citealt{BGT85}).}}
\label{fig:BGT85}
\end{figure}

  Notice that for dense systems, the natural scaling of the vertically
integrated pressure tensor components is no longer $\langle\sigma\rangle
v^2$, but $\rho h_0^3\Omega^2$. Therefore, the three fundamental
coefficients $t_1, t_2, a_{r\theta}$ defined in Eqs.~(5.28) through
(5.31) are most conveniently written in
dimensionless form as

\begin{plain}
$$t_i=\rho h_0^3\Omega^2 f{\cal T}_i,\eqno(5.69)$$
$$a_{r\theta}=\rho h_0^3\Omega^2 f{\cal A}_{r\theta},\eqno(5.70)$$
\end{plain}

\noindent where, consistently with Eqs.~(5.35) through (5.38),

\begin{plain}
$$f={\sigma_0 v^2\over (1-q^2)^{1/2}\rho h_0^3\Omega^2},\eqno(5.71)$$
\end{plain}

  The behavior of the three dimensionless quantities $f{\cal T}_1,
f{\cal T}_2$ and $f{\cal A}_{r\theta}$  for $F_1=F_2=1$ is displayed 
on Figure~\ref{fig:BGT85}, taken from \cite{BGT85}. The graph is represented for values of $q<0.5$, because otherwise $p_0$ becomes negative for some values of $M'$. Notice that ${\cal T}_2$ is negative as in the dilute approximation, but that ${\cal T}_1$ is
positive for $q$ smaller than some critical value (although this cannot
be seen on the graph, due to the poor resolution), in opposition to the 
dilute case. The implications of these results for the dynamics will be 
discussed in section 5.3 and sections 6 and 7. 

\subsection{Energy dissipation, and viscous flux of angular momentum}
  Energy and angular momentum budgets are important for the long term
dynamics of the rings. We will here have a look into two fundamental
features of energy and angular momentum exchanges, differing a more
complete discussion to section 6.

\subsubsection{Energy dissipation in planetary rings}
  The purpose of this section is to compute the rate of dissipation of
energy due to the inelasticity of the collisions, and to show that this
energy is drawn from the orbital motion.

  The orbital energy per unit mass is by definition

\begin{plain}
$$E=\left({1\over 2}{\bf u}^2+\phi_p\right).\eqno(5.72)$$
\end{plain}

  From the equation of continuity (3.13) and the equation of motion
(3.14), one obtains the equation of evolution 
of $E$ as

\begin{plain}
$$\rho{DE\over Dt}=-{\partial{\rm p}_{ij}u_i\over\partial x_j}
+{\rm p}_{ij}{\partial u_i\over\partial x_j}.\eqno(5.73)$$
\end{plain}

  In this equation, we have ignored the work of the ring self-gravity
and of the satellite perturbations, as they have no effect on the
argument. The two terms on the right-hand side are the rate of work of the internal stress of the ring during the motion. 

  On the other hand, the internal kinetic energy per unit ring mass is

\begin{plain}
$$U={1\over 2\rho}{\rm p}_{ii}.\eqno(5.74)$$
\end{plain}

  From Eq.~(3.15), one can derive the equation of evolution of $U$

\begin{plain}
$$\rho{DU\over Dt}=-{\rm p}_{ij}{\partial u_i\over\partial x_j}+
{1\over 2}\left(\partial{\rm p}_{ii}\over\partial t\right)_c,\eqno(5.75)$$
\end{plain}

\noindent where, consistently with the approximations made earlier, we
have neglected the heat flux terms. The first term on the right-hand side 
is the contribution of the internal stress (in
quasi-static transformations, it results in the well-known $pdV$ work of
thermodynamics), and the last term is the rate of loss of energy due to
collisions (if collisions are elastic, this term equals to zero,
by conservation of kinetic energy). This last term is obviously
negative. Comparing Eq. (5.73) and (5.75) shows that the last term of
the right-hand side of Eq. (5.73) represents the rate of transfer of
macroscopic energy into random motions.

  For dense systems, we can put Eq. (5.75) in a different form. From 
the definitions of the
stress and strain tensors Eqs. (5.39) and (5.40), one obtains

\begin{plain}
$$\rho{DU\over Dt}= 2\eta(u_{ij})^2+{1\over 2}
\left(\partial{\rm p}_{ii}\over \partial t\right)_c,\eqno(5.76)$$
\end{plain}

\noindent Eq.~(5.55) is identical to Eq. (5.76) except for $DU/Dt$,
which has been set equal to zero, because some of the (crude)
approximations which we have made earlier are not compatible with this
term (in particular, its vertical dependence), and because it represents
no essential piece of physics, as it averages to zero [see also the 
discussion after Eq.~(5.79)].

  Now, remember that the streamlines of the flow form closed curves in a
frame rotating at $\Omega_p$ (see
section 4) so that we can actually integrate Eqs.~(5.75) and (5.73)
over an arbitrary volume bounded by streamlines\footnote{This volume rotates with angular speed $\Omega_p$ in an inertial
frame, but this has no effect on the argument.}. Notice that these
volume integrals are in fact integrals over a given mass element of the
ring. Note also that the steady state
condition implies $DU/Dt=-(\partial U/\partial\theta)\Omega/m$, so that 
$DU/Dt$ does not contribute to the integral. Thus, one obtains:

\begin{plain}
$$\int dV\ \rho{DE\over Dt}=\int dV\ {\rm p}_{ij}{\partial
u_i\over\partial x_j}-\int dV{\partial {\rm p}_{ij}u_i\over\partial x_j}
,\eqno(5.77)$$
$$\int dV\ {\rm p}_{ij}{\partial u_i\over\partial x_j}-\int dV\ {1\over 2}
\left(\partial{\rm p}_{ii}\over\partial t\right)_c=0.\eqno(5.78)$$
\end{plain}

These two equations demonstrate the advertised result, i.e. that the average
energy lost per unit mass
during collisions is drawn from the energy of the orbital motion.
Therefore, on the average, the ring material tends to fall on the
planet.

We can compute the vertically integrated average rate of loss of energy of
the ring material bounded by two streamlines of radii $a_1$ and $a_2$,
by using Eqs.~(5.16) through (5.19), which yield

\begin{plain}
$$\eqalignno{\int rdr\int d\theta\ & {1\over 2}\left[{\partial\over\partial t}
(P_{rr}+P_{\theta\theta}+P_{zz})\right]_c \cr
& ={1\over 2}
\int_0^{2\pi}dM'\int_{a_1}^{a_2}da\ Ja\left[\Omega{\partial\over\partial
M'}(P_{rr}+P_{\theta\theta}+P_{zz})\right.\cr
&\left.\qquad +{\Omega q\sin M'\over J}(3P_{rr}+P_{\theta\theta}+P_{zz})+
\Omega P_{r\theta}\left({1\over J}-4\right)\right]\cr 
&=\pi\int_{a_1}^{a_2}da\ a\Omega (2qt_1-3a_{r\theta}).&(5.79)\cr}$$
\end{plain}
This result is also valid for dense systems, although Eqs.~(5.16) through
(5.19) were derived under the assumption that $u_z=0$. This is most easily
seen from Eq. (5.78) by noting that in the dense system approximation,
$P_{rz}=0$ and $P_{\theta z}=0$, so that the only possible difference in the first
term of this equation between the dilute and the dense approximations
comes from the contribution of ${\rm p}_{zz}\partial u_z/\partial z$, which vanishes upon vertical integration [see, e.g., Eq.~(B.12c) of \cite{SS85}]. Note that as a
consequence, Eq.~(5.55) is compatible with Eq. (5.79).

  We will show in section 6 that the rate of viscous loss of orbital
energy is also given by Eq.~(5.79), as can be expected from the previous
argument. This equation has an interesting consequence, which will be used in section 5.4: it implies that

\begin{plain}
$$2qt_1<3a_{r\theta},\eqno(5.80)$$
\end{plain}

\noindent because the collision terms are negative (the dissipation of
energy in collision reduces the internal kinetic energy) and because the
choice of the boundaries $a_1$ and $a_2$ is arbitrary.

\subsubsection{Viscous flux of angular momentum}
  We have argued in section 2 that the ring internal stress results in
angular momentum transport. Let us call $F_H^{vis}$ the vertically integrated
rate of transport of angular momentum across a unit length of streamline
(in short, the viscous flux of angular momentum). The components of the
pressure tensor being the components of the internal force of the ring
fluid per unit surface [see the discussion after Eq.~(3.9)], the
vertically integrated torque exerted per unit length of streamline due
to the material inside it is $aP_{r\theta}$, to lowest order in
eccentricity. Therefore,

\begin{plain}
$$F_H^{vis}=aP_{r\theta},\eqno(5.81)$$
\end{plain}

  Defining the viscous angular momentum luminosity, $L_H^{vis}$, as the
integral of the flux around the streamline, one has

\begin{plain}
$$L_H^{vis}=2\pi a^2 a_{r\theta}.\eqno(5.82)$$
\end{plain}

   It is instructive to derive general expressions for fluids with
constant kinematic viscosity $\nu=\eta/\rho$. From  Eqs.~(5.39) and
(5.40), one obtains

\begin{plain}
$$F_H^{vis}=2\nu\sigma_0\Omega a\left({1\over J}-{1\over
4J^2}\right),\eqno(5.83)$$
$$L_H^{vis}=\pi\nu\sigma_0\Omega
a^2{3-4q^2\over(1-q^2)^{3/2}},\eqno(5.84)$$
\end{plain}

  In the limit of axisymmetric flows ($q=0$), the viscous luminosity of
angular momentum reduces to

\begin{plain}
$$L_H^{vis}=3\pi\sigma\nu\Omega a^2,\eqno(5.85)$$
\end{plain}

  This result, derived by \cite{LP74} in the context of
the theory of accretion disks, was used in section 2.3 in the discussion of
axisymmetric viscous instabilities. Eq.~(5.85) shows that angular
momentum flows outwards in axisymmetric flows.

  Eqs.~(5.83) and (5.84) show that two values of $q$ are particularly
significant. First, for $q<q_1=3/4$, $F_H^{vis}>0$ for all azimuthal locations,
whereas for $q>q_1$, $F_H^{vis}<0$ for some interval of azimuth. This
inward flux of angular momentum arises because the angular velocity
increases outwards in this longitude interval. For $q>q_2={\sqrt 3}/2$, the
luminosity itself becomes negative: the perturbation of the flow is high
enough to make the angular momentum flow inwards, in opposition to the
unperturbed case.

\begin{figure}[th]
\centering
\includegraphics[width=0.5\linewidth]{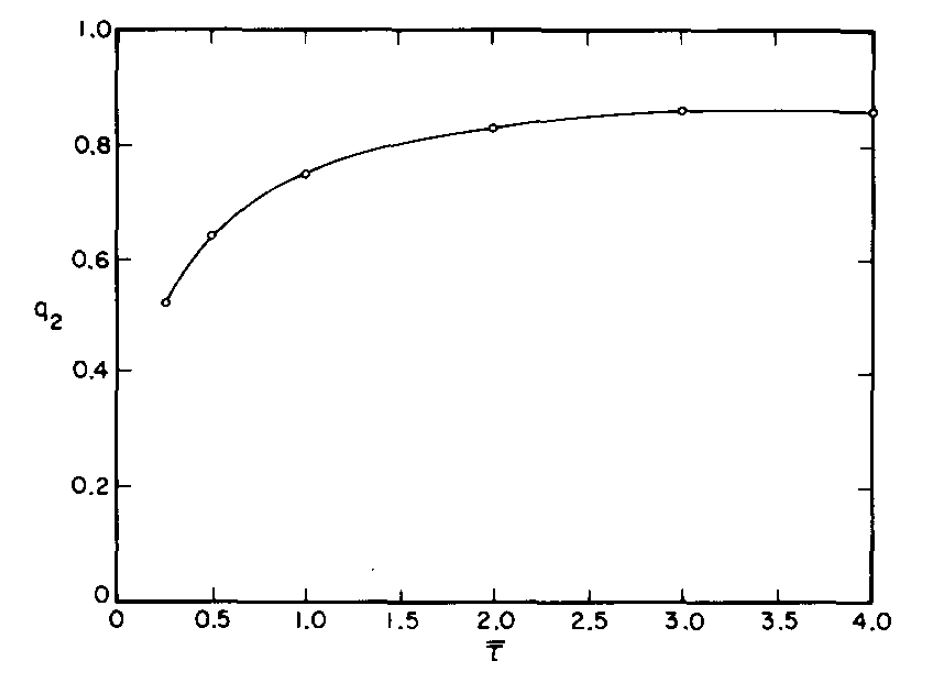}
\caption{\small{The value of q for which the angular momentum luminosity reverses direction as a function of the azimuthally averaged effective optical depth (from \citealt{BGT83b}).}}
\label{fig:BGT83}
\end{figure}

  One can wonder if this feature is a consequence of our assumption of
constant kinematic viscosity. This is not the case. \cite{BGT83b} in their analysis of dilute systems show that the angular
momentum flux and luminosity, which are both positive for axisymmetric
flows, both change sign in sufficiently perturbed regions. For example, the 
corresponding value of $q_2$ is represented as a function of $\langle 
\tau_e\rangle$ on Figure~\ref{fig:BGT83}.

  The same feature is also true of dense systems, although this cannot
be seen on Figure~\ref{fig:BGT85}. For example, for the particular 
choice\footnote{These
values are more appropriate for dense rings than the one chosen in Figure~\ref{fig:BGT85}. The value of $F_1$ is suggested by the analysis of dense systems by \cite{AT86} (N.\ Borderies, private communication) and the
value of $F_2$ is the smallest one found in Saturn's rings.} of $F_1=0.55$ 
and $F_2=7$, \cite{BGT86} quote $q_2=0.79$.

  In conclusion, it appears that the reversals of the angular momentum
flux and luminosity for sufficiently perturbed flows is a general
feature, the existence of which does not depend on the details of the
microphysics controlling the pressure tensor. This is fortunate, as angular momentum luminosity reversal
plays a central r\^ole in the ring confinement by satellites (the
so-called ``shepherding mechanism"; see \citealt{BGT84,BGT89}).

\subsection{Summary and parametrization of the pressure tensor}
  This whole section was mainly devoted to the derivation of the
pressure tensor under various sets of physical conditions which are likely
to be relevant to planetary rings. The pressure tensor influences the
mean motion only through the three coefficients $t_1$, $t_2$ and
$a_{r\theta}$ defined in Eqs.~(5.28) through (5.31). The analyses
presented in this section show that there exists a number of general
features characterizing these three quantities, as pointed out by
\cite{BGT86}.
\begin{enumerate}
\item Streamlines cross at $q=1$, so that the pressure tensor is
likely to diverge as $q\rightarrow 1$, or even for smaller values of $q$, as shown on Figure~\ref{fig:Shu} and in the related discussion.
\item $t_1$ and $t_2$ vanish as $q\rightarrow 0$, because the flow,
and therefore the pressure tensor components, become axisymmetric in
this limit. Hence, it is reasonable to assume that $t_{1,2}\propto q$ 
for small $q$, as this dependence is characteristic of both the dilute
and dense models. Notice that $t_1$ is negative in the
dilute approximation, whereas it is positive for small values of $q$ in
the dense model; $t_2$ is negative in both models.
\item We have shown that the energy dissipation due to inelastic
collisions implies $2qt_1<3a_{r\theta}$. This is a general result, which
does not depend on the choice of the collisional model [see Eq.~(5.80)].
\item We have just seen that $a_{r\theta}$ is in general positive
for small $q$, so that angular momentum flows outwards. However, as
$q\rightarrow 1$, the direction of the angular momentum flow is
reversed, so that there is some value $q=q_a(\sigma_0)$ for which
$a_{r\theta}=0$ and the transition between the two regimes occur. For
dense systems, $q_2$ is independent of $\sigma_0$.
\end{enumerate}

  From these general considerations, \cite{BGT86} were motivated to devise simple empirical formul\ae\ for the three
coefficients $a_{r\theta}, t_1, t_2$, of the form

\begin{plain}
$$a_{r\theta}=B_a\sigma_0^b{q_a-q\over(q_c-q)^c},\eqno(5.86)$$
$$t_1=B_1\sigma_0^b q{q_1-q\over(q_c-q)^c},\eqno(5.87)$$
$$t_2=B_2\sigma_0^b q{q_2-q\over(q_c-q)^c}.\eqno(5.88)$$
\end{plain}

  In these equations, $B_a, B_1$ and $B_2$ are positive quantities,
$0<q_a<1$, $q_1<q_a$ is either positive or negative, and $q_2$ is
negative. In the model for dense systems, $b=3$, $q_c=1$ and the 
rapid divergence of
the viscous coefficients is well represented by $c\simeq 3$. In
principle, $q_a$, $q_1$, $q_2$ and $q_c$ are functions of $\sigma_0$,
except for dense systems. Orders of magnitude for $B_a$, $B_1$ and
$B_2$ can be obtained from the comparison of these relations with Eqs.~(5.37) and (5.38) on one hand, and (5.69) and (5.70) on the other.

\section{Ring dynamics: perturbation equations and conserved quantities}
  
   We have already listed the perturbations
acting on a ring: the ring self-gravity, the gravitational action of
satellites, the disk pressure tensor. For the satellite perturbations,
we will restrict ourselves to one Fourier component in the expansion of
Eq.~(4.37), in order to investigate in some detail the dynamics of a
density wave near an isolated Lindblad resonance. A complete discussion 
of disk-satellite interactions (e.g. of the effect of such interactions on the
eccentricity evolution of narrow elliptic rings, or of the shepherding
of a ring by a satellite), would take us too far afield and is outside 
the scope of these notes. 

   A number of derivations are more easily performed when the
rings are represented as a discrete collection of streamlines. This point
of view is adopted in what follows. The continuum description will be
recovered by taking the appropriate limits when needed. Conversely, one can go from the continuum limit to the discrete form of the equations. 

It is well known in Celestial Mechanics that 
perturbing accelerations generate both
periodic and secular (or long period) terms. The periodic effects have
usually a small amplitude in ring dynamics and can safely be neglected. Also, the long term dynamics is easier to capture in simulations once the short-term effects are averaged out. We
will therefore average the perturbation equations in what follows. There are some added subtlety in this procedure for the fluid equations with respect test particles:

\begin{enumerate}
\item The perturbation equations are not used as is commonly done in Celestial Mechanics, by inserting an unperturbed solution in the right-hand side and deducing the magnitude of the perturbation. Instead, the perturbed motion is also inserted in the right-hand side, so that these equations provide self-consistency conditions for the existence of the assumed form of the perturbed motion. Because of this, $M_e=m(\varphi-\Omega_p t) + m\Delta$ is in fact always enforced. This implies that $M_e$ is not and independent phase, but a quantity depending on $a,\varphi$ and $t$.
\item The equations are phase-averaged. Any form of averaging can be freely chosen  --- this a matter of convenience, not a theoretical requirement --- and it turns out that this choice is more adequate than the usual time-averaging procedure, in particular because $M_e=M_e(a,\varphi,t)$. In the process, short-periodic terms of order $\epsilon g_{pert.}/g_{plan.}$ are averaged out ($g_{pert.}$ and $g_{plan.}$ stand for the accelerations due to the perturbing forces and the planet's, respectively). This is required for the self-consistency of the assumed form of perturbed motions and this self-consistency is ensured by the smallness of the short periodic term. In this sense, the perturbation equations are indeed handled perturbatively, albeit again in a slightly not usual way. 
\end{enumerate}

In what follows, the subscript $e$ of $a_e$, $M_e$ and $\varpi_e$
is dropped
for convenience. However, it should be kept in mind that the elements
involved in the discussion are epicyclic elements and not elliptic ones.

For definiteness, streamlines and associated quantities are labeled with their index $i$; $1\le i \le N_s$, where $N_s$ is the total number of streamlines. Streamlines width is noted $\delta a$ when needed. Streamlines semi-major axis are defined as the mid-position in the streamlines, and all quantities are evaluated their, except the stress tensor which should be evaluated at the streamline boundary; however this would introduce a staggered double-mesh and this refinement is avoided in these notes.

\subsection{Disk self-gravity}\label{sec:sg}

Let us consider two streamlines (indexed as streamline 1 and
streamline 2) with $m$ lobes and orbital elements $a_1,\epsilon_1,
\Delta_1$ and $a_2,\epsilon_2, \Delta_2$ respectively. We suppose
$a_1<a_2$ for definiteness. We are going to compute the gravitational
perturbation of streamline 2 on streamline 1. In fact it is sufficient
to compute the gravitational perturbation on a fluid particle of
streamline 1, because this perturbation is identical for all fluid
particles, once averaged over the orbital time-scale. For $\Delta
a_{ij}\equiv a_i-a_j$ small enough (that is, much smaller than the curvature
radius of the streamlines, i.e. $\Delta a_{ij}\ll a$) one can locally identify 
the streamline with
a straight line and find the perturbing acceleration with
the help of Gauss's theorem. 

  Let us call $\lambda_2$ the linear mass density of streamline 2. The gravitational acceleration ${\bf g}_{sg}$
on a fluid particle is given by \citep{BGT83c}

\begin{plain}
$${\bf g}_{sg}={2G\lambda_2\over \Delta_c}{\bf u},\eqno(6.1)$$
\end{plain}

\noindent where ${\bf u}$ is the unit vector perpendicular to
streamline 2 and directed outwards (from streamline 1 to streamline 2), and $\Delta_c$ the distance of the fluid
particle to streamline 2 along ${\bf u}$. Let us introduce the angle $\beta$
between the radial direction and ${\bf u}$, so that $\Delta_c=|r_1-r_2|\cos\beta$. 
  The radial and tangential components of the perturbing acceleration
read:

\begin{plain}
$$R_{sg}=-{2G\lambda_2\over\Delta r_{12}},\eqno(6.2)$$
$$S_{sg}=-{2G\lambda_2\over\Delta r_{12}\cos\beta}\sin\beta,\eqno(6.3)$$
\end{plain}

\noindent where $\Delta r_{ij}\equiv r_i-r_j$. Elementary geometric
considerations lead to the following expressions for $\sin\alpha$ and
$\cos\alpha$:

\begin{plain}
$$\cos\beta={r_2d\theta/d\varphi\over \sqrt{(r_2d\theta/d\varphi)^2+\left(dr_2/d\varphi\right)^2}}=1+
\mathcal{O}(\epsilon_2^2),\eqno(6.4)$$
$$\sin\beta=-{dr_2/d\varphi\over\sqrt{(r_2d\theta/d\varphi)^2+\left(dr_2/d\varphi\right)^2}}=
-{1\over a_2}{dr_2\over d\varphi}+\mathcal{O}(\epsilon_2^2).\eqno(6.5)$$
\end{plain}

The last equalities are evaluated from Eq.~(4.57).

Finally, the components of the perturbing acceleration at position ($r,\theta$) and at time $t$ read:

\begin{plain}
$$R_{sg}=-{GM_2\over\pi a_2(r-r_2)},\eqno(6.7)$$
$$S_{sg}={GM_2m \epsilon_2\sin(m(\theta-\Omega_p t)+m\Delta_2)\over
\pi a_2(r-r_2)},\eqno(6.8)$$
\end{plain}

\noindent where the linear mass density has been expressed in terms of
the total mass of the streamline\footnote{One might wonder if the possible dependence of the linear mass density with azimuth contributes to these results, to the order in $\epsilon$ aimed at. However, this is not the case. In any case, such contributions would not affect the results of Eq.~(6.12) to (6.14).} $M_2=2\pi a_2\lambda_2$ and where
$r_2$ is a function of $\theta$ through Eq.~(4.57); we have used
the fact that $\varphi=\theta$ to lowest order in $\epsilon$. 
The streamline mass is connected to the ring surface density by 
$M_2=2\pi a_2\delta a\sigma_0$ where $\delta a$ is the distance between
two neighboring unperturbed streamlines. Eq.~(6.7)
corrects an unimportant error in the corresponding equation of \cite{BGT85}.

  A simple trigonometric manipulation shows that $\Delta r_{12}=
J_{12}\Delta a_{12}$ where

\begin{plain}
$$J_{12}=1-q_{12}\cos[m(\theta-\Omega_p t)+m\Delta_1+\gamma_{12}],\eqno(6.9)$$
\end{plain}

\noindent and

\begin{plain}
$$q_{12}{\rm exp}(i\gamma_{12})={a_1\epsilon_1 - a_2\epsilon_2{\rm exp}
[im(\Delta_2-\Delta_1)]\over\Delta a_{12}}.\eqno(6.10)$$
\end{plain}

\noindent Note that $q_{12}=q_{21}$, but $\gamma_{12}\neq\gamma_{21}$. As
$\Delta a_{12}\rightarrow 0$, $J_{12}\rightarrow J$, $q_{12}\rightarrow
q$ and $\gamma_{12}\rightarrow \gamma$. With the same definitions, one 
also has:


\begin{plain}
$$\eqalignno{&a_2\epsilon_2\sin[m(\theta-\Omega_p t)+m\Delta_2]=a_1\epsilon_1\sin[m(\theta-\Omega_p t)+m\Delta_1]-\cr
& \kern3truecm \Delta a_{12}q_{12}\sin[m(\theta-\Omega_pt)+m\Delta_1+\gamma_{12}]
&(6.11)\cr}$$
\end{plain}

\noindent We can now compute the perturbations of $a_1,\epsilon_1$ and
$\varpi_1$ by inserting Eqs.~(6.6) through (6.11) into Eqs.~(4.20) through
(4.22), and averaging over $\varphi_1$, with the additional constraint that 
$M_1=m(\varphi-\Omega_p t) +m\Delta_1$ and $\theta=\varphi$. Furthermore, 
we take $a_1=a_2=a$ except in the difference $\Delta a_{12}$ (remember
$\Delta a_{12}\ll a$ in ring problems), and keep the results to lowest
nonvanishing order in $\epsilon$. Finally, we take $\Omega_a=\kappa_a=
(GM_p/a^3)^{1/2}$ in the perturbation equation (this leads to
negligible fractional errors of order $J_2$), except in the difference
\footnote{These approximations are made
in the computations of all perturbations in the rest of the paper.}
$\Omega_a-\kappa_a$.
The resulting averaged equations read:

\begin{plain}
$$ \left(da_1\over dt\right)_{sg}=-{2(m-1)n_a\over\pi}{M_2\over M_p}
{a\over\Delta a_{12}}a\epsilon_1
H(q_{12}^2)q_{12}\sin\gamma_{12},\eqno(6.12)$$ 
$$\left(d\epsilon_1\over dt\right)_{sg}={n_a\over\pi}{M_2\over M_p}
{a\over\Delta a_{12}}H(q_{12}^2) q_{12}\sin\gamma_{12},\eqno(6.13)$$
$$\left(d\varpi_1\over dt\right)_{sg}={n_a\over\pi}{M_2\over M_p}
{a\over\epsilon_1\Delta a_{12}}H(q_{12}^2)
q_{12}\cos\gamma_{12},\eqno(6.14)$$
\end{plain}

\noindent where $H(q^2)$ is defined by (the last equality is taken from
\citealt{GR80}):

\begin{plain}
$$H(q^2)\equiv {1\over 2\pi q}\int_{-\pi}^{\pi}{\cos u\over 1-q\cos u}du
= {1-(1-q^2)^{1/2}\over q^2(1-q^2)^{1/2}}.\eqno(6.15)$$
\end{plain}

  These equations are of course also valid in the case $m=0$.
The only restriction comes from the constraint $\Delta a_{12}\ll a$, so 
that they cannot be used to compute the axisymmetric contribution 
of the self-gravity of wide rings (like
Saturn's rings). However, this contribution can be included, if needed,
in the computation of $\Omega_a$ and $\kappa_a$, so that this limitation
is not essential.

\subsection{Satellite perturbations}
  We are now going to compute the averaged perturbation generated by one
Fourier component of the satellite potential Eq.~(4.37). The perturbing
potential of the satellite at location $r,\theta$ and time $t$ reduces
to:

\begin{plain}
$$\phi_s=\Phi_{mk}(r/a_s)\cos m(\theta-\Omega_p t).\eqno(6.16)$$
\end{plain}

  The derivatives of this potential give the components of the
perturbing acceleration, which can be evaluated along a ring streamline 
of elements $a,\epsilon,\Delta$, and expanded to first order in $\epsilon$
with the help of Eqs.~(4.62) and (4.63). After these manipulations, one
obtains:

\begin{plain}

$$\eqalignno{
R_s = &-{d\Phi_{mk}\over da}(\cos M\cos m\Delta +\sin M\sin m\Delta),&(6.17)\cr
S_s=&m{\Phi_{mk}\over a}(\sin M\cos m\Delta -\cos M\sin m\Delta)
\cr &- m\left({d\Phi_{mk}\over da}-{\Phi_{mk}\over a}\right)\epsilon(
\cos M\sin M\cos m\Delta -\cos^2 M\sin m\Delta)\cr &+
2m^2{\Phi_{mk}\over a}\epsilon(\sin M\cos M\cos m\Delta+\sin^2
M\sin m\Delta),&(6.18)\cr}$$
\end{plain}

\noindent where the relation $M=m(\varphi-\Omega_p t)+m\Delta$ has been
used, and where $\Phi_{mk}$ has been evaluated at $a$. Inserting these 
results into the perturbation equations and averaging over $\varphi$ yields:

\begin{plain}
$$\left(da\over dt\right)_s=n_a a(m-1)\epsilon{a\Psi_{mk}\over G
M_p}\sin m\Delta,\eqno(6.19)$$
$$\left(d\epsilon\over dt\right)_s=-n_a {a\Psi_{mk}\over 2 G
M_p}\sin m\Delta,\eqno(6.20)$$
$$\left(d\varpi \over dt\right)_s={n_a\over \epsilon} 
{a\Psi_{mk}\over 2 G M_p}\cos m\Delta,\eqno(6.21)$$
\end{plain}

\noindent where one has defined

\begin{plain}
$$\Psi_{mk}\equiv a{d\Phi_{mk}\over da} + 2m\Phi_{mk}.\eqno(6.22)$$
\end{plain}

The question of the effect of multiple resonant terms is complex and not addressed in these notes. The interested reader is referred to \cite{BGT83a} and \cite{GT81}.

\subsection{Pressure tensor}\label{sec:vis}
   The radial and tangential perturbing accelerations due to the ring
pressure tensor are given in Eqs.~(3.17) and (3.18). However, it is 
preferable for our purposes to derive the perturbing acceleration produced 
on a streamline of mass M by the material outside it. 

The pressure tensor captures the effect of collisions of one streamline on its neighbors. The vertically integrated acceleration produced on any streamline in the continuum limit reads: 

\begin{plain}
$$R_{vis}=-{1\over a\sigma_0}{\partial aP_{rr}\over \partial a} +\mathcal{O}(\epsilon),\eqno(6.23)$$
$$S_{vis}=-{1\over a^2\sigma_0}{\partial a^2 P_{r\theta}\over\partial a} +\mathcal{O}(\epsilon),\eqno(6.24)$$
\end{plain}

\noindent The only subtlety here is that the streamline $i$ boundaries lie at $(a_i+a_{i\pm 1})/2$; we have seen in section 5 that the pressure tensor depends on azimuth only through $M'$, which therefore differs from $M'_i$ at these two boundaries. In the continuum limit, a similar problem arises, and the angular average needs an integration by part before it can be performed $\langle R \cos M\rangle = \langle 1/\sigma_0\partial P_{rr}/\partial a \cos M \rangle = \langle 1/\sigma_0[\partial P_{rr}\cos M/\partial a + (dm\Delta/da) P_{rr}\sin M ] \rangle$. 

Both the discretized and continuum approaches lead to the same result and the phase-averaged perturbation equations read

\begin{plain}
$$\left(da_i\over dt\right)_{vis}=-{4\pi \over a n_a M_i}\Delta^{\pm} (a^2 a_{r\theta}),\eqno(6.25)$$
$$\left(d\epsilon_i\over dt\right)_{vis}={2\pi\over n_a M_i}\left[
-\Delta^{\pm} (t_1\cos\gamma+t_2\sin\gamma)+ (t^i_1\sin\gamma_i -t^i_2\cos\gamma_i)\Delta^{\pm} (m\Delta)\right],\eqno(6.26)$$
$$\left(d\varpi_i\over dt\right)_{vis}={2\pi\over n_a\epsilon_i M_i}\left[
\Delta^{\pm} (t_1\sin\gamma-t_2\cos\gamma)+ (t^i_1\cos\gamma_i +t^i_2\sin\gamma_i)\Delta^{\pm} (m\Delta)\right],\eqno(6.27)$$
\end{plain}

\noindent where $M_i$ is the mass of streamline $i$, $\Delta X \equiv X^{i,i+1}-X^{i-1,i}$ and $X^{ij}$ is evaluated at the boundary between streamlines $i$ and $j$ [i.e. in $(a_i+a_j)/2$, $|\Delta a|$ being the inter-streamline distance]; $t_1,t_2$ and $a_{r\theta}$ are the quantities introduced in Eqs.~(5.28) through (5.31).

This results differs from \cite{BGT83a}, \cite{BGT85} and \cite{BGT86} in several respects. First, $\sigma_0$ has been included here in the pressure tensor components. The rationale for this difference is that the full pressure tensor divergence defines the acceleration on a fluid particle and is therefore included in the difference $\Delta^{\pm} X$. This correction is unimportant in the WKB limit of \cite{BGT86} and in the nearly incompressible limit of \cite{BGT85}. Also, Eqs~(6.26) and (6.27) include a contribution --- the $\Delta^{\pm}(m\Delta)$ term --- that is missing from the various BGT papers. This term provides in fact the dominant stress tensor contribution in the tight-winding approximation that is relevant for density waves; BGT were nevertheless able to recover the correct result from the other term by an (incorrect) non-local approximation to $\gamma$. Finally, factors $a,a^2$ have been pulled out of Eqs.~(6.26) and (6.27), which are relevant only for perturbed flows, but not from Eq.~(6.25), which applies to circular flows as well and where this dependence is important.

\subsection{Mass, energy, and angular momentum budget of ring systems:}

The question of energy dissipation and viscous angular momentum
transport has been addressed in sec V. Here, we wish to derive more
general expressions, in Eulerian variables ($a,\varphi$). Furthermore,
as we are mostly interested in radial transport, all the following
expressions will be integrated over $\varphi$. For this purpose, we note
$\langle X\rangle$ the azimuthal average of any quantity $X$. We have

\begin{plain}
$$\langle X\rangle ={1\over 2\pi}\int_{0}^{2\pi}d\varphi\ X
={1\over 2\pi}\int_{0}^{2\pi}dM\ X
={1\over 2\pi}\int_{0}^{2\pi}dM'\ X,\eqno(6.28)$$
\end{plain}

\noindent so that this definition is in agreement with the one used in
section 5; note also that all the perturbation equations of the previous
sections should more properly have been written in this bracket
notation. 

  The three quantities we are interested in are the ring unperturbed
surface density $\sigma_0$, and the energy ${\cal E}$ and angular momentum 
${\cal H}$ per unit unperturbed radial length $a$,

\begin{plain}
$${\cal E}=2\pi a\sigma_0 E,\eqno(6.29)$$
$${\cal H}=2\pi a\sigma_0 H.\eqno(6.30)$$
\end{plain}

\noindent By definition, these are azimuthally averaged quantities. In this section we depart from the semi-Lagrangian approach adopted so far to adopt a semi-Eulerian one instead. As such, we are interested in the time evolution in ($a,\varphi$) coordinates instead of using these quantities as Lagrangian labels. The main reason for this departure is that conserved quantities provide not only constraints on the dynamics, but also lead to a powerful amplitude equation for density waves. Note that, as any quantity $X=X(a,\varphi,t)$, one can always write

\begin{plain}
$${dX\over dt}={\partial X\over\partial t}+ 
{da\over dt}{\partial X\over\partial a}+
{d\varphi\over
dt}{\partial X\over\partial\varphi}.\eqno(6.33a)$$
\end{plain}

  The equation of conservation of mass is obtained from Eq.~(3.16)
either by change of variables or by more direct means and reads: 

\begin{plain}
$${\partial\sigma_0\over\partial t}+{1\over a}{\partial\over\partial
a}\left(a\sigma_0{da\over dt}\right)+{1\over a}{\partial\over\partial
\varphi}\left(a\sigma_0{d\varphi\over dt}\right)=0,\eqno(6.31)$$
\end{plain}

\noindent and yields, after azimuthal average,

\begin{plain}
$${\partial\sigma_0\over\partial t}+{1\over a}{\partial\over\partial
a}\left(a\sigma_0\left\langle{da\over dt}\right\rangle\right)=0.\eqno(6.32)$$
\end{plain}
Note that the designation of unperturbed surface density for
$\sigma_0$ is somewhat improper because it evolves due to the radial drift that the perturbations (in particular the ring internal stress) induce.

  The equations for the ring energy ${\cal E}$ and angular momentum
${\cal H}$ are most easily derived in integrated form. First, note that
for any quantity $X$, using the mass conservation constraint Eq. (6.31), one has

\begin{plain}
$$a\sigma_0{dX\over dt}={\partial a\sigma_0 X\over\partial t}+ 
{\partial\over\partial a}\left(a\sigma_0 X{da\over dt}\right)+
{\partial\over\partial\varphi}\left(a\sigma_0 X{d\varphi\over
dt}\right),\eqno(6.33b)$$
\end{plain}

\noindent so that after integration between two unperturbed radii\footnote{This integration is very handy to tackle the change of variable ($r\theta$) to ($a,\varphi$) for the pressure tensor contribution.} $a_1$
and $a_2$ and azimuthal average, one obtains (assuming $X$  independent of $\varphi$ and defined per unit mass such as $E$ or $H$)

\begin{plain}
$$\eqalign{\int_{a_1}^{a_2}da{\partial\over\partial t}\left(2\pi a\sigma_0 X\right)
= & -\int_{a_1}^{a_2}da{\partial\over\partial a}\left(2\pi
a\sigma_0 X\left\langle{da\over dt}\right\rangle\right) \cr
& + \int_{a_1}^{a_2}da\ 2\pi a\sigma_0\left\langle{dX\over
dt}\right\rangle_{pert}.}\eqno(6.34)$$
\end{plain}

  We can apply this result to the computation of the equations of
evolution of ${\cal E}$ and ${\cal H}$, by noting that\footnote{$(\Omega_a/\kappa_a)^2\epsilon^2\simeq \epsilon^2$ consistently with earlier simplifications, but we keep the correct $J_k$ terms to zeroth order in eccentricity.}

\begin{plain}
$${dE\over dt}={\Omega^2 a\over 2}{da\over dt},\eqno(6.35)$$
$${dH\over dt}={1\over 2}\Omega a{da\over dt}-\Omega
a^2\epsilon{d\epsilon\over dt}=rS.\eqno(6.36)$$
\end{plain}

  Let us consider first the perturbations induced by the satellite. Note
that as the potential Eq. (6.16) is uniformly rotating at angular speed
$\Omega_p$, the change in specific energy and angular momentum are
related by Jacobi's constant, and 

\begin{plain}
$$\left\langle dE\over dt\right\rangle_{sat}-\Omega_p\left\langle  dH\over 
dt\right\rangle_{sat}=0. \eqno(6.37)$$
\end{plain}

\noindent From Eqs.~(6.35) and (6.19), one has

\begin{plain}
$$\int_{a_1}^{a_2} da\ 2\pi a\sigma_0
\left\langle{dE\over dt}\right\rangle_{sat}=
\Omega_p\int_{a_1}^{a_2} da\ \pi ma\sigma_0\epsilon\Psi_{mk}\sin
m\Delta=\Omega_p\int_{a_1}^{a_2}da\ {\cal T}_s,\eqno(6.38)$$
\end{plain}

\noindent where ${\cal T}_s\equiv 2\pi a\sigma_0\langle{dH/dt}\rangle_{sat}=\pi 
ma\sigma_0 \Psi_{mk}\epsilon\sin m\Delta$ is the torque density due to the 
satellite.

  Let us now consider the contribution of the ring self-gravity.
The self-gravitational potential is also uniformly rotating with
angular speed $\Omega_p$, so that the changes in specific energy and
angular momentum are again related by Jacobi's constant. It is useful to
introduce the rate of transfer of specific energy (resp. specific angular
momentum) $L_E^{sg}$ (resp. $L_H^{sg}$) from all streamlines with radii
$a_1<a$ on all streamlines of radii $a_2>a$ (which is the opposite of
the rate of transfer from $a_2>a$ on $a_1<a$). By definition, and from Eq.~(6.12) and (6.13), one has

\begin{plain}
$$\eqalign{L_E^{sg}=\Omega_p L_H^{sg}=-4\pi Gm\Omega_p\int_0^a da_1\ \sigma_0(a_1)
a_1\epsilon_1\int_a^{+\infty}da_2\ & \sigma_0(a_2)a_2\times \cr
& {H(q_{12}^2)q_{12}\sin \gamma_{12}\over a_1-a_2}.}\eqno(6.39)$$
\end{plain}

\noindent The first equality follows from Eq.~(6.12) by noting that $\Omega_p=(m-1)\Omega$\footnote{For $m\neq 1$.} or simply by noting that the self-gravitational potential is stationary in the rotating frame so that Jacobi's integral applies.

  With this definition, one easily checks that

\begin{plain}
$$\int_{a_1}^{a_2}da\ 2\pi a\sigma_0\left\langle{dE\over
dt}\right\rangle_{sg}=-L_E^{sg}(a_2)+L_E^{sg}(a_1)=-\int_{a_1}^{a_2}
{\partial\over\partial a}L_E^{sg},\eqno(6.40)$$
$$\int_{a_1}^{a_2}da\ 2\pi a\sigma_0\left\langle{dH\over
dt}\right\rangle_{sg}=-L_H^{sg}(a_2)+L_H^{sg}(a_1)=-\int_{a_1}^{a_2}
{\partial\over\partial a}L_H^{sg},\eqno(6.41)$$
\end{plain}

\noindent so that $L_E^{sg}$ and $L_H^{sg}$ are
energy and angular momentum luminosities. The preceding expressions are
correct only if $|a_1-a|,|a_2-a|\ll a$, a condition which is satisfied
in all cases of interest.

We can finally compute the contribution of the ring internal stress, which is somewhat more delicate to handle. First, we cannot use the discrete equations of section 6.3 as the equation for $a$ keeps only the leading order term, whereas, as we shall see now, energy dissipation depends on the next-to-leading order terms. Instead, we revert to the continuous versions
of components of the perturbing acceleration as given in Eqs.~(3.17) and
(3.18). Expressing the various derivatives in terms of
$\partial/\partial a$ and $\partial/\partial\varphi$, and after some
integrations by part, one obtains to leading orders in $\epsilon$

\begin{plain}
$$\int_{a_1}^{a_2}da\ 2\pi a\sigma_0\left\langle{dE\over
dt}\right\rangle_{vis}=
\int_{a_1}^{a_2}da\ \left[-{\partial L_E^{vis}\over\partial a}+\pi\Omega
a(2qt_1-3a_{r\theta})\right],\eqno(6.42)$$
$$\int_{a_1}^{a_2}da\ 2\pi a\sigma_0\left\langle{dH\over
dt}\right\rangle_{vis}=-
\int_{a_1}^{a_2}da\ {\partial L_H^{vis}\over\partial a},\eqno(6.43)$$
\end{plain}

\noindent where $L_E^{vis}$ and $L_H^{vis}$ are defined by

\begin{plain}
$$\eqalign{ L_E^{vis}=2\pi\Omega
a^2[a_{r\theta} & +m\epsilon(2c_{r\theta}-s_{\theta\theta})\cos\gamma+
m\epsilon(2s_{r\theta}+c_{\theta\theta})\sin\gamma \cr 
& +\epsilon(s_{rr}\cos\gamma -c_{rr}\sin\gamma)],}\eqno(6.44)$$
$$\eqalign{ L_H^{vis}=2\pi a^2[a_{r\theta} & +m\epsilon(2c_{r\theta}-s_{\theta\theta})
\cos\gamma+m\epsilon(2s_{r\theta}+c_{\theta\theta})\sin\gamma \cr 
& - 2\epsilon(c_{r\theta}\cos\gamma +s_{r\theta}\sin\gamma)].}\eqno(6.45)$$
\end{plain}

\noindent The terms of order $\epsilon$ are negligible in front of
$a_{r\theta}$ but note that their derivatives may be comparable to $2qt_1-3a_{r\theta}$ in Eq.~(6.42). It is nevertheless legitimate to neglect them, as these two contributions are of a different qualitative nature: the former conserve energy and angular momentum, while the latter dissipate energy. Small energy and angular momentum redistribution terms do not affect the long term evolution, but small dissipation does. These terms are kept however as they are needed for part of the concluding discussion of section 7.2.3.
 
When these terms are neglected, $L_H^{vis}$ reduces to the
quantity introduced in Eq.~(5.82) and $L_E^{vis}=\Omega L_H^{vis}$. 
Note that the second term on the right-hand side of Eq.~(6.42) 
represents the rate of dissipation of macroscopic energy, and 
is equal to the rate of dissipation of energy in collisions as 
computed in Eq.~(5.79).

  We are now in position to write down the equations of evolution of
energy and angular momentum in local form. Introducing

\begin{plain}
$$L_E^c=2\pi a\sigma_0 E\left\langle{da\over dt}\right\rangle,\eqno(6.46)$$
$$L_H^c=2\pi a\sigma_0 H\left\langle{da\over dt}\right\rangle,\eqno(6.47)$$
\end{plain}

\noindent one can finally express them as

\begin{plain}
$${\partial {\cal E}\over\partial t}+{\partial\over\partial a}\left(
L_E^c+L_E^{sg}+L_E^{vis}\right)=\Omega_p{\cal T}_s +\pi\Omega a(2qt_1-3
a_{r\theta}),\eqno(6.48)$$
$${\partial {\cal H}\over\partial t}+{\partial\over\partial a}\left(
L_H^c+L_H^{sg}+L_H^{vis}\right)={\cal T}_s.\eqno(6.49)$$
\end{plain}

 These equations express the fact that the change in ${\cal E}$ and
${\cal H}$ is due to flux terms on one hand (the advective, self-gravity
and viscous fluxes) and to source and sink terms (the satellite and the
ring internal stress).
 
 Note that Eq.~(6.47) is valid even for more general expressions of the
satellite torque than the one derived here. This equation allows us to
derive an important feature of the confinement of rings
by satellites. A complete discussion of the shepherding process is outside
the scope of this lecture, so that we will only briefly recall some essential
facts.

  It is well-known, for example, that the outer edges of the A and B rings 
of Saturn correspond to resonances with satellites, and it has been argued 
that the angular momentum exchanges between the satellite and the ring
material at the edge is responsible for the survival of the edge against
viscous diffusion \citep{BGT82}. The same thing 
is very likely to be true for all the known narrow rings \citep{GT79a,BGT89}. In the case 
of Saturn's F ring and Uranus'
$\epsilon$ ring at least, the two ``shepherd" satellites bracketing
each ring have been observed by the {\it Voyager} probes. 

  The process works as follows. As already mentioned in section 2, the 
gravitational interaction between a ring and a satellite results in 
an outward flow of angular momentum: from the ring to the
satellite if the satellite lies outside the ring (in which case $T_s<0$)
and from the satellite to the ring if it lies inside (in which case
$T_s>0$). We have seen also in section 5.3.2 that for not too strongly
perturbed rings, the viscous stress induces an outward transport of
angular momentum, from the inner edge to the outer edge of a narrow
ringlet, so that the ringlet tends to spread. The spreading will be
halted when the angular momentum fluxes induced by the satellites will
balance the flux due to the viscous stress. Such an equilibrium is
possible, because the satellite torques decrease in absolute value when
the distance between the ring and the satellite increases: therefore, if
the satellite torque is too small, the ring will spread until it reaches
the right magnitude; if it is too large, the ring will contract until
the satellite torque and the viscous torque have again adjusted. Of
course, this equilibrium is not permanent: the inner satellite looses
angular momentum, while the other gains some, and they are repelled by
the ring. However, as they are usually much more massive than the rings,
one can expect that the system will survive in quasi-equilibrium much
longer than the ring alone would. Nevertheless, the evolution
time-scales computed from these angular momentum exchanges are still in
general uncomfortably short.

  Let us now turn to the implication of Eq. (6.47) for this scenario.
Once the quasi-equilibrium just described is obtained, $\partial{\cal H}
/\partial t\simeq 0$ and $L_H^c\simeq 0$. Ignoring for simplicity
the self-gravity term, the integration of this equation from, say, the 
inner edge $a_i$ to some radius $a$ inside the ring yields $L_H^{vis}(a)=
\int_{a_i}^a T_sda$ (a similar result holds for the outer edge). Taking 
the limit $a\rightarrow a_i$, one obtains $L_H^{vis}(a_i)=0$, a natural
result considering that the torque has only a finite density (we have
assumed that there is no very low optical depth material outside the
ring). As unperturbed rings have a positive viscous luminosity, this
condition can only be satisfied if the rate of perturbation of the edge
is high enough so that the condition of reversal of the angular momentum
flux is reached. One sees again that such an equilibrium can be
obtained and is stable: if the satellite perturbation on the edge is
too small, the ring spreads and the edge moves towards the satellite,
whose perturbation increases as it gets closer, until the right value is
obtained. Note that angular momentum reversal is a necessary feature
of the shepherding process (the same conclusion is reached from a
discussion of ring energetics; see \citealt{BGT84}).

\section{Applications to ring dynamics}
 
  In this section, we wish to apply the apparatus of the previous
sections to actual ring dynamical problems: the model of self-gravity
for the rigid precession of elliptic rings, and the description of linear
and nonlinear density waves excited at Lindblad resonances with external
satellites. These applications are
provided only as examples of use of the formalism; therefore only the
major aspects of these questions will be addressed, in order to keep 
the emphasis on physical issues. Two of the major applications of the
formalism, the discussion of the evolution of ring and satellites
eccentricities, and the detailed discussion of the shepherding
mechanism are not described here, although they are of great importance
for ring-satellite system evolution. The reader is referred to
the specialized literature (see, e.g. \citealt{GT80,GT81}, and \citealt{BGT82,BGT89}) on these topics.

\subsection{The self-gravity model for elliptic rings}

  Most of the Uranian rings (in particular, the $\alpha$, $\beta$,
$\gamma$, $\delta$ and $\epsilon$) and some of the Saturnian ringlets (e.g.
the Titan and Huygens ringlets), which are all very narrow features --
from a few kilometers to a few tens of kilometers wide for radii of
the order of $10^5$ kilometers -- are known to be eccentric, as
discussed in sections\footnote{Many of these rings 
are also inclined, but for simplicity we set the inclination to zero.} 
IV.3.1 and IV.3.3. These rings share a number of interesting features:
their apses are almost aligned (a small apsidal shift has been detected
in some rings, in particular the $\alpha$ and $\beta$ rings of Uranus); 
they also exhibit a positive difference of eccentricity between the outer and 
inner edge (except, maybe, the $m=0$ mode of the $\gamma$ ring). The
basic observational and dynamical properties of the Uranian rings have been
reviewed by \cite{EN84}.

  These structures are submitted to various dynamical effects.
The planet quadrupole moment tends to destroy the observed apse, but
this tendency can be balanced by the self-gravity of the ring (\citealp{GT79a,GT79b}; see
below). The internal stress also acts on the apse alignment and on the
ring mean eccentricity, and produces some radial spreading of the ring.
This spreading can be overcome by the action of satellites (see section 6), which also influence the evolution of the mean eccentricity. In
this section, a simplified discussion of the dynamics of narrow ringlets
will be presented, mostly taken from the analysis by \cite{BGT83a}. First, the changes in epicyclic semimajor axes will
be ignored, under the assumption that an equilibrium between viscous
spreading and the action of the satellites has been reached. Second, the
ring will be represented with only two streamlines with equal masses
$M_1=M_2=M_0/2$. Small quantities are supposed to be ordered by two
small parameters, $p_1$ and $p_2$ such that $p_1\ll p_2\ll 1$. The two
streamlines have semimajor axes $a_1$ and $a_2$, with $a_1<a_2$.
Consistently with the approximations performed in the previous sections,
one defines $a=(a_1+a_2)/2$, and assumes $a_1=a_2=a$ in the perturbation
terms, except in the difference $\Delta a_{21}=\delta a$.
The difference of eccentricity between the outer and inner edge 
$\Delta\epsilon_{21}\equiv\delta\epsilon$ is taken to be $O(p_1)$,
whereas the mean eccentricity $\epsilon\equiv (\epsilon_1+\epsilon_2)/2$
is $O(p_2)$, so that $\delta\epsilon/\epsilon\ll 1$, as observed. We
also define $m\Delta=m(\Delta_1+\Delta_2)/2$ and
$\delta(m\Delta)=m(\Delta_2-\Delta_1)$. 
Furthermore, as $q\sim 1$, $\delta a/a$ is
$O(p_1)$, as is $\epsilon\delta(m\Delta)$. Typically, the eccentricities
range from a few times $10^{-4}$ to $10^{-2}$, whereas the eccentricity
differences $\delta\epsilon$ vary from a few times $10^{-5}$ to a few
times $10^{-4}$, so that the approximation $p_1\ll 1$ and $p_2\ll 1$ are
very well satisfied. On the other hand, $\delta\epsilon/\epsilon$ ranges
from $0.03$ to $0.5$, so that the approximation $p_1/p_2\ll 1$ is cruder
but not critical. With these orderings, to leading order in the various
small quantities, one has 

\begin{plain}
$$q_{ij}\cos\gamma_{ij}=a{\delta \epsilon\over\delta a},\eqno(7.1)$$
$$q_{ij}\sin\gamma_{ij}=a\epsilon_j{\delta(m\Delta)\over\delta
a},\eqno(7.2)$$
$$q^2_{12}=q^2_{21}=q^2=\left(a\delta\epsilon\over\delta a\right)^2+
\left(a\epsilon{\delta(m\Delta)\over\delta a}\right)^2,\eqno(7.3)$$
$$\gamma_{12}-\gamma_{21}=\delta(m\Delta)\sim O(p_1/p_2).\eqno(7.4)$$
\end{plain}

  We are now in position to derive the equations of evolution of our
simplified system. First, note that the equation for $m\Delta_i$ is
related to the equation of evolution of $\varpi_e$
by the streamline condition $M=m(\varphi-\Omega_p t)+m\Delta$, so
that

\begin{plain}
$${d(m\Delta_i)\over dt}=(1-m)\Omega_i+m\Omega_p-{d\varpi_i\over
dt}.\eqno(7.5)$$
\end{plain}

  Note that Eq.~(7.5) generalizes Eq.~(4.59) to the case where $\Delta$
is time dependent. By definition, $m\Delta_i$ contains only periodic terms; the
secular terms are accounted for by $\Omega_p$.
  From Eqs. (4.21), (4.22), (6.13), (6.14), (6.26) and (6.27)
applied to the evolution of the eccentricity and apsidal shift of the two
streamlines, one can derive the equations of the four quantities
$\epsilon$, $m\Delta$, $\delta\epsilon$, and $\delta(m\Delta)$. They
read, to leading order in the various small quantities:

\begin{plain}
$${d\delta\epsilon\over
dt}=(\Omega_{sg}-\lambda_2)\epsilon\delta(m\Delta)+\lambda_1\delta\epsilon
,\eqno(7.6)$$
$${d\delta(m\Delta)\over
dt}=\delta\Omega_{plan}-(\Omega_{sg}-\lambda_2){\delta\epsilon\over\epsilon}+
\lambda_1\delta(m\Delta),\eqno(7.7)$$
$${d\epsilon\over dt}=-{\Omega_{sg}-\lambda_2\over 4}\delta\epsilon\delta
(m\Delta),\eqno(7.8)$$
$${d(m\Delta)\over dt}=m\Omega_p-\Omega_{plan},\eqno(7.9)$$
\end{plain}

\noindent where $\Omega_{sg}$, $\lambda_1$, $\lambda_2$, $\Omega_{plan}$
and $\delta\Omega_{plan}$ are quantities homogeneous to frequencies and
defined by

\begin{plain}
$$\Omega_{sg}={n\over\pi}{M_0\over M_p}\left(a\over\delta a\right)^2
H(q^2),\eqno(7.10)$$
$$\lambda_1={2 t_1\over q n \sigma_0(\delta a)^2},\eqno(7.11)$$
$$\lambda_2=-{2 t_2\over q n \sigma_0(\delta a)^2},\eqno(7.12)$$
$$\Omega_{plan}=
(m-1)n+{3\over 2}J_2 n\left(R_p\over a\right)^2\left[1+{m-1\over
2}\right],\eqno(7.13)$$
$$\delta\Omega_{plan}= \left({3\over 2}(m-1)n+{21\over 4}J_2 
n\left(R_p\over a\right)^2\left[1+{m-1\over 2}\right]\right){\delta a\over a}.
\eqno(7.14)$$
\end{plain}

\noindent In these equations, $R_p$ is the equatorial radius of the planet, 
$n=(G M_p/a^3)^{1/2}$ and the surface density is related to the ring
mass by $2\pi a\delta a\sigma_0=M_0$. $\Omega_{sg}$ is a characteristic
frequency imposed by the self-gravity of the ring; for typical values of
the ring surface density ($\sigma_0\sim 50$ g/cm$^2$),
$\Omega_{sg}^{-1}$ is of the order of a few years to a few tens of years (e.g.
$\sim 9$ years for the $\alpha$ ring). $\lambda_1^{-1}$ and $\lambda_2^{-1}$ 
are characteristic time-scales imposed by the ring internal stress. They
are usually much longer than $\Omega_{sg}^{-1}$; e.g. assuming $t_1,
t_2\sim\sigma_0 v^2$ with $v\sim 1$ mm/s, $\lambda_1^{-1}\sim 90$ years
for the $\alpha$ ring. This reflects the fact that the ring internal
stress produces a very weak force, even compared to the ring
self-gravity. Finally, $\Omega_{plan}$ and $\delta\Omega_{plan}$ are
frequencies imposed by the planet. For simplicity, terms smaller than
$J_2$ have been neglected. Note that even the $J_2$ term is completely
negligible for $m\neq 1$, in which case $\Omega_{plan}\sim n$ and 
$\delta\Omega_{plan}\sim n \delta a/a$. For
$m=1$ (i.e. purely elliptic modes), the only remaining contribution is
that due to $J_2$: $\Omega_{plan}\sim J_2 n$ and
$\delta\Omega_{plan}\sim J_2 n \delta a/a$, so that these two quantities
are $10^2$ or $10^3$ times smaller for $m=1$ than for $m\neq 1$. It has
been implicitly assumed that the right-hand sides of Eqs.~(7.6) through
(7.9) are linear in the unknowns. This is
not true, as $\Omega_{sg}$, $\lambda_1$ and $\lambda_2$ depend on $q$.
However, $q$ is always different enough from unity that the dependence
of these three frequencies on $q$ can be ignored, and this assumption is
made in the remainder of this subsection. Note also that the
contribution of the satellites has not been included, as the equations
derived in section 6.2 apply for a single isolated resonance, whereas
the cumulative effect of all resonances should be considered.  

We are now in position to describe some simple consequences of Eqs.~(7.6) through (7.9). Note first that Eq.~(7.9) uncouples from the other
three, and implies that $dm\Delta/dt$ is a constant. This constant 
must be equal to zero, in accordance with the interpretation of $\Omega_p$ as
the rotation velocity of the pattern defined by the streamlines.
Therefore, Eq.~(7.9) is in fact the relation imposing $\Omega_p$:

\begin{plain}
$$\Omega_p=\Omega_{plan}/m.\eqno(7.15)$$
\end{plain}

This shows that a narrow ring described by an $m\neq 1$ mode precesses
at a rate $\sim n$, i.e. $\sim J_2^{-1}$ faster than a purely elliptic 
($m=1$) ring. This feature is a direct consequence of the dominance of
the planet on the motions, Eq.~(4.51).

Note also that the time-scale of evolution of $\epsilon$ is $\sim
(p_2/p_1)^2$ longer than the time-scale of evolution of $\delta\epsilon$
and $\epsilon\delta(m\Delta)$. Therefore, we can look into the evolution
of these last two quantities while assuming that $\epsilon$ is constant
in time, in first approximation, which effectively uncouples Eq (7.8)
from Eqs. (7.6) and (7.7). These two equations show that a narrow ring
has an equilibrium configuration (i.e., $d\delta\epsilon/dt=0$ and 
$d\delta(m\Delta)/dt=0$) for

\begin{plain}
$${\delta\epsilon_0\over\epsilon}={(\Omega_{sg}-\lambda_2)\delta\Omega_{plan}
\over \lambda_1^2+(\Omega_{sg}-\lambda_2)^2}\simeq{\delta\Omega_{plan}\over
\Omega_{sg}},\eqno(7.16)$$
$$\delta(m\Delta)_0=-{\lambda_1\over\Omega_{sg}-\lambda_2}{\delta\epsilon_0
\over\epsilon}\simeq -{\lambda_1\over
\Omega_{sg}}{\delta\epsilon_0\over\epsilon},\eqno(7.17)$$
\end{plain}

\noindent where $\Omega_{sg}\gg \lambda_1,\lambda_2$ has been used in
the second equalities. Furthermore, the general solution of Eqs. (7.6)
and (7.7) is given by

\begin{plain}
$$\delta(m\Delta)=\delta(m\Delta)_0+A\exp(\lambda_1
t)\cos[(\Omega_{sg}-\lambda_2)t+\varphi_0],\eqno(7.18)$$
$$\delta\epsilon=\delta\epsilon_0-\epsilon A\exp(\lambda_1
t)\sin[(\Omega_{sg}-\lambda_2)t+\varphi_0],\eqno(7.19)$$
\end{plain}

\noindent where $A$ is an arbitrary (but small) amplitude.
  The general solution consists of oscillations around equilibrium,
which are damped if the viscous coefficient $\lambda_1<0$, as has
usually been assumed in the literature. This is the case in the dilute
approximation, but we have seen in chapter V that the opposite is true
for dense systems, at least for small values of $q$. Let us assume for
the time being that $\lambda_1<0$, so that the system is driven to its
equilibrium point described by Eqs.~(7.16) and (7.17). Note that Eqs.~(7.18) and (7.19) show that $t_2$ is a pressure-like coefficient, while
$t_1$ has a viscous-like action on the system. 

  The meaning of
Eq.~(7.16) is not especially obvious in the compact form in which it is
displayed. Let us for definiteness consider the case of an $m=1$ mode.
This equation then reduces to

\begin{plain}
$$\delta\epsilon_0={21\pi\epsilon\over 4}J_2{M_p\over M_0}\left(R_p\over
a\right)^2\left(\delta a\over a\right)^3{1\over H(q^2)}.\eqno(7.20)$$
\end{plain}

  For $m\neq 1$, one obtains
  
\begin{plain}
$$\delta\epsilon_0={3\pi(m-1)\epsilon\over 2}
{M_p\over M_0}\left(\delta a\over a\right)^3{1\over H(q^2)}.\eqno(7.21)$$
\end{plain}

  One sees that these equations relate parameters describing the overall
shape of the ring, $a$, $\delta a$, $\epsilon$, $\delta\epsilon$ to the
mass of the ring $M_0$, and indeed this relation has been used in the
literature to estimate the mass of the $\alpha$, $\beta$, $\delta$ and 
$\epsilon$ rings of Uranus (the data analysis has not yield values for
$\delta\epsilon$ for the other rings, due to problems which are only
partly understood). Furthermore, as $\lambda_1\ll\Omega_{sg}$,
$\epsilon\delta(m\Delta)_0\ll\delta\epsilon$, so that $q\simeq
a\delta\epsilon/\delta a$, and $\gamma\simeq 0$. Therefore, the width of
the ring, which is given by $W=J\delta a$ as a function of azimuth, has
the same azimuthal behavior as the ring radius $r=(r_1+r_2)/2$, so that
$W\propto r$, as observed. Note also that these two relations predict
$\delta\epsilon_0>0$ for $m>0$, as observed. Because of these successes,
the self-gravity model was for a long time widely regarded as the
correct explanation of the rigid precession of narrow elliptic rings.

In what concerns the evolution of the eccentricity, note that Eq.~(7.8) implies that $d\epsilon/dt\simeq -\Omega_{sg}\delta\epsilon_0
\delta(m\Delta)_0\sim \lambda_1(\delta\epsilon_0)^2/\epsilon
<0$ if $\lambda_1<0$, so that\cite{BGT83a} were motivated to assume that the observed
eccentricities were the result of a balance between viscous damping and
the excitation by the ring ``shepherd" satellites.

  Let us conclude this subsection with a number of comments. 
First, if $\lambda_1$ is positive, the oscillations of $\delta\epsilon$
and $\delta(m\Delta)$ are amplified instead of damped, and the
eccentricity grows, as can be seen from our estimate $d\epsilon/dt\sim
\lambda_1 (\delta\epsilon_0)^2/\epsilon$. This behavior is characteristic
of what is commonly referred to as viscous overstabilities\footnote{Viscous instabilities are axisymmetric motions; viscous overstabilities involve an oscillatory response.}, and this
argument was used by \cite{BGT85} to argue that
such instabilities could as well be responsible for the eccentricities of
the narrow rings. However, the outcome of the evolution of the
eccentricity depends critically on the existence of the oscillations of
$\delta\epsilon$ and $\delta(m\Delta)$, and on the necessary change of
sign of $\lambda_1$ (see section 5), so that, in fact, an initially 
circular ring cannot
reach eccentricities $\epsilon\gg\delta\epsilon$, at least in the
framework of the two-streamline model for narrow rings \citep{LR95}; note that both effects were ignored in the analysis
by \citealt{PL88}). It remains to be seen if the same conclusion 
holds for more general models of narrow rings.

  But the most important problem encountered by the self-gravity model
arose after the Voyager II encounter with Uranus, which revealed
that the mass ot the rings deduced from
Eqs.~(7.20) and (7.21) was roughly underestimated by a factor of $\sim 10$.
The problem is particularly acute for the $\alpha$ and $\beta$ ring
\citep{GP87}. The first piece of evidence comes from
the radio data, the analysis of which gives estimates of the ring
surface densities. The other piece of evidence is connected to the
existence of an unexpectedly extended hydrogen atmosphere around
Uranus; this atmosphere is the source of an extra torque acting on the
ring, that the shepherd satellites have to balance (in addition to the
viscous torque) in order for the
narrow rings to survive, but it turns out that this requirement also
can only be satisfied for ring surface densities about ten times larger
than the ones derived from Eqs.~(7.20) and (7.21) (for more details
about this point, see \citealt{GP87}). 

  On the other hand, the analysis of the radio data relies on standard
radiative transfer theory, which, applied to the Uranian ring, is known
to give inconsistent results, most probably because the mean separation 
between ring particles is not small compared to the radio wavelengths.
Thus, the derivation of the ring surface densities can
possibly be in error, although this does not seem to be a likely
possibility. Similarly, the density of hydrogen in Uranus atmosphere at
the ring location is extrapolated from the density at inner locations,
but major errors in the extrapolation procedure seem rather unlikely.
On the theoretical side, the dynamical agents which can enforce rigid
precession are not very numerous: the precession rates due to the satellites 
and smooth pressure terms seem far too small, and the possibility 
put forward by \cite{GT79a} that the rigid precession
might alternatively be enforced by shock-like phenomena appears to 
be inconsistent with the small observed apsidal shifts. In
conclusion, the issue of the rigid precession of the narrow rings is
still open at the time of writing. 

\subsection{Density waves at Lindblad resonances with external satellites}

\begin{figure}[th]
\centering
\includegraphics[width=0.7\linewidth]{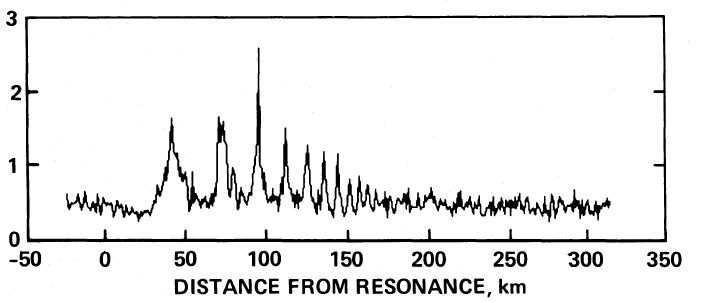}
\caption{\small{The radial variation of optical depth of the Mimas 5:3 density wave as a function of the distance to resonance (reproduced from \citealt{SDLYC85}).}}
\label{fig:Mimas53}
\end{figure}

The two {\it Voyager} spacecrafts have discovered tens of density (and
bending) wavetrains in Saturn's rings, mainly the A and B rings. These
wavetrains share a number of striking characteristics. First, they are
all associated with resonances with satellites, suggesting
that density waves are stable in Saturn's rings (see below). Note that 
the reverse is
not true, i.e. a number of the Lindblad resonances in Saturn's rings are not
associated with density waves: some are associated with gaps, some others
don't exhibit any peculiar behavior at all. The reason of this disparity
is only partly understood (see below). For 
simplicity, and
because of their practical importance, we will only consider density
waves excited at inner Lindblad resonances with Saturn's satellites.
These waves propagate from the resonance outwards (whereas bending waves
propagate inwards). The wavelengths, of the order of a few kilometers or
tens of kilometers, are much smaller than the mean radii of the waves
(of the order of 10$^5$ kilometers). Because the wavelength is so
short compared to the radii, the winding of the density variations 
schematically displayed
on Figure~\ref{fig:ecc} is very high (much higher than represented) so that these
waves appear as quasi-circular features on the {\it Voyager} images.
Generally, the waves propagate over a few (or a few tens) of cycles
before they are damped by the ring internal stress. The radial
wavelength varies like the inverse of the distance to the resonance, so
that it becomes shorter as the wave propagates outwards. For example,
the Mimas 5:3 density wave radial optical depth profile is displayed on
Figure~\ref{fig:Mimas53}. The form of the surface density Eqs.~(4.69) and (4.73) is able to
reproduce correctly this optical depth profile at constant azimuth and
time for an appropriate (and
in fact uniquely determined) choice of the functions involved \citep{LB86}. The propagation of this wave is obviously nonlinear, 
i.e. the
density contrast is large compared to the background density. Assuming 
$\tau\propto\sigma$, Eq.~(4.69) yields $\tau_{p}/\tau_{t}=1-q/1+q$
where $\tau_p$ and $\tau_t$ are the optical depths at the peaks and troughs
of the wave. From the data of Figure~\ref{fig:Mimas53} one sees that $q\lesssim 1$; a
linear wave propagation would require $q\ll 1$. Note that by definition,
the wavenumber $k$ is given by

\begin{plain}
$$k=m{d\Delta\over da}+{d\gamma\over da}.\eqno(7.22)$$
\end{plain}

\noindent The surface density varies radially as
$\sigma=\sigma_0/[1-q\cos(\int kda+cst)]$, so that
$\int_{peak}^{peak}k da\simeq k\lambda=2\pi$, where $\lambda$ is the
wavelength.

   The following discussion is self-contained, and no prior knowledge on
the dynamics of density waves is required. However, some background on
the usual Eulerian density wave linear theory at the introductory level  
of \cite{Sh84} is certainly helpful. The material of this section is mostly
taken from the papers by \cite{SYL85,SDLYC85,BGT85,BGT86}.

   The analysis of density waves is performed in the framework of the
WKBJ approximation (or tight-winding limit), 
which relies on the fact that the dominant
contributions to the radial variations are due to the phase and not to the
amplitude of the quantities under consideration. In this approximation,
the dominant contribution to $J=\partial r/\partial a$ is due to the
derivative of $\Delta$ and not to the derivative of $\epsilon$, so that

\begin{plain}
$$ma\epsilon{d\Delta\over da}\gg{d a\epsilon\over da},\eqno(7.23)$$
\end{plain}

\noindent and, as a consequence,

\begin{plain}
$$\gamma\simeq{\pi\over 2}.\eqno(7.24)$$
\end{plain}

\noindent The wavenumber reduces to $k=md\Delta/da$, and is related to
$q$ and $\epsilon$ by\footnote{Note that {\textit q}
is positive by definition, but that {\textit k} can be either positive
or negative.}

\begin{plain}
$$q|\sin\gamma|=q=|k|a\epsilon.\eqno(7.25)$$
\end{plain}

  Note that for the wave of Figure~\ref{fig:Mimas53}, the wavelength is of the
order of 10 km, so that $\epsilon\sim q/ka\sim 1/ka \sim 10 ^{-5}$ for
$a\sim 10^5$ km, and
the approximation $\epsilon\ll 1$ is remarkably good. It is also
possible to check the validity of the WKBJ approximation from these
data. The wave propagates over a radial distance $\Delta a_w\simeq 200$
km, so that $d a\epsilon/da\sim -a\epsilon/\Delta a_w$. Therefore, 

\begin{plain}
$$\left|{d a\epsilon/da\over ma\epsilon d\Delta/da}\right|\sim 
{a\epsilon/\Delta a_w\over ka\epsilon}={\lambda\over 2\pi\Delta a_w}
\ll 1.\eqno(7.26)$$
\end{plain}

\noindent A general condition of validity of the WKBJ approximation will be
derived below.

  Our purpose is to describe how the wave is excited by the satellite and 
damped by the ring internal stress, and to estimate the exchanges of angular 
momentum between the wave and the satellite, which can be done in principle
once a wave equation has been derived. However, let us derive first the
wave dispersion relation, as quite a number of
aspects of the wave propagation can be understood directly from it.

\subsubsection{Dispersion relation}

  The dispersion relation is an equation relating the wave temporal
frequency, here $\omega=m\Omega_p$ to its spatial frequency $k$. This
relation is most directly obtained from the condition of existence of
the wave, Eq.~(4.59), which states that the resonance condition must be
satisfied throughout the wave region. Eq. (4.59) can be recast as:

\begin{plain}
$$\left(d\varpi\over dt\right)_{sg}+\left(d\varpi\over dt\right)_{vis}+
\left(d\varpi\over
dt\right)_{sat}=\kappa-m(\Omega-\Omega_p).\eqno(7.27)$$
\end{plain}

  Let us evaluate the various terms on the left-hand side in the
tight-winding limit. From Eq. (6.14), one has

\begin{plain}
$$\left(d\varpi\over dt\right)_{sg}={2 n_a a^2\over M_p\epsilon}
\int_0^{+\infty}da'{\sigma_0(a')H(q_{aa'}^2)q_{aa'}\cos\gamma_{aa'}\over
a-a'}.\eqno(7.28)$$
\end{plain}

\noindent Furthermore, we expect that the largest contribution to the
self-gravity integral comes from regions with $|a-a'|\ll a$, so that
we may take $m\Delta(a')=m\Delta+k(a'-a)$; in this approximation 
Eq.~(6.10) reads

\begin{plain}
$$q_{aa'}\exp(i\gamma_{aa'})=ika\epsilon\exp(-iu){\sin u\over
u},\eqno(7.29)$$
\end{plain}

\noindent where we have defined

\begin{plain}
$$u={k(a-a')\over 2}.\eqno(7.30)$$
\end{plain}

 Inserting this result in Eq.~(7.28) yields, treating $\sigma_0$ as a
constant over the region of contribution of the integrand (with $n_a=(GM_p/a^3)^{1/2}$)

\begin{plain}
$$\left(d\varpi\over dt\right)_{sg}={\pi G \sigma_0 |k|\over n_a}C(q),
\eqno(7.31)$$
\end{plain}

\noindent where

\begin{plain}
$$C(q)={4\over \pi}\int_0^\infty du\ {\sin^2 u\over
u^2}H\left(q^2{\sin^2 u\over u^2}\right).\eqno(7.32)$$
\end{plain}

\noindent Note that $C(q)\rightarrow 1$ as $q\rightarrow 0$, i.e. for
linear waves. In general, $C(q)$ is a factor of order unity. Note also that
most of the contribution to this integral comes from regions with
$u\lesssim 1$, i.e. with $(|a-a'|/a\lesssim 1/|k|a\ll 1$, as
expected.

The contribution of the pressure term is computed from the results of section 6.3.; the dominant term in the tight-winding approximation comes from $\Delta^{\pm}(m\Delta)$ term with $\gamma_i\simeq \pi/2$, i.e.:

\begin{plain}
$$\left(d\varpi_i\over dt\right)_{vis}={2\pi\over n_a\epsilon_i M_i} \left[ t^i_2\Delta^{\pm} (m\Delta)\right],\eqno(7.33)$$
\end{plain}

\noindent where $t_i^{jk}(q,\langle\tau\rangle)$ is evaluated $q=q_{jk}$ [i.e.\ at $a=(a_j+a_k)/2$]. Taking the limit 
$\Delta a\rightarrow 0$ and using $q=a\epsilon k = a\epsilon(dm\Delta/da)$, one finally obtains

\begin{plain}
$$\left(d\varpi\over dt\right)_{vis}={k^2 t_2\over n_a \sigma_0
q}.\eqno(7.34)$$
\end{plain}

\noindent Finally, let us argue that the contribution of the satellite can be
neglected. From Eq.~(6.21), and from the tight-winding condition
$|k|a\epsilon=q\sim 1$, one has

\begin{plain}
$$\left|{{\dot\varpi_{sat}}\over{\dot\varpi_{sg}}}\right|\lesssim
{\left|\Psi_{mk}\right|\over 2\pi G\sigma_0 a}.\eqno(7.35)$$
\end{plain}

\noindent Typically in Saturn's rings, this quantity is $\sim 0.1$ to
$0.5$, so that neglecting the satellite contribution is not quite
correct, but not too bad.

  Finally, the nonlinear dispersion relation reads

\begin{plain}
$$n_a [\kappa-m(\Omega-\Omega_p)]-\pi G\sigma_0|k|C(q)-{k^2t_2\over
\sigma_0 q}=0.\eqno(7.36)$$
\end{plain}

\noindent The dispersion relation expresses the fact the the ring
self-gravity and internal stress adjust the fluid particles precession
rate in order to maintain the resonance relation,
$\kappa=m(\Omega-\Omega_p)$, throughout the wave region. Note that by allowing negative values of $m$, this dispersion relation is valid for both inner ($m>0$) and outer ($m < 0$) Lindblad resonances.

Although $q$ reaches values of order unity, it is always small
enough that $C(q)\gtrsim 1$. It is therefore interesting to look into
the linear limit of Eq.~(7.36). For simplicity, we will assume that
the ring behaves as a Newtonian fluid obeying
an isothermal equation of state $p=c_0^2\sigma$ where $c_0$ is the
isothermal sound speed (of the order of the velocity dispersion). This
approximation for the pressure tensor is rather crude, but we will see
shortly that the viscous term can be neglected in the dispersion
relation for the problem of interest here, so that the exact form of
$t_2$ is not essential. In this approximation, $t_2$ can be computed
from Eqs.~(5.29) and (5.39), and reads

\begin{plain}
$$t_2=-\sigma_0 c_0^2{1-(1-q^2)^{1/2}\over q(1-q^2)^{1/2}},\eqno(7.37)$$
\end{plain}

\noindent which reduces to $t_2=-\sigma_0 c_0^2 q/2$ in the linear
limit (note that $2k^2t_2/\Omega\sigma_0 q\sim k^2 c^2$, so
that the following results are relevant even if the isothermal equation
of state does not apply). In this approximation, the dispersion 
relation reads

\begin{plain}
$$2 \Omega [\kappa-m(\Omega-\Omega_p)]-2\pi G\sigma_0
|k|+k^2c_0^2=0,\eqno(7.38)$$
\end{plain}

\noindent which is the standard density wave dispersion relation,
except for the first term on the left-hand side which should read
$\kappa^2 -m^2(\Omega-\Omega_p)^2$ (for a simple fluid mechanical derivation,
see \citealt{Sh84}). This difference is due to the approximations 
we have made in the preceding sections, where
$\kappa\simeq\Omega\simeq n_a \simeq m(\Omega-\Omega_p)$ was assumed for a nearly
Keplerian disk in the vicinity of an inner Lindblad resonance with a
satellite, except in terms involving the difference between $\kappa$ and
$m(\Omega-\Omega_p)$. 

  It is interesting to rederive Toomre's stability 
criterion from the exact linear dispersion relation. Stability is
insured if the temporal frequency of the wave $\omega$ contains no
imaginary part, i.e. if $(\omega-m\Omega)^2>0$, which implies
$\kappa^2-2\pi G\sigma_0|k|+k^2 c_0^2>0$. This expression has a minimum
for $|k|=\pi G\sigma_0/c_0^2$, and it is easy to check that this minimum
is positive if

\begin{plain}
$$Q\equiv{c\kappa\over\pi G\sigma_0}>1.\eqno(7.39)$$
\end{plain}

\noindent A very similar criterion was originally derived by \cite{T64}
for the dynamics
of spiral galaxies. Toomre's $Q$ parameter expresses the fact that the
ring (or stellar) velocity dispersion can stabilize the medium against
the spontaneous generation of density waves. In rings, $Q\sim 1$ but the
exact value is not very well known due to the uncertainty on the
magnitude of the velocity dispersion. Shu 1984 has pointed out that the
ring finite thickness and the existence of a finite size of particles
can also have a stabilizing effect, and the associated criteria are
closely related to Eq.~(7.39). Indeed, as the ring thickness $H\sim
c/\kappa$, Eq.~(7.39) can be recast as $H>\pi G\sigma_0/\kappa^2$ (but
this argument is {\it at most} dimensional: the ring can have a finite 
thickness even in the absence of velocity dispersion, due to the finite
particle size); the
criterion on the particle size is the same (note that the size of the
largest particles of the distribution in Saturn's rings is comparable to
the ring thickness). Presumably, the rings are stable due to all these
effects, although we are still unable to decide on the basis of the
available data if the stability critera are satisfied. 

  An interesting consequence can be derived from the linear
dispersion relation, Eq.~(7.38). Solving for $|k|$, one obtains

\begin{plain}
$$|k|={\pi G\sigma_o\over c_0^2}\pm\left[\left(\pi G\sigma_0\over c_0^2
\right)^2-{2\Omega(\kappa-m(\Omega-\Omega_p)]\over
c_0^2}\right]^{1/2}.\eqno(7.40)$$
\end{plain}

\noindent The $+$ sign corresponds to the short waves and the $-$ sign
to the long ones. Note that the loci of constant surface density
(for example the wave crests) correspond to constant values of
$m(\theta-\Omega_p t)+ \int kda$, so that the isodensity curves
$r(\theta)$ are solutions of the differential equation
$k(dr/d\theta)+m=0$. Consider inner Lindblad resonances for definiteness ($m>0$); thus, for $k>0$, $dr/d\theta<0$ and the wave
is trailing (it curves in the counter-rotation direction), while the opposite
is true for $k<0$ and the wave is leading. A similar conclusion holds at outer resonances ($k<0$ corresponds to trailing waves there).

The short waves have a very high wavenumber at resonance and
therefore oscillate quite rapidly, preventing an efficient coupling with
the satellite potential, which varies smoothly with radius at the
resonance. As a consequence, the satellite is not expected to excite short
waves\footnote{A more
formal argument can be found in \cite{GT79c}, pp. 861 and
862.}; on the contrary, as long waves have $|k|\rightarrow 0$ at the
resonance, they are expected to couple much more efficiently with the
satellite. On the basis of these arguments, the short waves are ignored 
in the remainder of this section. In the region of propagation, $|k|>0$, 
which implies $\kappa-m(\Omega-\Omega_p)>0$. In other words, denoting $a_R$ is the resonance radius implicitly defined by the resonance 
relation $\kappa=m(\Omega-\Omega_p)$, the long trailing waves propagate
outside the ILR (inner Lindblad resonance) and are evanescent inside, whereas the long trailing
waves propagate inside the OLR (outer Lindblad resonance) and are evanescent outside\footnote{Furthermore, long trailing waves propagating from Lindblad resonances are reflected on the corotation radius as a short
leading waves (\citealt{GT78b}, and \citealt{LL79}). This process is not relevant to planetary rings, because the waves damp much before they reach the corotation radius.}. This is consistent with the directions of propagation derived from the wave group velocity. \cite{T69} and \cite{D72} have shown in the context of the
linear density wave theory of spiral galaxies that the group velocity 
of the waves is given by\footnote{The demonstration of this seemingly natural result would take us too far afield to be repeated here. The most elegant derivation is based on the use of a phase-averaged Lagrangian density of the wave, as in \cite{D72}, which shows at the same time that the conserved quantity associated with the invariance with rotation is (not surprisingly) the phase averaged angular momentum density, and that this quantity is transported radially at the group velocity of the wave. For a general introduction to the use of averaged Lagrangian densities in wave theory, see \cite{W74}, chapters 11 and 14.} $c_g=\partial\omega/\partial k$, where $\omega=
m\Omega_p$. Consequently, long leading waves are not expected to
be excited at the Lindblad resonance by the satellite, as they would 
enter a region of evanescence (see also \citealt{GT79c} and \citealt{Sh84}). Therefore, only long trailing waves are discussed in the
remainder of these notes; $|m|\neq 1$ is also assumed for definiteness\footnote{Extending the following results to $|m|=1$ is straightforward but requires to maintain the $\mathcal{O}(J_2)$ difference between $\kappa$ and $\Omega$.}.

  In the vicinity of the resonance, $\kappa-m(\Omega-\Omega_p)\simeq
[3(m-1)\Omega_R/2](a-a_R)/a_R$, so that $(\pi G\sigma_0/c_0^2)^2\gg
2\Omega[\kappa-m(\Omega-\Omega_p)]$ occurs for

\begin{plain}
$${|a-a_R|\over a_R}\ll{\pi^2 G^2\sigma_0^2\over 3c_0^2|m-1|\Omega_R^2}
={1\over 3Q^2|m-1|}\sim 1,\eqno(7.41)$$
\end{plain}

\noindent a condition very largely satisfied. Therefore, keeping the 
leading order term in a Taylor
expansion of Eq.~(7.40) for long waves yields $2\pi G\sigma_0|k|=
2\Omega[\kappa-m(\Omega-\Omega_p)]$, i.e., the pressure term 
can be neglected for long waves near a Lindblad
resonance. The same feature is obviously true of the nonlinear
dispersion relation, as $C(q)\gtrsim 1$, so that Eq.~(7.36) reduces to ($m\neq 1$)

\begin{plain}
$$|k| C(q)={3\over 2\pi}(m-1)\left(M_p\over\sigma_0 a_R^2\right)
\left(a-a_R\over a_R\right){1\over a_R}.\eqno(7.42)$$
\end{plain}

\noindent As the wavelength $\lambda\simeq 2\pi/|k|$, this relation
predicts that $\lambda\propto 1/(a-a_R)$, as observed. As the
coefficient of proportionality depends on $\sigma_0$, the dispersion
relation has been used in the analysis of density wave profiles to
estimate the ring surface density. 

Notice finally that $|k|\rightarrow 0$ as $a\rightarrow a_R$, so that
the tight-winding condition $|ka|\gg 1$, which was used in the
derivation of the dispersion relation, breaks down too close to the
resonance. To estimate quantitatively the region of validity of the
tight-winding condition, let us first compute the value $\lambda_1$
of the first wavelength from Eq.~(7.42), and from the constraint 
$\int_{a_R}^{a_R+\lambda_1}|k|da=2\pi$, taking $C(q)=1$. One obtains

\begin{plain}
$${\lambda_1^2\over a_R^2}={8\pi^2\over 3|m-1|}\left(\sigma_0
a_R^2\over M_p\right).\eqno(7.43)$$
\end{plain}

\noindent It is customary to introduce a small
parameter\footnote{Usually, this small parameter is
denoted by $\epsilon$, but we have changed notation in order to prevent
confusions with the epicyclic eccentricity.} $\delta$, defined by

\begin{plain}
$$\delta\equiv {2\pi\sigma_0 a_R^2\over 3(m-1) M_p}\sim
{M_{ring}\over M_p}.\eqno(7.44)$$
\end{plain}

\noindent This parameter is typically $\sim 10^{-8}$. In terms of
$\delta$, one has $\lambda_1/a_R=(4\pi|\delta|)^{1/2}\sim 10^{-4}$, which
implies $\lambda_1\sim 10$ km, as observed. Note that for disks in
general, the tight-winding condition, $|k|a\gg 1$, or equivalently,
$\lambda_1/a\ll 1$, is valid as long as $(M_{disk}/M_*)^{1/2}\ll 1$,
where $M_*$ is the mass of the central object (star, black hole...), a
condition which is not very well satisfied in spiral galaxies. Now, 
from the dispersion relation and from the expression of
$\lambda_1$, the constraint $|ka|\gg 1$ reads

\begin{plain}
$$4\pi \left(a_R\over\lambda_1\right)\left(|a-a_R|\over\lambda_1\right)\gg
1,\eqno(7.45)$$
\end{plain}

\noindent or equivalently $|a-a_R|/\lambda_1\gg 10^{-4}$, so that the
tight-winding condition is in fact satisfied very close to the resonance, 
well inside the first wavelength. Neglecting the satellite
contribution is not so good an approximation. We will see in the next
subsection that in the linear limit, the satellite contribution is
negligible only outside the first wavelength. This shows however
that the satellite is not directly responsible for
the existence of the wave throughout the propagation region: for most of its
radial extent, the wave propagates essentially as a free wave due to the 
ring self-gravity.

\subsubsection{Forced amplitude and wave damping}

We are interested in this subsection in the description of stationary
density waves, i.e. waves for which $a,\epsilon$ and $\Delta$ do not
vary with time. Therefore, the wave is completely described by the
knowledge of $\epsilon(a)$ and $\Delta(a)$. Following \cite{SYL85}, we
achieve this purpose by deriving an equation for the quantity\footnote{This quantity is in fact the opposite of
the complex conjugate of theirs, unscaled.}.

\begin{plain}
$$Z=\epsilon\exp im\Delta.\eqno(7.46)$$
\end{plain}

  With this definition, $q_{ij}\exp \gamma_{ij}=(a_i Z_i -a_j
Z_j)\exp(-im\Delta_i)/(a_i-a_j)$, which suggests that the desired
equation can be obtained from the computation of $(\epsilon{\dot\varpi}
+i{\dot\epsilon})\exp im\Delta$. This procedure yields\footnote{The contribution from the stress-tensor are given in the tight-winding approximation, for simplicity.}

\begin{plain}
$${2\over\pi}\int_{-\infty}^{+\infty}dx'\ H(q^2_{xx'}){Z(x)-Z(x')\over
(x-x')^2} + {a_R k^2(t_2+it_1)\over\pi G\sigma_0^2 q}Z-{Zx\over\delta}=
-{\Psi_{mk}\over2\pi G\sigma_0 a_R},\eqno(7.47)$$
\end{plain}

\noindent where $\delta$ is the small parameter introduced in Eq.~(7.44)
and where $x\equiv (a-a_R)/a_R$ has been used instead of $a$ as a
distance parameter; $|x|$ measures the fractional distance to the
resonance. The first term on the left-hand side represents the
contribution of the ring self-gravity, the second is due to the ring
internal stress, the last reflects the precession rate required by the
existence of the wave pattern, and the term on the right-hand side is
the satellite forcing (cf Eq.\ 6.22, with both signs of $m$ allowed). This last term is a dimensionless quantity which
ranges from $\sim 0.1$ to $\sim 0.5$ for the strongest resonances in
Saturn's rings. Remember that it is of order $e_s^{|k|}$, where $k$ is
an integer (see section 4.3.2). All smoothly varying quantities have 
been pulled out of the
self-gravity integral. Eq.~(7.47) is the starting point of the analyses
of \cite{SYL85} and \cite{SDLYC85}. This equation has no analytic
solution in the nonlinear case, but quite a number of its features can
be understood in the linear limit, which we are therefore going to
discuss next, following the solution developed by \cite{SYL85}. 
In this limit, $q\rightarrow 0$ and one can take
$H(q^2)=1/2$. Furthermore, we will ignore the viscous terms for the time
being. Neglecting $t_2$ is equivalent to neglecting the short-waves, a
simplification  which was justified in the previous subsection.
Neglecting $t_1$ results in the suppression of the wave damping,
but we will return to this question shortly. With these approximations, Eq.~(7.47) reduces to

\begin{plain}
$${1\over\pi}\int_{-\infty}^{+\infty}dx'{dZ/dx'\over
x-x'}+{Zx\over\delta}=\ff,\eqno(7.48)$$
\end{plain}

\noindent where the self-gravity integral has been integrated by parts.
The remaining integral can be computed with the residue theorem\footnote{Because the wave complex amplitude $Z$ contains an exponential phase term, there is \textit{no guarantee} that one can find a contour at infinity where the contribution vanishes, so that the resulting differential equation is not always valid (it would actually be remarkable that the self-gravity integral equation can be exactly replaced by a differential equation, even in the linear limit). In fact, for free waves, Eq.~(7.49) is clearly wrong as it implies the absence of evanescence inside the resonance radius; however one can argue that the sign of $k$ should be present as a factor for the first term (see \citealt{Sh84}, pp.\ 540 -- 541, as well as the Appendix). In any case, it is shown in the Appendix that this equation is a rather accurate approximation for forced density waves throughout the disturbed region, and a very precise one as soon as $|x| \gtrsim (2|\delta|)^{1/2}/3$, i.e., nearly everywhere except in a rather limited band around the resonance radius. It is probably worth pointing out that the standard fluid treatment of linear density waves (e.g. \citealt{Sh84}), suffers from a similar mild limitation. In this fluid approach, the problem stems from the relation between the surface density and self-gravitational potential; the standard solution based on analytic continuation \citep{Sh70} and used in most linear analyses of density waves performed in the 70s and 80s, is only valid in the tight-winding approximation and does not apply in the evanescent region.}, and one finally obtains

\begin{plain}
$$i{dZ\over dx}+{Zx\over\delta}=\ff,\eqno(7.49)$$
\end{plain}

\noindent which is the usual equation of linear density wave theory [see \citealt{Sh84}, Eq.~(44)].

One can encompass both ILR and OLR solutions by introducing a new scaled radial coordinate $\xi= s x/|\delta|^{1/2}$ where $s=\mathrm{sgn}(\delta)$. From the dispersion relation, we expect that the forced
(i.e.\ particular) solution to this equation is a long wave, at least in
the far wave region. Therefore, we wish to find the particular solution
which is evanescent for $\xi\rightarrow -\infty$. As $\exp(is\xi^2/2)$
is an integrating factor, the desired solution is found to be\footnote{This result also follows from an elementary variation of the constants technique.}

\begin{plain}
$$Z=-i s \ff\exp\left(i s{\xi^2\over 2}\right)\int_{-\infty}^\xi 
dy\ \exp\left(-i s{y^2\over 2}\right).\eqno(7.50)$$
\end{plain}

This solution is depicted in Figure~\ref{fig:torque} (see Appendix~\ref{app:dw}), but it has two interesting asymptotic expansions \citep{GT79c,Sh84} that capture most of its behavior. First, in the
limit $\xi\ll -1$ (in the far 
evanescent region), replacing $\exp(-isy^2/2)$ by $(is/y)
d\exp(-isy^2/2)/dy$ and integrating by part yields

\begin{plain}
$$Z\simeq{\delta\over x}\ff.\eqno(7.51)$$
\end{plain}

\noindent On the other hand, for $\xi\gg 1$ (i.e. into the
propagation region), using $\int_{-\infty}^\xi=\int_{-\infty}^{+\infty}-
\int_\xi^{+\infty}$, and performing a similar integration by part, one
obtains

\begin{plain}
$$Z\simeq\ff (2\pi|\delta|)^{1/2}\exp\left(i{x^2\over 2\delta}-i s {3\pi\over
4}\right)+{\delta\over x}\ff,\eqno(7.52)$$
\end{plain}

\noindent where $\int_{-\infty}^{+\infty}dx\exp (-ix^2/2\delta)=
(2\pi|\delta|)^{1/2}\exp(-i s\pi/4)$ has been used. The last term represents 
the non-wavy part of the response,
and is negligible.

From this last result we can derive the expression of
the eccentricity along the wave

\begin{plain}
$$\epsilon={\left|\Psi_{mk}\right|\over 2\pi G\sigma_0 a_R}
(2\pi|\delta|)^{1/2}\sim 10^{-4},\eqno(7.53)$$
\end{plain}

\noindent which is seen to be independent of $a$; the wavenumber $k=
md\Delta/da$ reads

\begin{plain}
$$k={x\over a_R\delta}={3(m-1)M_p\over 2\pi\sigma_0 a_R^3}{a-a_R\over
a_R}.\eqno(7.54)$$
\end{plain}

\noindent which, by comparison with Eq.~(7.42), is seen to correspond to
a long trailing wave, as expected. Finally, from
$q=|k|a\epsilon$, one sees that $q\propto |x|$, so that the wave becomes
nonlinear for

\begin{plain}
$$|x|=x_{nl}\equiv{2\pi G\sigma_0 a_R\over \left|\Psi_{mk}\right|}
\left(|\delta|\over 2\pi\right)^{1/2}\sim {\lambda_1\over a_R}.\eqno(7.55)$$
\end{plain}

\noindent Therefore, density waves become nonlinear in the first
wavelength or so, as observed. It is to be noted that although the
linear theory predicts streamline crossing within a wavelength,
the nonlinear contributions of the self-gravity integral prevent this to happen.

Let us now investigate the effects of the neglected viscous terms, and
reintroduce them in Eq.~(7.49). Using the fact that in the propagation
region, $|dZ/dx|=|ika_R Z|\gg |\Psi_{mk}|/2\pi G\sigma_0 a_R$, this equation
reads

\begin{plain}
$$-ka_R Z+{Zx\over\delta}-{a_R k^2\over\pi
G\sigma_0^2q}(t_2+it_1)Z=0.\eqno(7.56)$$
\end{plain}

\noindent As $t_i\sim \sigma_0 c^2 q$ for small $q$, the ratio of the
first to the last term is of order $Q^2(m-1)x\ll 1$, where the velocity
dispersion has been expressed in terms of Toomre's $Q$ parameter. The
viscous terms are indeed very small and balancing the two first terms 
gives back the dispersion relation for long trailing wave. However, if
the contribution of $t_2$ is small and brings no qualitatively new
information, the contribution of $t_1$ is qualitatively different 
because it is a pure imaginary number, indicating that $k$ must contain
an imaginary part: $k=k_r+ik_i$, where $k_r$ is given Eq.~(7.54) and 

\begin{plain}
$$k_ia_R=-{a_Rk_r^2 t_1\over\pi G\sigma_0^2 q}.\eqno(7.57)$$
\end{plain}

\noindent Due to this new contribution, $Z$ is now given by

\begin{plain}
$$Z=Z_{nv}\exp\left(-\int k_ia_Rdy\right),\eqno(7.58)$$
\end{plain}

\noindent in the far wave zone, where $Z_{nv}$ is the inviscid
value of $Z$, Eq.~(7.52). As a consequence, $\epsilon$ now
reads

\begin{plain}
$$\epsilon=\epsilon_{nv}\exp\left(-\int_0^x k_ia_Rdy\right),
\eqno(7.59)$$
\end{plain}

\noindent where $\epsilon_{nv}$ is the inviscid value of $\epsilon$, Eq.~(7.53). This shows that the wave is damped if $t_1<0$, an assumption
which is made in the remainder of these notes\footnote{One sees also that the wave can be viscously unstable if the viscous
coefficient is positive. However, one could expect, as for narrow rings,
that oscillations of the eccentricity and phase shift would also be
unstable, and possibly prevent the growth of the wave. Such oscillations
are excluded by assumption in the present analysis. As the question
is still open, we will not push it further in these notes. An analysis
of the behavior of nonlinear density waves when the viscous coefficient
is positive has been performed by \cite{BGT86}, but these results are incomplete, as the possibility of oscillations of
the eccentricity and phase of the wave has been excluded at the onset of
their analysis.}. An estimate of the damping length scale $x_{vis}$ can be
obtained by setting $\int_0^|x| k_ia_Rdy=1$, which yields
$Q^2|m-1||x|_{vis}^3/2|\delta|=1$, i.e. $|x_{vis}|\simeq |\delta|^{1/3}$: this
implies that density waves can propagate over about ten cycles, as
observed. In their more quantitative analysis of wave damping, \cite{SDLYC85} have pointed out that the results displayed on Figure~\ref{fig:Shu} play a central
r\^ole. Indeed, as $q$ increases with $x$, if the unperturbed optical
depth $\tau_0$ is smaller than some critical value, the coefficient of
restitution $\epsilon_r\rightarrow 0$, and $c$ diverges (remember that
for all known materials $\epsilon_r$ is a decreasing function of $c$):
as $t_1\propto c^2$, this implies short damping length scales. On the
other hand, if the unperturbed optical depth is larger than this
critical value, $\epsilon_r$ depends much less on $q$, and the wave
could propagate much farther. These authors have attributed to this
effect the observed fact that density waves propagate much farther in
the B ring than in the A ring of Saturn. However, they have
also argued that in order to reproduce the observed short damping length
scale of the A ring density waves, rather elastic materials were needed, 
which would
rule out ``dynamical ephemeral bodies" (DEBs) as a likely model candidate for
ring particles. Unfortunately, DEBs are the most natural outcome of
the ring collisional processes \cite{WCDG84,L89}. Either this particle model is incorrect, or other unmodelled processed are at work wave such as the scattering of the wave by large particles,
as has been suggested at least by \cite{SDLYC85} and by Peter Goldreich
(private communication). At the time of writing, this issue is still
unresolved.

\subsubsection{Satellite torque and angular momentum transport}

  The question of angular momentum exchanges between the rings and the
satellites have been widely debated for the past ten years, for reasons
that will now be exposed. First, note that for a steady-state wave
propagating in an inviscid medium, Eq~(6.49) reduces to

\begin{plain}
$$L_H^{sg}=\int_0^a da{\cal T}_s.\eqno(7.60)$$
\end{plain}

\noindent Thus, for undamped density waves, all the angular momentum
deposited by the satellite is carried away by the wave\footnote{This result is a direct consequence of the conservation of the wave action; see \cite{D72}.}. 

  It is interesting to compute $L_H^{sg}$ in the tight-winding
approximation. From Eqs.~(6.39) and (7.29), and assuming $k>0$, one obtains

\begin{plain}
$$L_H^{sg}=-4\pi Gm\sigma_0^2 a_R^3\epsilon^2\int_{-\infty}^v dv_1
\int_{v-v_1}^{+\infty}du\ H\left(q^2{\sin^2u\over u^2}\right){\sin
2u\over u^2}.\eqno(7.61)$$
\end{plain}

\noindent Interchanging the order of the integrals and performing the
integration over $v_1$ yields

\begin{plain}
$$L_H^{sg}=-\pi^2 Gm\sigma_0^2 a_R^3\epsilon^2 B(q),\eqno(7.62)$$
\end{plain}

\noindent where $B(q)$ is defined by

\begin{plain}
$$B(q)={4\over\pi}\int_0^{+\infty}du\ H\left(q^2{\sin^2 u\over
u^2}\right){\sin 2u\over u}.\eqno(7.63)$$
\end{plain}

\noindent Note that $B(q)\rightarrow 1$ as $q\rightarrow 0$. This asymptotic form of the self-gravity angular momentum luminosity is similar to the approximation performed to obtain Eq.~(7.49) from (7.48). Notice also that Eq.~(6.39) predicts correctly $L_H^{sg}$ to order $\epsilon^2$, although the basic streamline parametrization on which the whole analysis relies is valid only to order $\epsilon$. Note finally
that for $k<0$ (leading wave) the luminosity Eq.~(7.62) changes sign. 

  Let us now rederive the expression of the
torque exerted on a wave excited at an inner Lindblad 
resonance in the limit of linear undamped waves \citep{GT78c,GT79c}. For definiteness we focus on an inner Lindblad resonance ($s=1$); the torque has the same magnitude but opposite sign at an outer resonance. 

From Eq.~(6.38), (7.46) and (7.50), one finds

\begin{plain}
$$\eqalign{T_s=\int_0^{+\infty}da\ {\cal T}_s=- & {m\Psi_{mk}^2 a_R\over 2G}\times \cr
& {\rm Im}\left[i\int_{-\infty}^{+\infty}dx\ \exp\left(i{x^2\over
2\delta}\right)\int_{-\infty}^x dy\ \exp\left(-i{y^2\over
2\delta}\right)\right],}\eqno(7.64)$$
\end{plain}

\noindent where Im($z$) designates the imaginary part of a complex
number $z$. Following \cite{SYL85}, let us call $N$ the complex
number whose imaginary part is to be evaluated in Eq. (7.64) and compute
the complex conjugate $N^*$. Interchanging the order of the integrals in
$N^*$ yields

\begin{plain}
$$N^*=- i\int_{-\infty}^{+\infty}dx\ \exp\left(i{x^2\over
2\delta}\right)\int_x^{+\infty} dy\ \exp\left(-i{y^2\over
2\delta}\right),\eqno(7.65)$$
\end{plain}

\noindent so that

\begin{plain}
$${\rm Im} N= {N-N^*\over 2i}={1\over 2}\int_{-\infty}^{+\infty}dx 
\int_{-\infty}^{+\infty} dy\ \exp\left(i{x^2-y^2\over
2\delta}\right)=\pi\delta,\eqno(7.66)$$
\end{plain}

\noindent Plugging this result into Eq. (7.64) gives back the standard 
expression of the linear torque\footnote{Because the linear equation is not exact, as discussed in an earlier footnote, one may at first glance question the validity and relevance of this result. However, it is well-known that the linear torque is independent of the details of the physics of disks; the same linear torque obtains in disks without self-gravity but pressure and/or damping instead (\citealt{MVS87} have produced what is probably the most generic justification of this result). As a consequence, the integrated linear torque is correct, but of course the details of the torque density are not in the near-resonance region are not, where most of the torque is deposited.}

\begin{plain}
$$T_s=-{\pi^2m\Psi_{mk}^2\sigma_0\over {\cal D}},\eqno(7.66)$$
\end{plain}

\noindent where 

\begin{plain}
$${\cal D}\equiv\left(a{d\over
da}\left[\kappa^2-m^2(\Omega-\Omega_p)^2\right]\right)_{a_R}={3(m-1)GM_p
\over a_R^3}.\eqno(7.67)$$
\end{plain}

\noindent The same expression is obtained from Eqs.~(7.53) and (7.62),
as expected from Eq.~(7.60). It is also instructive to see how the torque accumulates with radius; to this effect, the cumulative integrand $N(x)$ the imaginary quantity in Eq.~(7.64) is illustrated on Figure~\ref{fig:torque} (see Appendix~\ref{app:dw}).
 
Actual waves are affected by nonlinear effects and by viscous damping.
The effect of the ring viscosity (or more correctly, internal stress)
are not expected to be very important, as the damping occurs in the far
wave region, whereas a numerical evaluation of $\int_{-\infty}^x{\cal
T}_s da$ as a function of $x$ shows that most of the torque is deposited
within a few wavelengths of the resonance\footnote{This conclusion is robust although the details of the torque deposition implied by Eq.~(7.49) are not).}. \cite{SDLYC85} argue that nonlinear effects don't reduce the torque much
below its linear value, but their conclusion is dependent on the details of their model equation; however, a similar conclusion was reached by \cite{LB86} from a nonlinear kinematic analysis of the Mimas 5:3
density wave\footnote{This analysis suffers also from some limitations, but this conclusion is most probably robust. It is fair to say though, that the nonlinear torque magnitude is still somewaht uncertain to this date}. Therefore, Eq.~(7.66) should give a correct estimate of the
angular momentum exchange between rings and satellite at Lindblad
resonances. 

  As the wave is damped, the angular momentum flux due to the ring
self-gravity falls below the value implied by Eq.~(7.60). The variation
of $L_H^{sg}$ with $a$ can be related to the viscous coefficient $t_1$
in the following way, slightly adapted from \cite{BGT85}. From
Eq.~(6.47), we can eliminate $da/dt$ in Eq. (6.32) in favor of $L_H^c$. 
This yields

\begin{plain}
$$2\pi\Omega a^3{\partial \sigma_0\over\partial t}+{\partial L_H^c\over
\partial a}-{L_H^c\over 2a}=0.\eqno(7.68)$$
\end{plain}

\noindent Furthermore, from $L_E\equiv L_E^c+L_E^{sg}+L_E^{vis}$ (with a
similar definition for $L_H^{sg}$), and from $L_E^{sg}=\Omega_pL_H^{sg}$ 
and Eqs.~(6.44) through (6.47), one obtains

\begin{plain}
$$L_E=\Omega L_H+(\Omega_p-\Omega)L_H^{sg}-{3\over 2}\Omega L_H^c+
2\pi a^2\Omega\epsilon[t_1\cos\gamma+t_2\sin\gamma].\eqno(7.69)$$
\end{plain}

\noindent Computing $\partial L_H^{sg}/\partial a$ from there, and using Eqs.~(6.48) and (6.49) finally yields

\begin{plain}
$${\partial\over\partial a}(L_H^{sg}-2\pi a^2\epsilon
m[t_2\sin\gamma+t_1\cos\gamma])=-2\pi m a q t_1+{\cal T}_s,\eqno(7.70)$$
\end{plain}

\noindent which is the fundamental equation describing wave damping\footnote{From the linear theory of density waves, it is known that the wave amplitude is fixed either from a second order WKB analysis (in the wave propagation region) or from the equation for the wave action, i.e., from the equation of angular luminosity conservation (see, e.g., \citealt{D72}). The same consideration follows in the nonlinear regime, and in the wave propagation region, where the tight-winding condition applies, the satellite contribution to the amplitude in Eq.~(7.47) is negligible; similarly, in the same approximation, the self-gravity contribution to $de/dt$ cancels, and one needs an alternative way to constrain the wave amplitude. This is provided by the wave-damping equation, supplemented by the nonlinear dispersion relation and the conservation of the total (viscous and self-gravitational) angular momentum luminosity once the torque is totally deposited, which provide enough constraints to determine the wavenumber $k$, the nonlinearity parameter $q$ and the surface density $\sigma_0$ (see \citealt{BGT86} for details; the only uncertainty in this analysis, as pointed out earlier, is the magnitude of the nonlinear torque, although it should emerge self-consistently if one starts from the exact equations and avoids the tight-winding approximation).}. In
the tight-winding limit ($\gamma=\pi/2$), $t_1$ disappears from the
left-hand side. The contribution of the remaining $t_2$ term is usually 
small in planetary rings, but \cite{SDLYC85} have pointed out that 
it might generate a phenomenon referred to as a ``Q barrier" in the
linear theory of density waves in spiral galaxies (see, e.g., \citealt{T69}): as the wave propagates away from the resonance, the increase in 
macroscopic energy dissipation due to the presence of the wave (with
respect to unperturbed regions) results in an increase of the velocity 
dispersion. If the
increase is important enough, the two roots of the dispersion relation
(short and long waves) will merge, and the long wave becomes evanescent,
until the velocity dispersion has dropped again. Therefore, the long
wave could be partly ``reflected" as a short wave on the newly formed
evanescent region, and partly transmitted as the observed long wave. In
order to avoid this interesting, but rather complex possibility, \cite{SDLYC85} have neglected the contribution of $t_2$, an approximation made at
the onset by \cite{BGT86}. Whether Q barriers
actually occur in Saturn's rings is still an open question. 

In any case, Eq.~(7.70) relates the radial variation of the amplitude
of the motion, $\partial\epsilon/\partial a$ to the viscous coefficient $t_1$.
One sees again that the wave is damped if $t_1<0$, and that viscous
overstabilities can take place in the opposite case (but see the comment
on footnote 43). 

Let us come back now to the discussion of the effects of the satellite
torque on the evolution of the ring. Let us first assume for the sake of the argument that the wave is in steady state. Due to the ring
viscosity, and according to Eq.~(6.49), Eq. (7.60) now reads
$L_H^{sg}+L_H^{vis}\simeq T_s+L_{vis}^-$ in the propagation region,
where $L_{vis}^-$ is the unperturbed angular momentum flux inside the
wave region. Denoting the unperturbed flux outside the wave region by
$L_{vis}^+$, the steady state condition implies $L_{vis}^+=L_{vis}^-+
T_s$. First, notice that $L_{vis}^-$
and $L_{vis}^+$ are positive and $T_s$ is negative. If the satellite
torque is larger than the unperturbed angular momentum flux, 
$L_H>0$ outside the wave region and $L_H<0$ inside, 
resulting in a steady loss of angular momentum, an inward mass drift and 
the formation of a gap in the end. \cite{GT78c} have argued that this process was responsible for the
formation of the Cassini division by Mimas. Once a gap is
open, the torque can be reduced from its full value (the resonance
region is not completely occupied with ring material, so that the torque
integral is truncated), and another equilibrium or quasi-equilibrium can
be reached in which the satellite confines the inner edge of the gap (see
section 6). On the other hand, if $L_{vis}^-+T_s>0$, an enhancement of
the surface density in the ring region might result, a feature apparent
in the observed wave profiles. \cite{BGT86}
provide the following explanation for this effect. If a strong enough
density wave is launched in a medium of constant background surface
density, $q$ will quickly reach values such that $a_{r\theta}<0$ in the
wave zone, so that $L_H<0$ and an inward drift again takes place. As the
surface density increases, the angular momentum flux needed to evacuate
the torque can be obtained for lower values of q, which 
decreases until it becomes $\lesssim q_2$, the critical value
for viscous angular momentum flux reversal (see section 5). 
Then $L_H$ is positive again, and a quasi steady-state can be maintained. 
These authors have also used this argument to show that strong waves
have their amplitudes limited to $q\lesssim q_2<1$.  

   As satellite torques are negative, satellite orbits expand in
time. Calculations based on the cumulative effects of all torques
excited in the rings show that the related time-scales of satellite
recession are remarkably short, e.g. $\sim 10^7$ years for the shepherd
satellites of the F ring \citep{GT82,BGT84}. In the same time, a substantial inward drift of
ring material should have taken place \citep{LPC84}, and the
analysis of the situation in 1984 was that either the rings were young 
or that some essential piece of physics was overlooked, e.g. 
that nonlinear saturation might reduce 
the torque magnitude  below its linear value. This consideration
actually motivated the development of a nonlinear theory of density waves; however, nonlinear effects do not seem to reduce significantly the satellite torques. Conversely, the idea that the rings are young is now somewhat
more popular; \cite{D91}, e.g., has given some support to a recent
cometary origin for the rings, which appears to be the most likely
late mechanism of formation of ring systems. However, the issue is not
yet completely settled.

\section{Conclusion}
  A general formalism for the analysis of the dynamics of major ring
systems (in the sense defined in the introduction) has been presented
here. This formalism draws on two complementary approaches, one which deals 
with the macroscopic ring motions, and one related to the microphysical
collisional processes\footnote{In ring dynamics,
``microphysical'' has a rather strange meaning, as the individual ring
particles are macroscopic in size.}. It should not be forgotten that the
formalism relies on a fluid approach, which is strictly valid only for
phenomena whose characteristic length-scale is larger than the typical
particles' mean size and mean separation. Also, the dynamics is
treated in the one-fluid approximation.

  Obviously the collisional behavior
is much less understood than the global dynamics, limiting our
interpretation of the finest dynamical effects, although some important
general features of the ring internal stress have been uncovered in the
past few years. Note
that it might be difficult to improve systematically on this point, at
least in the framework of the Boltzmann collision term, which relies on
the assumption of ``molecular chaos", and neglects velocity and position
correlations of the particles. The condition of validity of this
assumption has been extensively discussed in Plasma Physics on the basis
of the so-called BBGKY hierarchy, and a criterion has been derived,
which states that correlations are negligible whenever the mean kinetic
energy of individual particles is much larger than their mean energy
of mutual interaction. In ring systems, the velocity dispersion is
comparable to the mean two-body gravitational potential, so that this
criterion is not satisfied, although it is not strongly violated. It
seems important however not to disregard heuristic approaches at the
profit of purely formal developments, as many breakthroughs have already
been performed in this way. Also, it is likely that direct N-body numerical simulations will bring useful information on this front in the future.

  On the side of the global dynamics, there are still some major
unresolved problems. The most critical is probably the question of 
the rigid precession
of narrow elliptic rings. As argued in section 7.1, the results of
the recent encounter of Voyager II with the Uranian system have cast some
doubts on the validity of the self-gravity model put forward by \cite{GT79a}. It has also appeared that the standard radiative
transfer theory is not appropriate for the analysis of the radio data on the Uranian rings, due to the high density and particle close packing that is prevalent in these systems.
   
  The analyses of the damping of density waves on one hand, and of the
collisional evolution of the ring particle size distribution on the
other, suggest two contradictory models of ring particle structure and
collisional properties. This issue might be alleviated if as yet
unmodelled damping mechanisms are found, e.g. the scattering of the wave
by large particles (see section 7.2.2); alternatively, the DEBs modeled might be disproved in the end. Also, the absence of
perturbation at some resonances is not yet understood; similarly, a
detailed criterion for gap opening remains to be established.

  The Uranian rings are apparently correctly described as elliptic
modes. However, the mechanism of selection of the observed modes is not
yet understood (why, e.g., does the $\alpha$ ring have an $m=1$ mode and the
delta ring an $m=2$ ?). Also, the kinematic analysis of the $\gamma$ ring
still exhibits unexplained kinematic residuals (see \citealt{Fetal88}).
 
  For a long time, no convincing mechanism of confinement of the inner
edges of Saturn's rings (like the A and B rings, for example) had been
proposed. However, it has recently been argued that such edges could be
maintained by ballistic transport processes (see \citealt{Detal92}).
Further studies are needed to confirm or disregard this proposal.

  The question of the confinement of the Neptunian ring arcs is still
open, although some interesting progress has been made recently (see
\citealt{P91}).

  Finally, let us point out that some interesting questions concerning
the dynamics of charged particles are still open, as, e.g., questions
related to the formation, propagation (if any) and disparition of
spokes.

  All these issues, as well as a number of others, are the object of
active ongoing research.

\section*{Acknowledgements\markboth{Acknowledgements}{}}
\label{sec:acknow}
\addcontentsline{toc}{section}{\nameref{sec:acknow}}

The material presented here represents mostly the work of a
collaboration between N. Borderies, P. Goldreich and S. Tremaine over
the past ten years or so. I am indebted to N. Borderies and P. Goldreich of many discussions on all aspects of the content of these notes, and I wish to thank them for these fruitful exchanges.

\clearpage

\section*{Appendix\markboth{Appendix}{}}
\label{sec:appendix}
\addcontentsline{toc}{section}{\nameref{sec:appendix}}

\appendix

\section*{On the validity of the complex plane integration of the self-gravity integral}\label{app:dw}
\renewcommand{\theequation}{\thesection\arabic{equation}} 
\setcounter{equation}{0}
\renewcommand{\thesubsection}{\Alph{subsection}}
\setcounter{subsection}{0}
\renewcommand{\thefigure}{\thesubsection\arabic{figure}}
\setcounter{figure}{0}

The objective of the present Appendix is to discuss in some detail the identity 

\begin{equation}
\frac{2}{\pi}\int_{-\infty}^{+\infty}dx'\ \frac{Z(x)-Z(x')}{(x-x')^2} = -i\frac{dZ}{dx},\label{residu}
\end{equation}

\noindent that has been used in the derivation of Eq.~(7.49) by \cite{SYL85} on the basis of complex contour integration theory.

\subsection{Free density waves}\label{sec:free}

It is certain that this relation is not correct independently of the functional form of $Z$. Indeed, the solution of Eq.~(7.49) for free waves (right-hand side set to zero) reads

\begin{equation}\label{free}
Z_f= \exp \pm i x^2/2\delta.
\end{equation} 

\begin{figure}[ht]
\centering
\includegraphics[width=0.7\linewidth]{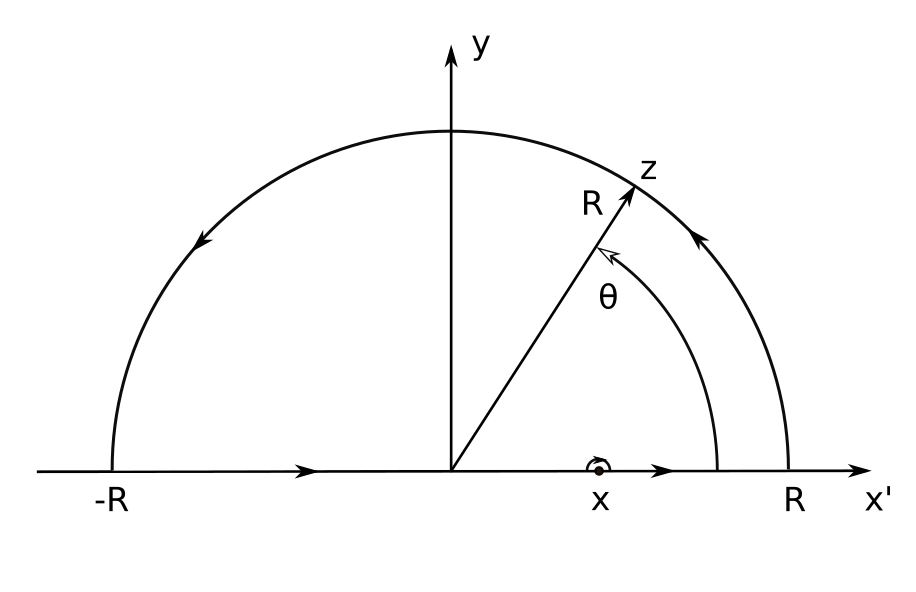}
\caption{\small{Contour of complex integration for free waves. As usual, the choice of direction around the pole in $x'=x$  (upper or lower small half-circle) does not influence the end result.}}
\label{fig:contour}
\end{figure}

For this functional form, the analytic continuation of $Z$ in the complex plane $z=x+i y$ never gives a vanishing integral on any contour joining the real line at $\pm\infty$. Consider the $\exp +i x^2/\delta$ solution with $\delta >0$ for example. On the upper half-circle (see Figure~\ref{fig:contour}), $z=R e^{i\theta}$ and $\exp +i x^2/\delta = \exp(iR^2\cos2\theta)\exp(-R^2\sin2\theta)$, so that the exponential diverges for $R\rightarrow\infty$ and $\theta \in [\pi/2,\pi]$; the same conclusion follows for a lower half-circle for $\theta \in [-\pi/2,0]$. Therefore the contribution of the closing contour is diverging for $R\rightarrow +\infty$, whereas Eq.~\eqref{residu} assumes it vanishes. At least for this form of $Z$, Eq.~\eqref{residu} is plainly wrong.

This conclusion is not surprising. The quadratic phase of Eq.~\eqref{free} implies that the tight-winding approximation is valid for both positive and negative $x$ ($Z_f$ is symmetric), except for an extremely narrow range around $x=0$. But $-idZ/dx=kZ$ with $k=x/\delta$ everywhere whereas in the tight-winding limit, the left-hand side of Eq.~\eqref{residu} is $|k|Z$ [see the derivation of Eq.~(7.31)] implying the presence of an evanescent zone inside the resonance radius, in contradiction with Eq.~\eqref{free}. 

In fact, if one assumes \textit{a priori} that the wave is evanescent outside the ILR and OLR radii and propagating with wavenumber $k$ inside, one can recover this result from contour integration, as the sign of the wave number fixes which semi-circle (upper or lower half-plane) must be chosen to complete the contour (see \citealt{Sh84}, Eqs.\ 78 through 81). This argument assumes the form of the solution, but does not help us to quantify the error associated with this assumption. This question is addressed in the remainder of this Appendix.

\subsection{Forced density wave}

One might ask if Eq.~\eqref{residu} is valid or at least approximately valid for the form (7.50) of $Z$, which would give an \textit{a posteriori} justification for the use of this relation in the case of forced density waves. This problem is examined here and the associated error is analyzed in the process.

For definiteness, the analysis focuses on inner Lindblad resonances, but outer resonances can be treated in the same way.

\subsubsection{Conclusions in a nutshell}\label{sec:nut}

\begin{figure}[ht]
    \centering
    \begin{subfigure}[h]{0.48\textwidth}
        \includegraphics[width=\textwidth]{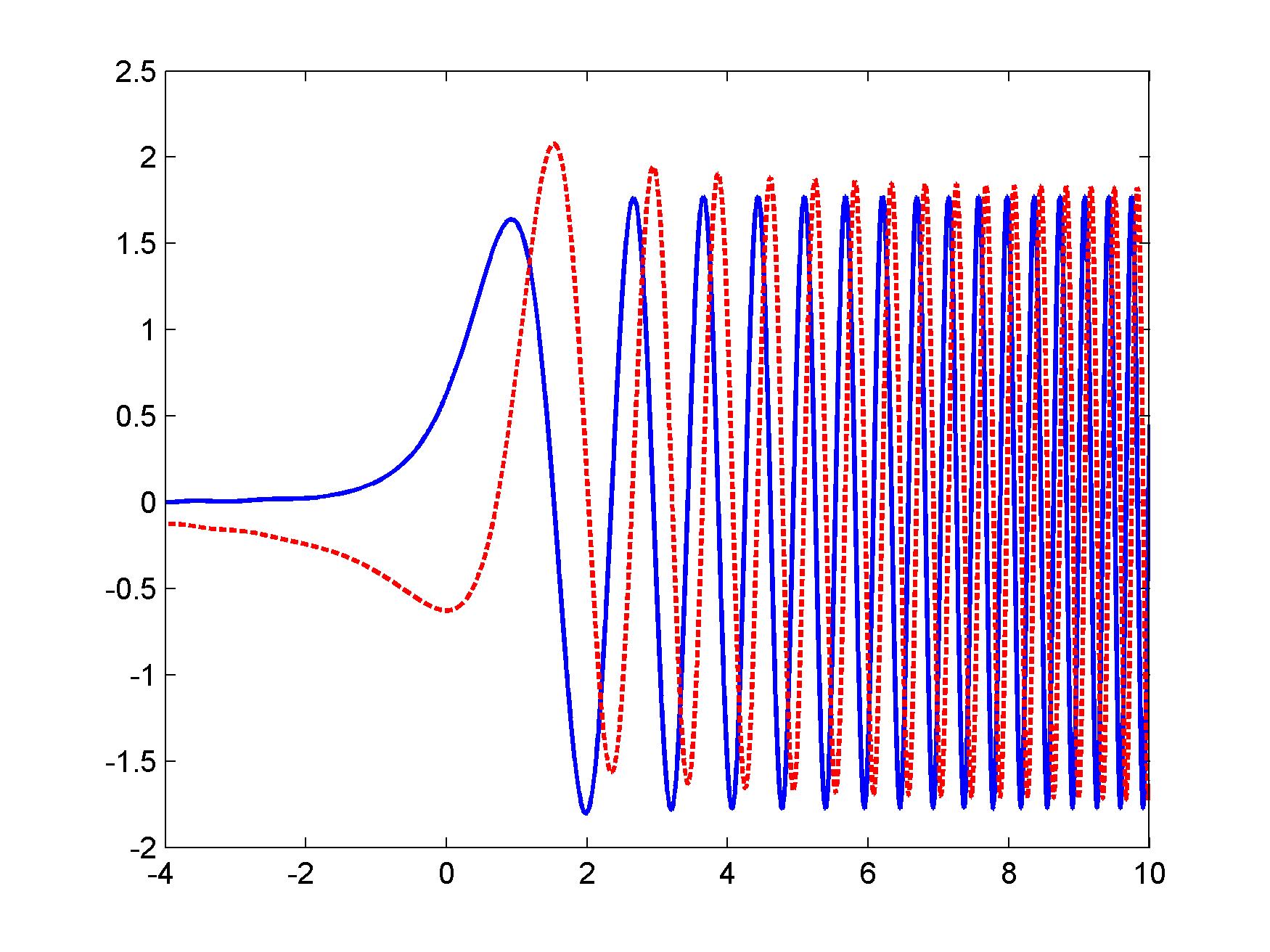}
        \caption{$Z(x)$}
        \label{zed}
    \end{subfigure}
    \begin{subfigure}[h]{0.48\textwidth}
        \includegraphics[width=\textwidth]{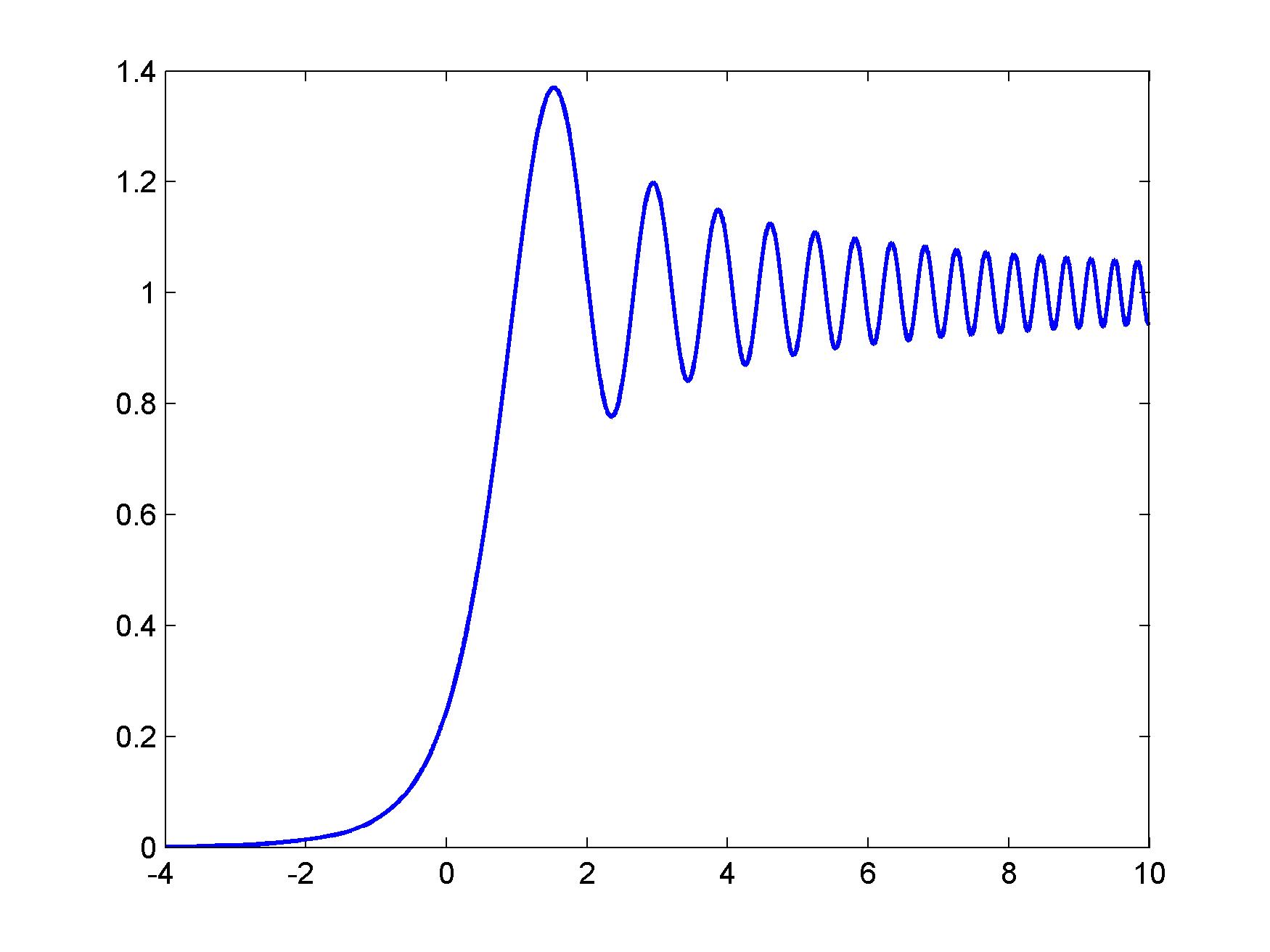}
        \caption{$N(x)$}
        \label{torque}
    \end{subfigure}
\caption{\small{The real (solid line, $\propto \epsilon\sin m\Delta$) and imaginary (dotted line, $\propto \epsilon\cos m\Delta$) part of $Z(u)/(2\delta)^{1/2}$ [Eq.~\eqref{forced}] as a function of the normalized distance $u=x/\sqrt{2\delta}$ are shown on the left. On the right, the normalized cumulative torque as a function of $x/\sqrt{2\delta}$, i.e., the cumulative integral defined as $Im(N)$ in Eqs.~(7.64) and (7.65), normalized by $\pi\delta$.}}
\label{fig:torque}
\end{figure}

For readers who want to skip the details of the soon-to-be-performed contour integration, let me just jump to the conclusion derived in the remainder of this section:

\begin{enumerate}
\item Eq.~(7.49), albeit not exact throughout the wave region for forced linear density waves, nevertheless captures the correct asymptotic behavior in the evanescent and propagation zone. In fact, the error becomes negligible for $|x| \gtrsim \sqrt{2\delta}/3$ (from Figure~\ref{fig:JR}), i.e., almost everywhere except in a rather limited band around the resonance radius\footnote{For comparison, the first wavelength is $\lambda=2\sqrt{\pi\delta}$, i.e. nearly a factor of $10$ larger than the preceding limit.} ($\delta$ is defined in Eq.\ 7.44). Moreover, within this narrow band, Eq.~(7.50) produces only a mild logarithmic divergence of the self-gravity integral, indicating that a minor alteration of this function in this range would result in a nearly exact solution.
\item Because the magnitude of the torque (one the most important aspects of the linear theory of density wave) does not depend on the details of the physics (see, e.g., \citealt{MVS87}), this model equation gives the correct integrated torque no matter what. Furthermore, the normalized torque accumulation as a function of $x/\sqrt{2\delta}$ is shown on Figure~\ref{fig:torque} for $Z$ given by Eq.~(7.50); this indicates that most of the torque accumulates in the domain of accuracy of Eq.~\eqref{residu}, so that the detail of this accumulation is also rather accurate.
\end{enumerate}

As a consequence, Eq.~(7.49) appears to be a remarkably good model equation for linear density waves throughout the wave region. This conclusion may seem hardly unexpected, but is worth pointing out nonetheless, as the problem addressed in this Appendix is easily overlooked and in any case not mentioned in \cite{SYL85} where Eq.~\eqref{residu} is stated without further justification. And indeed, because Eq.~(7.49) recovers a known form of linear density wave theory, the problem was either overlooked or deemed unimportant.

\subsubsection{Setting up the problem}

To tackle this problem, let us first rescale $x,x'$ by $1/(2\delta)^{1/2}$ and ignore the normalization factor of $Z$ which simplifies Eq.~(7.50) in the complex plane to

\begin{equation}\label{forced}
Z(u)=(2\delta)^{1/2} \exp( +i u^2 )\int_{-\infty}^{u} du' \exp (-i u'^2), 
\end{equation}

\noindent where the $-\infty$ limit of the complex integral is on the $x$ axis and $(u,u')=(z/(2\delta)^{1/2},z'/(2\delta)^{1/2})$. Note that the rescaling $x,x',z,z'$ by $1/(2\delta)^{1/2}$ does not change the normalization of integrals of the form $\int_C dz' [Z(x)-Z(z')]/(x-z')^2$ on any contour $C$ due to the invariance of $dzdz'/(x-z')^2$. Because of this rescaling invariance, we keep the notation $z,z',x,x'$ for the rescaled quantities, and forget about the normalization factor in Eq.~\eqref{forced}, except when needed to state relevant scalings in actual physical variables.

\medskip

It is convenient to first focus on the complex integral in Eq.~\eqref{forced}. 

Because $\int_{-\infty}^{z} dz' \exp (-i z'^2)$ is analytic, the integral can be evaluated on any path and one can write

\begin{equation}\label{integral}
\int_{-\infty}^{z}dz' \exp (-i z'^2) = \int_{-\infty}^{0} dx' \exp (-i x'^2) + e^{i\theta}\int_{0}^{R} dr \exp \left(-ir^2 e^{2i\theta}\right).
\end{equation}

The first integral has a principal value expression:

\begin{equation}\label{halfaxis}
\int_{-\infty}^{0} dx' \exp (\pm i x'^2) = \int_{0}^{+\infty} dx' \exp (\pm i x'^2) =\frac{\pi^{1/2}}{2}e^{\pm i\pi/4},
\end{equation}

\noindent and so does the second one in the limit $R\rightarrow\infty$ and for $\sin 2\theta < 0$:

\begin{equation}
\int_{0}^{+\infty} dr \exp \left(-ir^2 e^{2i\theta}\right) = \pm \frac{\pi^{1/2}}{2}e^{-i\theta}.
\end{equation}

\noindent The sign on this last relation depends on the choice of the branch of the square root function, which is implicitly involved in the result\footnote{This can be formally seen through a change of variable in the integral from $z$ to $(\exp 2i\theta)^{1/2} z$.}: the square root of $ e^{2i\theta}$ is $\pm e^{i\theta}$. It is customary to chose the $+$ sign but it is not possible to make this choice consistently throughout the integration domain. Let us discuss this for $\theta \in ]\pi/2,\pi[$ to illustrate the point:

\begin{enumerate}
\item For $\theta \rightarrow \pi/2^+$, the integral should converge to $\int_{0}^{-\infty} dx' \exp (- i x'^2)$ so that the minus sign should be chosen, consistently with Eq.~\eqref{halfaxis}.
\item Conversely, for $\theta \rightarrow \pi^-$, the integral should converge to $\int_{0}^{+\infty} dx' \exp (- i x'^2)$ and consistency now implies to choose the plus sign instead.
\end{enumerate} 

\begin{figure}[ht]
\centering
\includegraphics[width=0.7\linewidth]{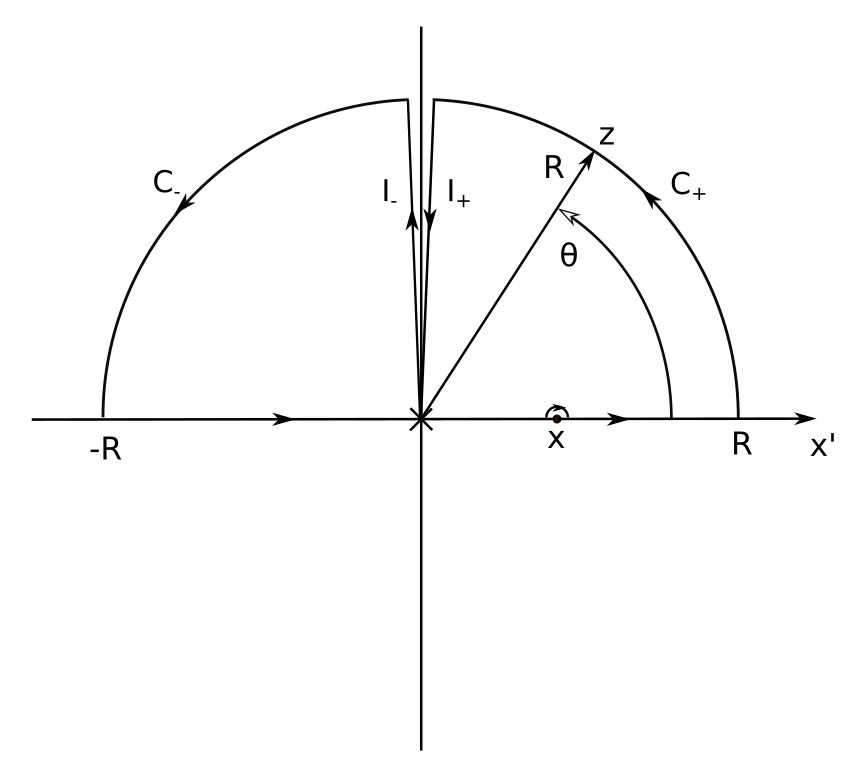}
\caption{\small{Contour of complex integration for forced wave. The whole y axis except $z=0$ is a cut in the complex plane where the second integral in the expression of $Z$ Eq.~\eqref{integral} changes sign. $I_+$ and $I_-$ are located at vanishing angular distance from the $y$ axis.}}
\label{fig:contour2b}
\end{figure}

The same remarks apply to $\theta \in ]-\pi/2,0[$. Note that because the integrals in Eq.~\eqref{halfaxis} are evaluated on the real axis, their sign is unambiguous. As a consequence, a cut has to be made in complex plane to allow for this change of sign convention in the integral. This cut can be freely chosen, but it turns out that the most convenient place is along the $y$ axis. Note that the cut must be made both along positive and negative $y$ values ($2\theta$ increases by $2\pi$ each time $\theta$ increases by $\pi$). At first sight, this seems to imply that the two half-planes separated by the $y$ axis become disconnected, but in fact a contour can go from one half-plane to the other through $z=0$, where the integral in Eq.~\eqref{integral} is zero and its sign can be left unspecified. 

As a consequence of this discussion, the contour sketched on Figure~\ref{fig:contour2b} is chosen to evaluate the left-hand side of Eq.~\eqref{residu} (this evaluation can also be performed by choosing a similar contour in the lower half-plane). More specifically, and defining the contour $C$ as $C=C_+ + C_- + I_+ + I_-$, the objective is to provide relevant bounds to the modulus of the integral

\begin{equation}\label{keycontour}
\mathcal{I}_R(x)=\int_C dz' \frac{Z(x)-Z(z')}{(x-z')^2}.
\end{equation}

In conclusion of this preliminary, by defining the square root of $\exp 2i\theta$ as $\exp i\theta$ for $\theta \in ]-\pi,\pi]$, one can rewrite Eq.~\eqref{integral} as

\begin{equation}\label{integral2}
\int_{-\infty}^{z}dz' \exp (-i z'^2) = \frac{\pi^{1/2}}{2}e^{i\pi/4} +\varepsilon e^{i\theta}\int_{0}^{R} dr \exp \left(-ir^2 e^{2i\theta}\right),
\end{equation}

\noindent with $\varepsilon=+1$ for $\theta\in [0,\pi/2^-]$ and $\varepsilon=-1$ for $\theta\in [\pi/2^+,\pi]$. 

\subsubsection{Evaluation on $\bm{C_+}$ and $\bm{C_-}$}

For any given $x$, the $Z(x)/(z'-x)^2$ term obviously has vanishing contribution as $R\rightarrow +\infty$ on $C_+$ and $C_-$ and is ignored. Similarly, $(z-x)^2\simeq z^2$ in the same limit. One is therefore interested in estimating the asymptotic behavior of $\int_{C_+,C_-} dz\ Z(z)/z^2$.

Note first the following relevant relation, already used in Eq.~(7.51) and (7.52), which follows from an integration by part:

\begin{equation}\label{intpart}
\int dr e^{pr^2} =\left[\frac{e^{pr^2}}{2pr}\right] + \int dr \frac{e^{pr^2}}{2pr^2}.
\end{equation}

\noindent This relation is valid independently of the convergence of the integral on any interval where the integral is finite. Note also that the last term is of the order of $e^{pr^2}/r^3$ for $r >> 1$ (as can be seen from a further integration by part). This result is valid for any complex number $p$.

Let us start with $C_-$. In this range of $\theta$, the integral in Eq.~\eqref{integral2} converges at $R=\infty$ and repeating the reasoning used for Eq.~(7.51) with the help of Eq.~\eqref{intpart} leads to (setting $p=-i e^{2i\theta}$ and remembering that $\varepsilon=-1$)

\begin{equation}
e^{i\theta} \int_{-\infty}^{z}dz' \exp (-i z'^2) = i\frac{e^{-iz^2}}{2z} + \mathcal{O}\left(\frac{e^{-iz^2}}{z^3}\right)
\end{equation}

\noindent and $Z= i/2z$ for $R\gg 1$ (consistently with the analytic continuation of Eq.~[7.51]). As a consequence, $\left|\int_{C_-} dz\ Z(z)/z^2 \right| \rightarrow 0$ as $R\rightarrow +\infty$.

The contribution of $C_+$ requires a little more care due to the divergence of the integral in Eq.~\eqref{integral2} as $e^{-iz^2}$ now diverges at large $R$, but can nevertheless be handled in a similar way. Let us first chose an arbitrary radius $R_0$ such that $1\ll R_0 \ll R$ to rewrite Eq.~\eqref{integral2} as

\begin{eqnarray}\label{integral3}
\int_{-\infty}^{z}dz' \exp (-i z'^2) & =& \frac{\pi^{1/2}}{2}e^{i\pi/4} + e^{i\theta}\int_{0}^{R_0} dr \exp \left(-ir^2 e^{2i\theta}\right)\nonumber\\
& & \qquad + e^{i\theta}\int_{R_0}^{R} dr \exp \left(-ir^2 e^{2i\theta}\right)\nonumber\\
& = & \frac{\pi^{1/2}}{2}e^{i\pi/4} + C_0\nonumber\\
& & \qquad - i\frac{e^{-iz_0^2}}{2z_0} + i\frac{e^{-iz^2}}{2z} + \mathcal{O}\left(\frac{e^{-iz^2}}{z^3}\right),
\end{eqnarray}

\noindent where $C_0$ stands for the integral from $0$ to $R_0$ and $z_0=R_0 e^{i\theta}$. From this expression, it follows again that $Z= i/2z$ to leading order (the other terms are exponentially suppressed by the $e^{iz^2}$ factor in this range of $\theta$). This again leads to $\left|\int_{C_+} dz\ Z(z)/z^2 \right| \rightarrow 0$ as $R\rightarrow +\infty$.

\subsubsection{Evalution on $\bm{I_+}$ and $\bm{I_-}$}

These are the only remaining contributions that may invalidate Eq.~\eqref{residu}. Defining $\mathcal{I}^{\pm}_R(x)$ as the contributions of $I_{\pm}$ to $\mathcal{I}_R(x)$, one has:

\begin{eqnarray}\label{I+}
\mathcal{I}^+_R(x) & = & i \int_R^0 dy \frac{Z(x)-Z(iy)}{(x-iy)^2}\nonumber\\ 
& = & -i \int_0^R dy \frac{Z(x)}{(x-iy)^2} + i \frac{\pi^{1/2}}{2}e^{i\pi/4} \int_0^R dy \frac{e^{iy^2}}{(x-iy)^2} +\nonumber\\
& & \qquad i\int_0^R dy \frac{e^{iy^2}}{(x-iy)^2}\int_0^y dy' e^{-iy^2}
\end{eqnarray}

\noindent Similarly:

\begin{eqnarray}\label{I-}
\mathcal{I}^-_R(x) & = & i \int_0^R dy \frac{Z(x)-Z(iy)}{(x-iy)^2}\nonumber\\ 
& = & i \int_0^R dy \frac{Z(x)}{(x-iy)^2} - i \frac{\pi^{1/2}}{2}e^{i\pi/4} \int_0^R dy \frac{e^{iy^2}}{(x-iy)^2} +\nonumber\\
& & \qquad + i\int_0^R dy \frac{e^{iy^2}}{(x-iy)^2}\int_0^y dy' e^{-iy^2},
\end{eqnarray}

\noindent so that

\begin{equation}\label{JR}
 \mathcal{J}_R(x)\equiv \mathcal{I}^+_R(x) + \mathcal{I}^-_R(x) = 2i \int_0^R dy \frac{e^{iy^2}}{(x-iy)^2}\int_0^y dy' e^{-iy^2}.
\end{equation}

\noindent It is not possible to evaluate this integral with the theorem of residues, as we would be facing the very same problem we are trying to solve. However something has been gained with respect to the initial integral in Eq.~\eqref{residu}, as the pole is no longer on the axis (except for the single point $x=0$) so that this integral can be bounded. Let us first derive an upper bound to the modulus:

\begin{equation}
\left| \mathcal{J}_{+\infty}(x) \right| \le  2\int_0^{+\infty} \frac{dy}{x^2+y^2}\left| \int_0^y dy' e^{-iy^2} \right|.
\end{equation}

\begin{figure}
    \centering
    \begin{subfigure}[h]{0.48\textwidth}
        \includegraphics[width=\textwidth]{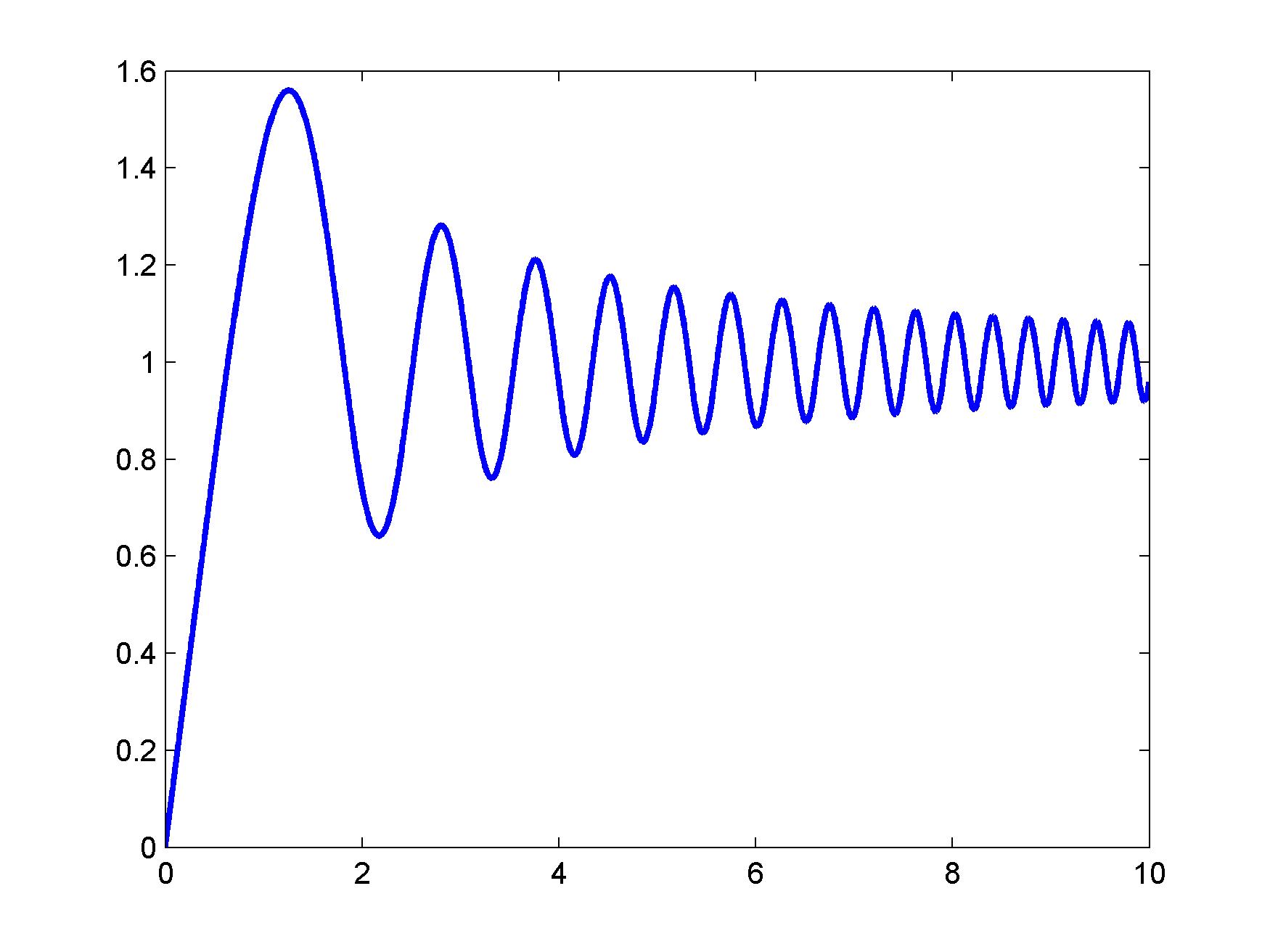}
        \caption{Real part}
        \label{Reexp}
    \end{subfigure}
    \begin{subfigure}[h]{0.48\textwidth}
        \includegraphics[width=\textwidth]{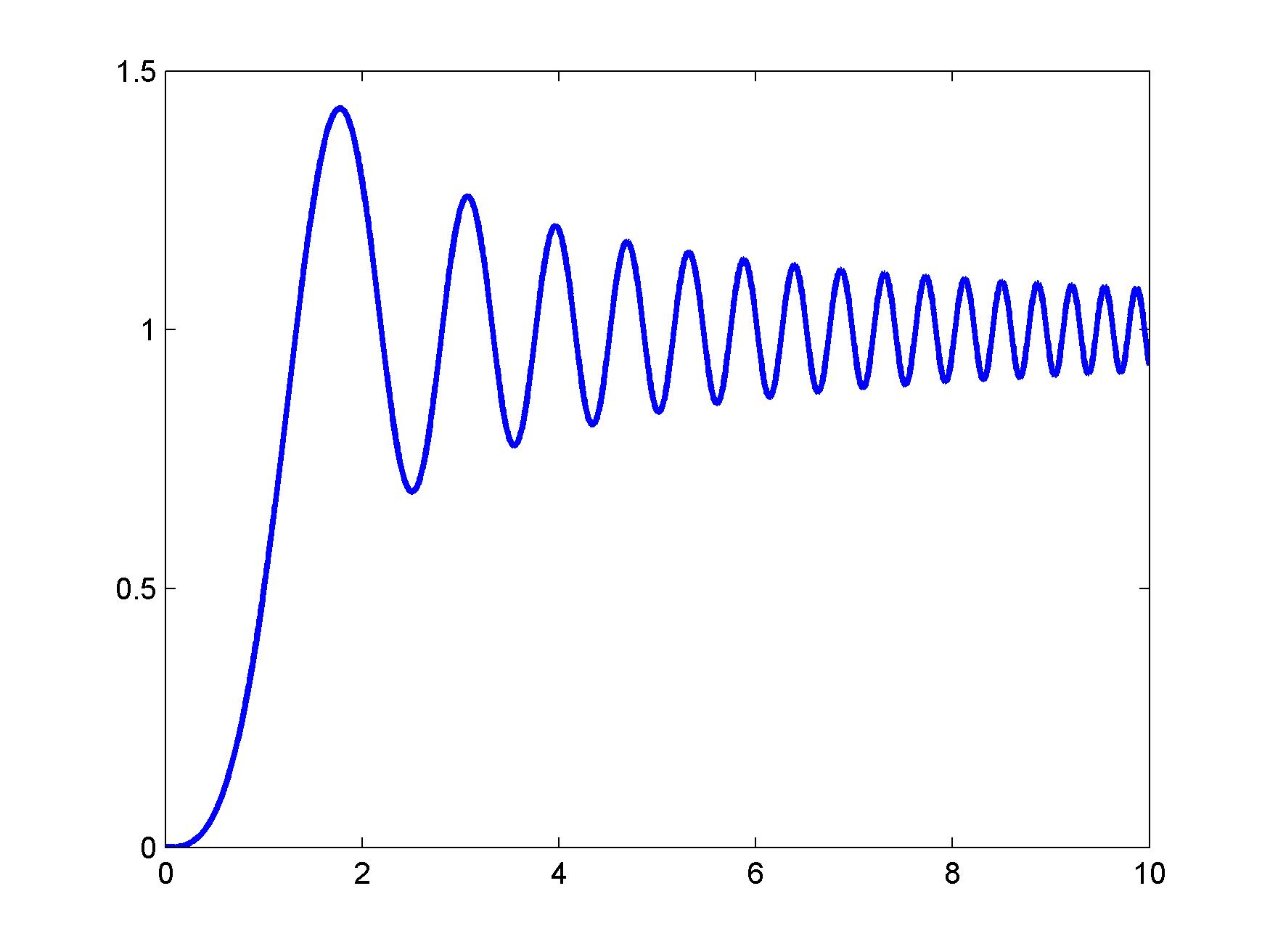}
        \caption{Imaginary part}
        \label{Imexp}
    \end{subfigure}
    \caption{\small{Real and imaginary part of $\int_0^y dy' e^{-iy^2}$ as a function of $y$, normalized to their common asymptotic value $\sqrt{2\pi}/4$.}}\label{fig:erfc}
\end{figure}

Figure~\ref{fig:erfc} shows the behavior of the real and imaginary part of $\int_0^y dy' e^{-iy^2}$ as a function of $y$. This shows that both the real and imaginary part of $\int_0^y dy' e^{-iy^2}$ are bounded by $1.6\times (2\pi)^{1/2}/4 \simeq 3\sqrt{2\pi}/8$ (i.e., about $3/2$ their limiting values at $R\rightarrow +\infty$), so that 

\begin{equation}
\left| \int_0^y dy' e^{-iy^2} \right| \lesssim \frac{3\sqrt{\pi}}{4}\sim 1.
\end{equation}

\noindent From this, on gets

\begin{equation}\label{bound}
\left| \mathcal{J}_{+\infty}(x) \right| \le  \frac{3\sqrt{\pi}}{4}\int_0^{+\infty} \frac{dy}{x^2+y^2}= \frac{3\pi^{3/2}}{8|x|}\simeq \frac{2}{|x|}
\end{equation}

\noindent This shows that the contributions of $I_+ + I_-$ to $ \mathcal{I}_{+\infty}(x)$ are negligible in Eq.~(7.49) when $ |x| \gtrsim 1$.

\begin{figure}[ht]
\centering
    \begin{subfigure}[h]{0.48\textwidth}
        \includegraphics[width=\textwidth]{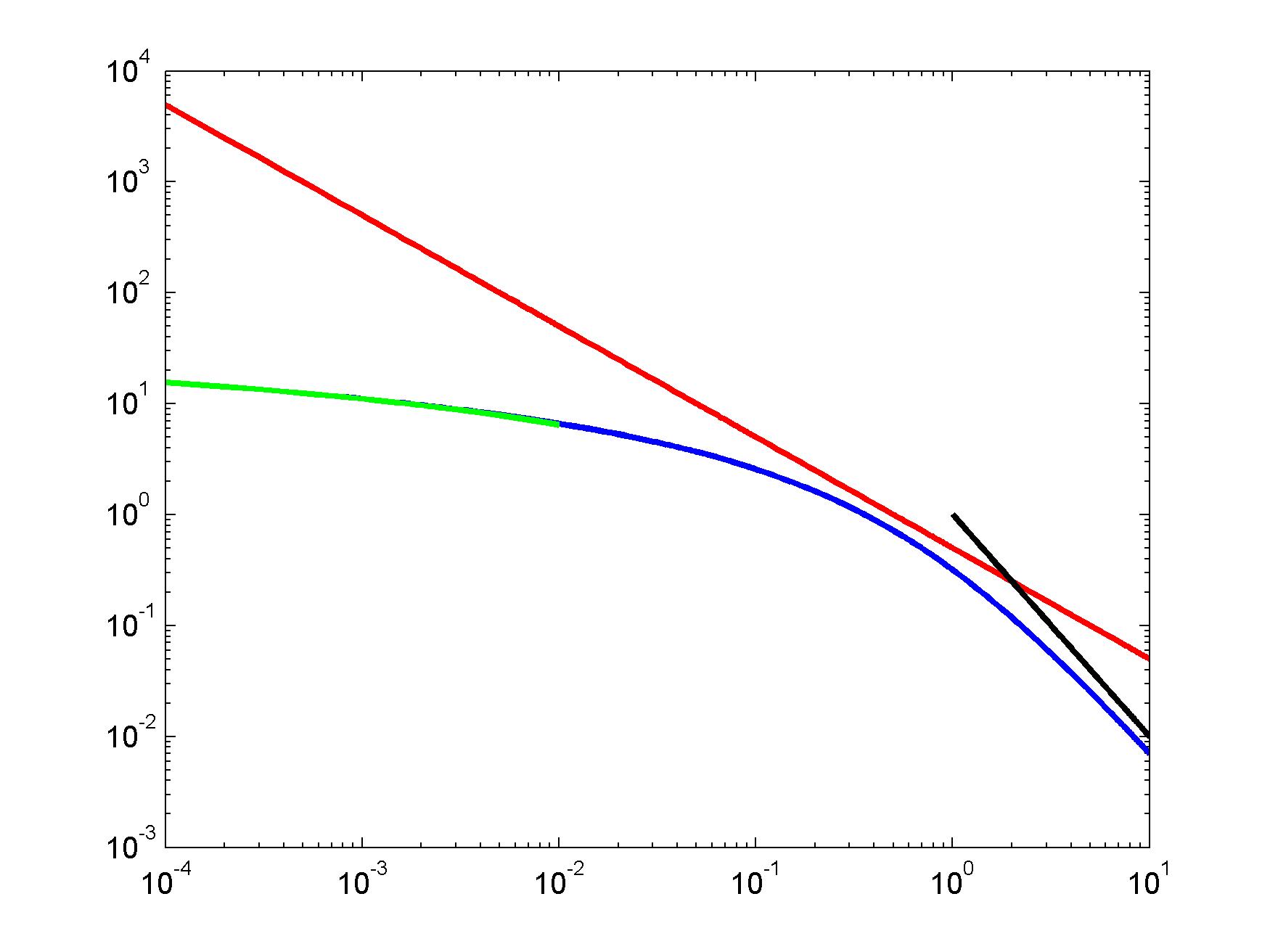}
        \caption{Global behavior}
        \label{loglog}
    \end{subfigure}
    \begin{subfigure}[h]{0.48\textwidth}
        \includegraphics[width=\textwidth]{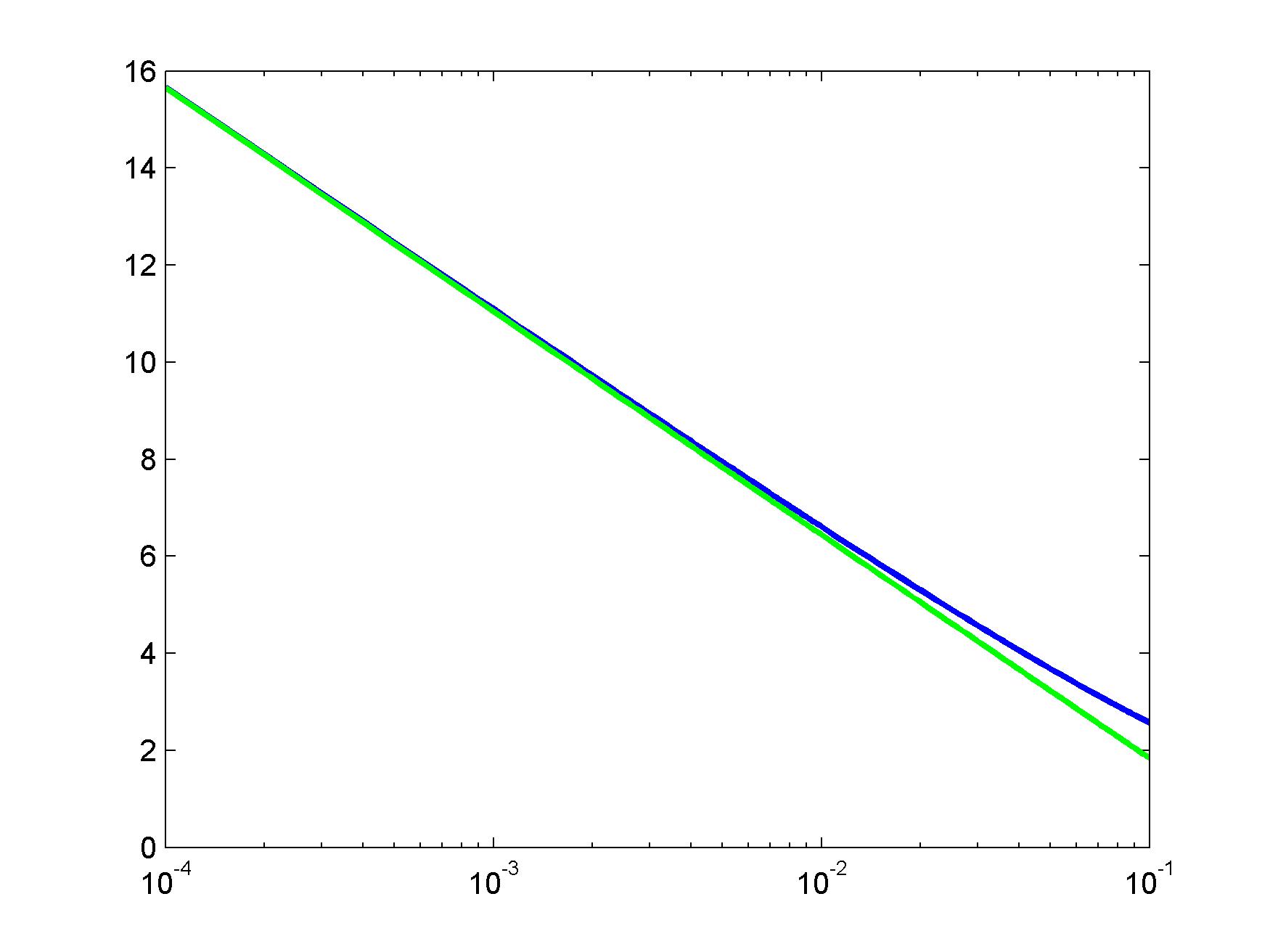}
        \caption{Semilog zoom}
        \label{logx}
    \end{subfigure}
\caption{\small{The left graph shows the behavior of $|\mathcal{J}_{+\infty}(x)|$ as a function of $x$ in log scale (only $x>0$ is shown as the integral depends only on $|x|$). The red line corresponds to $1/2x$, which gives a tighter strict upper bound to this quantity. The black line corresponds to $x^{-2}$, which captures more closely the $x\gg 1$ asymptotic behavior. The green line corresponds to $-2\ln 4x$. The right graph zoom shows that $-2\ln 4x$ is nearly exactly superposed onto the graph of $|\mathcal{J}_{+\infty}(x)|$, confirming the logarithmic divergence as $x\rightarrow 0$.}}
\label{fig:JR}
\end{figure}

Conversely, for $ |x| \ll 1$, one can see that $\mathcal{J}_{+\infty}(x)$ diverges logarithmically with $x$. Indeed, in this limit, the largest contribution to $\mathcal{J}_{+\infty}(x)$ comes from $|y| \lesssim |x|$. Defining $c$ such that $1\gg c \gg |x|$, one has

\begin{equation}\label{JRb}
\left| \mathcal{J}_R(x)\right| \simeq 2 \left| \int_0^c  \frac{ydy}{x^2+y^2}\right| \simeq \left| \ln c^2 - \ln x^2\right| = - 2\ln \left | \frac{x}{c} \right |.
\end{equation}

\noindent It must be noted that one can always find a value of $c$ such that Eq.~\eqref{JRb} is exact for $|x|\rightarrow 0$. Figure~\ref{fig:JR} shows that $c=1/4$ with a high precision.

More generally these two scalings ($|x| \gg 1$ and $|x| \ll 1$) can be tightened with the help of a numerical quadrature of the integrals in Eq.~\eqref{JR}, the results of which are shown on Figure~\ref{fig:JR}.

\bigskip

This somewhat tedious analysis of the domain of validity of Eq.~\eqref{residu} for forced density waves has reached its end. The conclusions stated in section \ref{sec:nut} follow in a straightforward manner, once reverting to the original scaling of $x$ and $Z$ and comparing the magnitude of the neglected contribution of $C$ to the satellite forcing term. 

\clearpage 
\newpage\null\thispagestyle{empty}\newpage
\addcontentsline{toc}{section}{References}
\bibliography{ringbiblio}
\bibliographystyle{plainnat}

\clearpage
\newpage\null\thispagestyle{empty}\newpage
\thispagestyle{empty}
\vfill
\begin{figure}[h]
    \centering
    \includegraphics[scale=0.4]{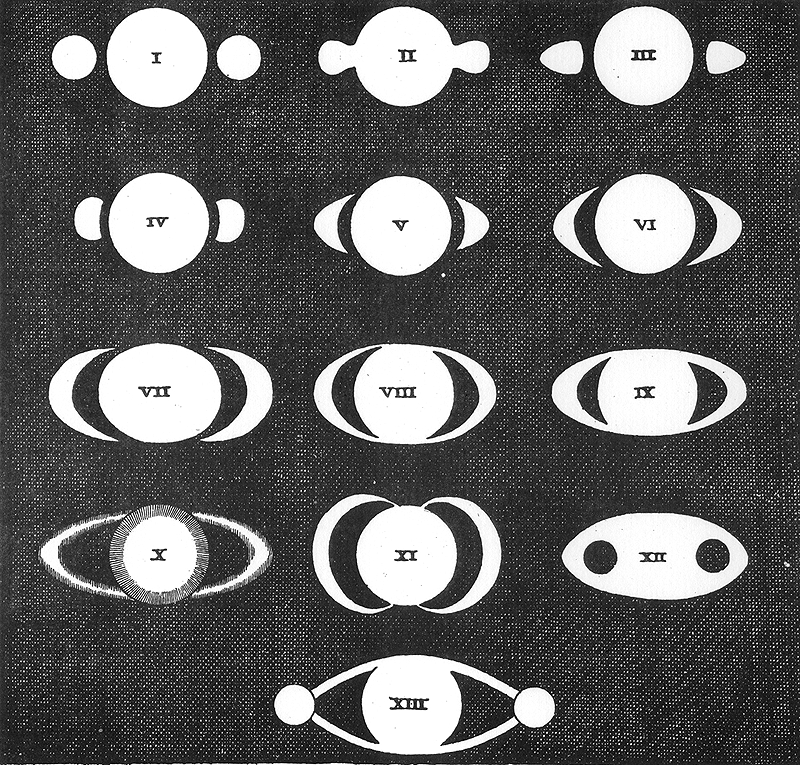}
\end{figure}
\vskip 3truecm
\begin{figure}[h]
    \centering
    \includegraphics[scale=0.4]{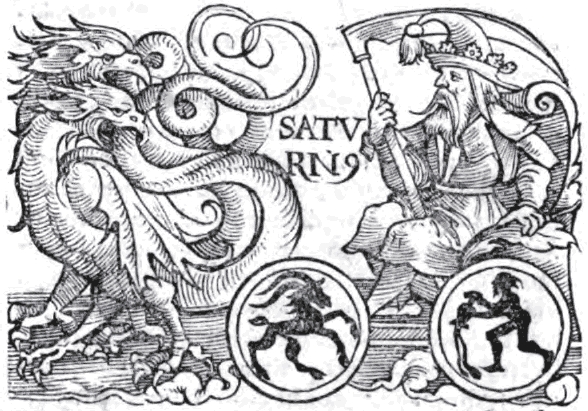}
\end{figure}
\vfill
\end{document}